\documentclass[a4paper,11pt]{article}
\usepackage{jcappub} 
\usepackage{lineno}
\usepackage{booktabs}

\newcommand{\Euclid}{Euclid}
\newcommand{\LSST}{LSST}
\newcommand{\Planck}{Planck}
\newcommand{\threextwo}{$3\times2$pt}
\newcommand{\fourxthree}{$4\times3$pt}
\newcommand{\lcdm}{$\Lambda\textrm{CDM}$}
\newcommand{\SO}{{\em Simons Observatory}}
\newcommand{\ccl}{\texttt{CCL}}
\newcommand{\fastpt}{\texttt{FAST-PT}}

\title{\boldmath From \threextwo\ to \fourxthree: calibrating galaxy bias and intrinsic alignments with galaxy clustering and weak lensing bispectra}

\author[a,b,1]{Raphaël Kou\note{Corresponding author.}}
\author[a]{and Antony Lewis}
\affiliation[a]{Department of Physics \& Astronomy, University of Sussex, Brighton BN1 9QH, UK}
\affiliation[b]{Univ Toulouse, CNES, CNRS, IRAP, Toulouse, France}

\emailAdd{raphael.kou@utoulouse.fr}
\emailAdd{antony@cosmologist.info}

\abstract{Combining galaxy clustering and weak gravitational lensing beyond two-point statistics offers a way to extract additional cosmological information while calibrating astrophysical nuisance parameters. We investigate this approach by combining 1-loop power spectra with tree-level bispectra, including all auto- and cross-bispectra of galaxy clustering and weak lensing, a combination that we refer to as \fourxthree\ by analogy with the well-known \threextwo\ analysis. Modelling two- and three-point statistics consistently, with galaxy bias and intrinsic alignments described by the tidal alignment and tidal torquing model, we perform Fisher forecasts of dark energy constraints for a \Euclid-like survey, with conservative scale cuts. When combined with cosmic microwave background (CMB) data, adding \fourxthree\ bispectra to the \threextwo\ improves the dark energy figure of merit by about $30\%$, a gain that remains stable when using the three, four or five lowest redshift bins of our ten. To extend the analysis to ten tomographic bins, we compress the bispectra into skew-spectra, which improve the figure of merit by $20\%$ with the CMB and by $37\%$ without. We show that most of this improvement comes from the calibration of second-order galaxy bias parameters by three-point statistics, mainly by the galaxy-auto and the galaxy--galaxy--galaxy lensing bispectra. We also find that baryonic feedback is correlated with second-order intrinsic alignment parameters once the amplitude of the matter power spectrum is constrained by the CMB, and that bispectra reduce this correlation by calibrating intrinsic alignments, almost entirely through the cross-bispectra between two galaxy clustering fields and one weak lensing field, a mechanism that could become more valuable with more flexible baryonic feedback models. These results demonstrate the potential of combined galaxy clustering and weak lensing bispectra to improve cosmological constraints from stage-IV galaxy surveys through the calibration of galaxy bias and intrinsic alignments.}

\begin{document}
\maketitle
\flushbottom

\section{Introduction}\label{sec:introduction}

The large-scale distribution of matter in the Universe is one of the most powerful probes of cosmology. On sufficiently large scales, cosmological fields remain close to Gaussian and are accurately described by linear perturbation theory. In this regime, two-point statistics capture most of the available information and provide robust observables for parameter inference. However, a substantial fraction of the cosmological information resides on smaller scales, where gravitational collapse couples Fourier modes and makes the fields significantly harder to model. Exploiting these scales requires going beyond linear theory.

Higher-order perturbation theory provides a natural framework to model mildly non-linear scales, by computing non-linear corrections to the matter and galaxy power spectra. For galaxy clustering, this requires extending the bias expansion relating galaxies to the underlying matter distribution to higher-order, which introduces additional nuisance parameters. These parameters are necessary for accurate modelling, but they degrade cosmological constraints unless they can be tightly calibrated.

Combining galaxy clustering and weak gravitational lensing, in what are commonly referred to as \threextwo\ analyses, partly mitigates this limitation. Weak lensing traces the total matter distribution and is therefore insensitive to galaxy bias, so that cross-correlations between the two probes help calibrate the bias parameters. This complementarity has made \threextwo\ analyses the standard approach of current photometric galaxy surveys. Weak lensing, however, comes with its own astrophysical nuisance: the intrinsic alignment of galaxy shapes, which must be modelled and marginalised over, and whose description also involves higher-order terms when non-linear models are used.

Beyond two-point statistics, non-linear gravitational evolution generically induces non-Gaussian features in cosmological fields, leading to non-vanishing higher-order correlation functions. These statistics arise from the coupling between Fourier modes induced by gravitational evolution and galaxy biasing, and therefore carry information complementary to two-point observables. In this context, bispectra and trispectra, the harmonic-space counterparts of the three- and four-point correlation functions, have emerged as promising probes of cosmology and structure formation.

The combination of power spectra with bispectra, and in some cases trispectra, has been extensively investigated for individual probes. In galaxy clustering, higher-order statistics have been studied both theoretically and observationally~\cite[e.g.][]{Gil-Marin:2014baa,Gil-Marin:2016wya,DiDio:2018unb,Philcox:2021kcw,DAmico:2022osl,Spaar:2023his,NovellMasot:2025fju,Montandon:2025uul,Verdiani:2025jgz,Lu:2025gki,Bakx:2025pop,DESI:2026haa,Euclid:2026glh,Chudaykin:2025aux,Novell-Masot:2026sgg}. Similar developments have been carried out for weak lensing, where both forecasts and analyses of observational data have demonstrated the constraining power of higher-order statistics beyond standard two-point analyses~\cite[e.g.][]{Pyne:2020ijd,Jung:2021weh,Munshi:2021kth,Halder:2021itp,DES:2022oqz,Burger:2023qef,Frugte:2025afv,DES:2025llp,Sugiyama:2025dfc,Porth:2023dzx,Silvestre-Rosello:2025rmt,Porth:2025pem,Sugiyama:2024uqo,Samario-Nava:2025auc}.

Cross-bispectra between different probes have received considerably less attention. Cross-bispectra involving galaxy clustering and weak lensing were first considered in ref.~\cite{Schneider:2005cs}, and the three-point correlation between the positions of two lens galaxies and the shear of a background source galaxy, known as galaxy--galaxy--galaxy lensing (G3L), has since been detected in several analyses~\cite{Simon:2007yg,Simon:2013ps,Linke:2020ssu}. More recently, the modelling and estimation of mixed four-point statistics involving galaxy clustering and weak lensing have been investigated in ref.~\cite{Oel:2026eni}. Cross-bispectra involving other probes have also been measured, including CMB lensing--galaxy~\cite{kou:tel-04732792,Farren:2023yna,Harscouet:2025ksm}, CMB lensing--Lyman-$\alpha$ forest~\cite{Karacayli:2024udo}, and CMB lensing--line intensity mapping~\cite{CHIME:2026rbk} cross-bispectra.

Only a few works have quantified the cosmological constraining power gained by combining galaxy clustering and weak lensing cross-bispectra with the corresponding two-point statistics. To our knowledge, refs.~\cite{Heinrich:2020xub,Halder:2023kfy} are the only studies of this kind: ref.~\cite{Heinrich:2020xub} forecast constraints on modified gravity models combining tree-level power spectra and bispectra, while ref.~\cite{Halder:2023kfy} studied dark energy constraints using integrated three-point correlation functions, defined as the correlation between the mean density within a sky patch and the two-point correlation function measured in the same patch. More recently, the combination of galaxy clustering and weak lensing bispectra has also been considered to constrain baryonic feedback~\cite{Bera:2026ykq}. Combining galaxy clustering and weak lensing at the bispectrum level gives rise to four types of auto- and cross-bispectra, $GGG$, $GGL$, $GLL$ and $LLL$, where $G$ and $L$ denote galaxy clustering and weak lensing respectively. By analogy with the \threextwo\ framework, we refer to this set of observables as \fourxthree\ (as also in ref.~\cite{Bera:2026ykq}).

In this work, we combine 1-loop power spectra with tree-level bispectra of galaxy clustering and weak lensing. At these orders, both sets of observables depend on the same second-order nuisance parameters, but with different sensitivities: second-order terms enter power spectra only through loop corrections, whereas they contribute to bispectra already at tree-level. Bispectra can therefore calibrate these parameters efficiently, and power spectra can in turn benefit from this calibration. This complementarity is well established for galaxy clustering, in particular in spectroscopic surveys, where tree-level bispectra have been shown, both in forecasts and in data analyses, to break the degeneracies between second-order bias parameters and cosmology affecting 1-loop power spectra~\cite{Saito:2014qha,Gil-Marin:2014sta,Chudaykin:2019ock,Agarwal:2020lov,Philcox:2021kcw,Ivanov:2021kcd}. A similar approach was followed in ref.~\cite{Chen:2021vba}, which combined 1-loop power spectra and tree-level bispectra of galaxy clustering and CMB lensing. For intrinsic alignments, this possibility has received much less attention. Intrinsic alignments have generally been treated as a contaminant to three-point weak-lensing statistics. Their contribution has been estimated using simulations and analytical models~\cite{Semboloni:2008xa,Merkel:2014wma}, and methods to mitigate or self-calibrate this contamination have been proposed~\cite{Shi:2010yu,Troxel:2011za,Troxel:2012mx,Troxel:2012pu}. More recently, joint analyses of two- and three-point weak-lensing statistics have been shown to help self-calibrate intrinsic alignments alongside other weak-lensing systematic effects~\cite{Pyne:2020ijd}. Perturbative models of intrinsic-alignment bispectra have also been developed and tested against simulations~\cite{Gomes:2026wki,Bakx:2025uvf}. These developments open the possibility of using bispectra to constrain second-order intrinsic-alignment parameters, in a similar way to how they constrain second-order galaxy-bias parameters. However, this possibility has not yet been explored in a joint analysis of galaxy clustering and weak lensing.

Here, we apply this idea to photometric galaxy clustering and weak lensing, using all their auto- and cross-bispectra. When intrinsic alignments are described by a non-linear model, such as the tidal alignment and tidal torquing (TATT) model, the bispectra depend on both second-order galaxy bias and second-order intrinsic alignment parameters. The cross-bispectra between galaxy clustering and weak lensing, in particular, are sensitive to both sets of parameters at once. Combining the four types of bispectra with the \threextwo\ therefore allows us to calibrate galaxy bias and intrinsic alignments simultaneously, and to quantify how this calibration improves cosmological constraints.

A practical obstacle to such analyses is the size of the data vector. Bispectra depend on three multipoles, and the number of probe combinations and tomographic bin triplets grows rapidly, so that the full \fourxthree\ data vector, and above all its covariance, quickly become prohibitively expensive to compute, even after multipole binning. We therefore investigate skew-spectra, which correlate the product of two fields with a third one, as a compression scheme for the \fourxthree\ data vector, and quantify the information lost in this compression relative to the full bispectrum analysis, with particular emphasis on the dark energy equation of state.

In this work, we present a unified framework in which all auto- and cross-bispectra of galaxy clustering, weak lensing and intrinsic alignments follow from a single tree-level expression, given the linear and quadratic kernels of each tracer, including those of the TATT model. We determine scale cuts for both power spectra and bispectra from the bias that the error of our perturbative model induces on the amplitude of each observable, given the precision of a \Euclid-like survey. We then perform full \fourxthree\ Fisher forecasts with up to five tomographic bins, and show that bispectra improve the dark energy figure of merit by about $30\%$ relative to CMB and \threextwo\ observables, mostly by calibrating second-order galaxy bias parameters. Finally, we propose a strategy to select a small set of skew-spectra that recovers most of this information, and to generalise it to an arbitrary number of tomographic bins, which allows us to forecast the gain for ten tomographic bins. More broadly, this work is a step towards a multi-probe framework that consistently incorporates higher-order statistics, in order to fully exploit the information content of upcoming large-scale structure surveys.

The paper is organised as follows. In section~\ref{sec:modelling}, we present the modelling of the power spectra, bispectra and skew-spectra of galaxy clustering and weak lensing, with the derivation of the intrinsic alignment kernels given in appendix~\ref{appendix:IA_kernels}. Section~\ref{sec:forecast_setup} describes the survey specifications, covariances and Fisher formalism used in our forecasts. In section~\ref{sec:scale_cuts}, we compare our model to non-perturbative predictions and determine the scale cuts used in the analysis. We present the results of the full \fourxthree\ analysis with three tomographic bins in section~\ref{sec:results_3bins}, introduce and validate our skew-spectrum compression strategy in section~\ref{sec:skew_compression}, and apply it to ten tomographic bins in section~\ref{sec:results_10bins}, where we also discuss the limitations of our analysis. We conclude in section~\ref{sec:conclusion}.

\section{Modelling of two- and three-point statistics}\label{sec:modelling}

In this section, we present the modelling of the observables entering our analysis: the angular power spectra of galaxy clustering and weak lensing (\threextwo), their auto- and cross-bispectra (\fourxthree), and the skew-spectra that we use to compress the latter. We first define the projection kernels (section~\ref{sec:kernels}) and the galaxy bias and intrinsic alignment models (section~\ref{sec:bias_ia}), which are common to all observables, before describing power spectra (section~\ref{sec:power_spectra}), bispectra (section~\ref{sec:bispectra}) and skew-spectra (section~\ref{sec:skew_spectra}). The survey specifications, including the tomographic redshift distributions, are given in section~\ref{sec:forecast_setup}.

\subsection{Tracers and projection kernels}\label{sec:kernels}

We denote by $G_i$ and $L_i$ the observed galaxy clustering and weak lensing fields in tomographic bin $i$. The galaxy clustering field is the projected galaxy overdensity $g$. The weak lensing field is the E-mode of the observed shear, which is the sum of the cosmic shear $\gamma$ and of the intrinsic alignment contribution $I$, such that $L_i=\gamma_i+I_i$. Throughout this work, we neglect magnification bias, which we would otherwise need to model consistently in both power spectra and bispectra; we leave its inclusion in bispectra for future work.

Each projected field $X_i$, with $X\in\{g,\gamma,I\}$, is related to a three-dimensional field through a redshift kernel $W_X^i(z)$. The three-dimensional fields are the galaxy overdensity $\delta_g$ for $g$, the matter overdensity $\delta$ for $\gamma$, and the E-mode of the intrinsic shear, $\gamma^I_E$, for $I$. The kernels are
\begin{align}
    W_g^i(z) &= W_I^i(z) = n_i(z), \label{eq:kernel_g}\\
    W_\gamma^i(z) &= \frac{3}{2}\Omega_\mathrm{m}\frac{H_0^2}{c\,H(z)}\,(1+z)\,\chi(z)\int_z^{z_\mathrm{max}}\mathrm{d}z'\,n_i(z')\,\frac{\chi(z')-\chi(z)}{\chi(z')}, \label{eq:kernel_gamma}
\end{align}
where $\chi(z)$ is the comoving distance to redshift $z$, $H(z)$ and $H_0$ are the Hubble rate at redshift $z$ and today, $\Omega_\mathrm{m}$ is the matter density parameter, $c$ is the speed of light, and $n_i(z)$ is the redshift distribution of tomographic bin $i$, normalised to unity and vanishing beyond the maximum redshift of the survey, $z_\mathrm{max}$.
\subsection{Galaxy bias and intrinsic alignment models}\label{sec:bias_ia}

We relate the galaxy overdensity to the matter overdensity through the Eulerian bias expansion~\cite{McDonald:2009dh,Saito:2014qha} implemented in the perturbation theory module of the Core Cosmology Library (\ccl)\footnote{\url{https://github.com/LSSTDESC/CCL}}~\cite{LSSTDarkEnergyScience:2018yem}, which computes the 1-loop terms with \fastpt~\cite{McEwen:2016fjn,Fang:2016wcf,Blazek:2017wbz},

\begin{equation}\label{eq:bias_expansion}
    \delta_g = b_1\delta+\frac{b_2}{2}\delta^2+\frac{b_s}{2}s^2+\frac{b_{3\mathrm{nl}}}{2}\psi_\mathrm{nl}+\frac{b_{k^2}}{2}\nabla^2\delta,
\end{equation}
where $s^2=s_{ij}s_{ij}$, with $s_{ij}$ the tidal tensor, and $\psi_\mathrm{nl}$ is the third-order non-local operator of refs.~\cite{McDonald:2009dh,Desjacques:2016bnm}. The parameters $b_1$, $b_2$, $b_s$, $b_{3\mathrm{nl}}$ and $b_{k^2}$ are free and independent in each tomographic bin.

We model intrinsic alignments with the tidal alignment and tidal torquing (TATT) model~\cite{Blazek:2017wbz}, in which the intrinsic shape tensor is
\begin{equation}\label{eq:tatt}
    \gamma_{ij}^I = C_1s_{ij}+C_{1\delta}\,\delta\, s_{ij}+C_2\left(s_{ik}s_{kj}-\frac{1}{3}\delta^\mathrm{K}_{ij}s^2\right),
\end{equation}
where $\delta^\mathrm{K}_{ij}$ is the Kronecker symbol. Keeping only the first term gives the linear alignment model~\cite{Catelan:2000vm,Hirata:2004gc}, or the non-linear alignment model~\cite{Bridle:2007ft,Hirata:2007np} when the linear matter power spectrum is replaced by its non-linear counterpart. The two other terms are second-order in the density field and therefore contribute to power spectra at 1-loop order and to bispectra at tree-level. We note, however, that the TATT model is not a complete second-order description, as it neglects operators involving the velocity shear tensor. Readers interested in these additional contributions are referred to refs.~\cite{Schmitz:2018rfw,Bakx:2023mld,Bakx:2025uvf,Gomes:2026wki}. We parametrise the redshift dependence of the coefficients as
\begin{align}
    C_1(z)&=-A_1\frac{\bar{C}_1\rho_\mathrm{crit}\Omega_\mathrm{m}}{D(z)}\left(\frac{1+z}{1+z_\mathrm{piv}}\right)^{\eta_1}, \label{eq:C1}\\
    C_{1\delta}(z) &= A_\delta\,C_1(z), \label{eq:C1_delta}\\
    C_2(z) &= 5A_2\frac{\bar{C}_1\rho_\mathrm{crit}\Omega_\mathrm{m}}{D^2(z)}\left(\frac{1+z}{1+z_\mathrm{piv}}\right)^{\eta_2}, \label{eq:C2}
\end{align}
where $D(z)$ is the linear growth factor normalised to unity today, $\rho_\mathrm{crit}$ is the critical density today, $z_\mathrm{piv}=0.62$ is a pivot redshift, and $\bar{C}_1$ is a normalisation constant calibrated on SuperCOSMOS observations~\cite{Hambly:2001yi}, such that $\bar{C}_1\rho_\mathrm{crit}=0.0134$~\cite{Brown:2000gt,Hirata:2004gc,Bridle:2007ft,Krause:2015jqa}. The five parameters $A_1$, $\eta_1$, $A_2$, $\eta_2$ and $A_\delta$ are free and shared by all tomographic bins.

\subsection{Power spectra}\label{sec:power_spectra}

The \threextwo\ data vector contains the angular power spectra $C_{\ell,ij}^{GG}$, $C_{\ell,ij}^{GL}$ and $C_{\ell,ij}^{LL}$, where the first and second tracers are in tomographic bins $i$ and $j$ respectively. Using $L=\gamma+I$, they read
\begin{align}
    C_{\ell,ij}^{GG}&=C_{\ell,ij}^{gg}, \label{eq:cl_GG}\\
    C_{\ell,ij}^{GL} &= C_{\ell,ij}^{g\gamma}+C_{\ell,ij}^{gI}, \label{eq:cl_GL}\\
    C_{\ell,ij}^{LL} &= C_{\ell,ij}^{\gamma\gamma}+C_{\ell,ij}^{\gamma I}+C_{\ell,ij}^{I \gamma}+C_{\ell,ij}^{II}. \label{eq:cl_LL}
\end{align}
Under the Limber approximation~\cite{Limber:1954zz}, each term is given by
\begin{equation}\label{eq:cl_limber}
    C_{\ell,ij}^{XY} = \int_0^{z_\mathrm{max}} \mathrm{d}z\,\frac{H(z)}{c\,\chi^2(z)}\,W_X^i(z)\,W_Y^j(z)\,P_{XY}\!\left(k=\frac{\ell+1/2}{\chi(z)},z\right),
\end{equation}
where $P_{XY}$ is the three-dimensional power spectrum of the fields associated with $X$ and $Y$. We compute angular power spectra with \ccl.

For the cosmic shear auto-spectrum, we use the non-linear matter power spectrum $P_\mathrm{NL}$ predicted by \texttt{HMcode}~\cite{Mead:2020vgs}, which allows us to model this observable down to smaller scales than perturbation theory. We include the suppression of the matter power spectrum by baryonic feedback with the \texttt{HMcode} feedback model, whose strength is set by a single parameter, the AGN heating temperature $T_\mathrm{AGN}$~\cite{Mead:2020vgs}.
All other power spectra, i.e.\ those involving at least one galaxy or intrinsic alignment field, are computed at 1-loop order in Eulerian perturbation theory, using the \fastpt\ interface of \ccl. Expanding the matter overdensity as $\delta=\delta^{(1)}+\delta^{(2)}+\delta^{(3)}+\dots$, the 1-loop matter power spectrum is
\begin{equation}\label{eq:pk_1loop}
    P_{\delta\delta}(k) = P_\mathrm{L}(k)+P_{22}(k)+P_{13}(k),
\end{equation}
where $P_\mathrm{L}$ is the linear matter power spectrum, and $P_{22}$ and $P_{13}$ arise from the correlation of two second-order fields and of a first-order with a third-order field, respectively. The galaxy--galaxy and galaxy--matter power spectra follow from the bias expansion of eq.~\eqref{eq:bias_expansion}, in which second- and third-order operators contribute to the loop corrections.  Power spectra involving intrinsic alignments, $P_{\delta I}$, $P_{II}$ and $P_{gI}$, are computed at 1-loop order with the TATT model of eq.~\eqref{eq:tatt}, but we note that in the current \ccl\ implementation, the galaxy--intrinsic alignment cross-spectrum only includes linear galaxy bias.

\subsection{Bispectra}\label{sec:bispectra}

By analogy with the \threextwo, we refer to as \fourxthree\ the combination of the auto-bispectra $B^{GGG}$ and $B^{LLL}$ and of the cross-bispectra $B^{GGL}$ and $B^{GLL}$. Omitting tomographic indices for clarity, the full-sky bispectrum of three projected fields $X$, $Y$ and $Z$ is defined by
\begin{equation}\label{eq:bispectrum_def}
    \left\langle X_{\ell_1m_1}Y_{\ell_2m_2}Z_{\ell_3m_3}\right\rangle =
    \begin{pmatrix}
      \ell_1 & \ell_2 & \ell_3 \\
      m_1 & m_2 & m_3
    \end{pmatrix}
    B^{XYZ}_{\ell_1\ell_2\ell_3},
\end{equation}
where $X_{\ell m}$ are the spherical harmonic coefficients of $X$ and the matrix is a Wigner 3-$j$ symbol. The reduced bispectrum $b^{XYZ}_{\ell_1\ell_2\ell_3}$ is defined by
\begin{equation}\label{eq:reduced_bispectrum}
    B^{XYZ}_{\ell_1\ell_2\ell_3} = h_{\ell_1\ell_2\ell_3}\,b^{XYZ}_{\ell_1\ell_2\ell_3}, \qquad h_{\ell_1\ell_2\ell_3}=\sqrt{\frac{(2\ell_1+1)(2\ell_2+1)(2\ell_3+1)}{4\pi}}
    \begin{pmatrix}
      \ell_1 & \ell_2 & \ell_3 \\
      0 & 0 & 0
    \end{pmatrix},
\end{equation}
and, under the Limber approximation~\cite{LoVerde:2008re,DiDio:2018unb}, it is related to the three-dimensional bispectrum $B_{XYZ}$ through
\begin{equation}\label{eq:bispectrum_limber}
    b^{XYZ}_{\ell_1\ell_2\ell_3} = \int_0^{z_\mathrm{max}} \mathrm{d}z\,\frac{H^2(z)}{c^2\,\chi^4(z)}\, W_X(z)\,W_Y(z)\,W_Z(z)\, B_{XYZ}\big(k_1,k_2,k_3,z\big),
\end{equation}
with $k_a=(\ell_a+1/2)/\chi(z)$ for $a\in\{1,2,3\}$. As for power spectra, the observed bispectra are obtained by summing over the components of each weak lensing field, e.g.\ $B^{GGL}=B^{gg\gamma}+B^{ggI}$.

We model all three-dimensional bispectra at tree-level in perturbation theory. The bispectrum of any three fields can therefore be expressed using linear and quadratic kernels, $K_1^X$ and $K_2^X$, such that,
\begin{equation}\label{eq:tree_level_bispectrum}
    B_{XYZ}(\mathbf{k}_1,\mathbf{k}_2,\mathbf{k}_3) = 2K_1^X(\mathbf{k}_1)K_1^Y(\mathbf{k}_2)K_2^Z(\mathbf{k}_1,\mathbf{k}_2)P_\mathrm{L}(k_1)P_\mathrm{L}(k_2) + \text{2 cyclic perms.},
\end{equation}
where $\mathbf{k}_1+\mathbf{k}_2+\mathbf{k}_3=\mathbf{0}$ and, in each term, the quadratic kernel is that of the field carrying the third wavevector, evaluated at the two other wavevectors.

We denote by $\mu_{ab}=\hat{\mathbf{k}}_a\cdot\hat{\mathbf{k}}_b$ the cosine of the angle between two wavevectors. Since weak lensing traces the matter distribution, the cosmic shear kernels are those of matter,
\begin{align}
    K_1^\gamma(\mathbf{k}) &= 1, \\
    K_2^\gamma(\mathbf{k}_a,\mathbf{k}_b) &= F_2(\mathbf{k}_a,\mathbf{k}_b) = \frac{5}{7}+\frac{1}{2}\left(\frac{k_a}{k_b}+\frac{k_b}{k_a}\right)\mu_{ab}+\frac{2}{7}\mu_{ab}^2,
\end{align}
and eq.~\eqref{eq:tree_level_bispectrum} reduces to the usual tree-level matter bispectrum. For galaxies, the kernels follow from the bias expansion of eq.~\eqref{eq:bias_expansion}~\cite{Baldauf:2012hs,Saito:2014qha,Desjacques:2016bnm},
\begin{align}
    K_1^g(\mathbf{k}) &= b_1, \\
    K_2^g(\mathbf{k}_a,\mathbf{k}_b) &= b_1F_2(\mathbf{k}_a,\mathbf{k}_b)+\frac{b_2}{2}+\frac{b_s}{2}S_2(\mathbf{k}_a,\mathbf{k}_b), \qquad S_2(\mathbf{k}_a,\mathbf{k}_b) = \mu_{ab}^2-\frac{1}{3}.\label{eq:K2_g}
\end{align}
The third-order galaxy bias parameter $b_{3\mathrm{nl}}$ does not appear in those kernels since the third-order operator $\psi_\mathrm{nl}$ does not contribute at tree-level. The higher-derivative term $b_{k^2}\nabla^2\delta$ is linear in $\delta$ and would formally modify the linear kernel. However, its contribution is suppressed by a factor $(kR)^2$, where $R$ is the typical size of the region from which galaxies form, and becomes relevant on scales similar to those where 1-loop corrections do~\cite{Eggemeier:2018qae}. As is common in tree-level bispectrum analyses, we therefore neglect it, consistently with neglecting 1-loop corrections to the bispectrum. Neither $b_{3\mathrm{nl}}$ nor $b_{k^2}$ therefore enters our bispectrum model. For intrinsic alignments, we derive the kernels of the TATT model in appendix~\ref{appendix:IA_kernels} and obtain
\begin{align}
    K_1^I(\mathbf{k}) &= C_1, \label{eq:K1_IA}\\
    K_2^I(\mathbf{k}_a,\mathbf{k}_b) &= C_1F_2(\mathbf{k}_a,\mathbf{k}_b)+\left(C_{1\delta}+\frac{1}{3}C_2\right)\left(\mu_{ac}^2+\mu_{bc}^2-1\right), \label{eq:K2_IA}
\end{align}
where $\mathbf{k}_c=-(\mathbf{k}_a+\mathbf{k}_b)$ is the wavevector of the intrinsic alignment field. Equivalent bispectra of intrinsic alignments in the TATT model were derived in ref.~\cite{Gomes:2026wki} to model the intrinsic alignment contamination of the cosmic shear three-point correlation function; their expressions differ in form from ours mainly because of different normalisation choices. These bispectra can also be recovered as a restricted case of the tree-level effective field theory (EFT) of intrinsic alignments~\cite{Vlah:2019byq,Vlah:2020ovg,Bakx:2025siw}, which includes an additional second-order operator absent from the TATT model considered here. 

As can be seen from eq.~\eqref{eq:K2_IA}, at a given redshift, the density-weighting and tidal torquing terms only contribute to the bispectra through the combination $C_{1\delta}+C_2/3$. Disentangling $A_\delta$ and $A_2$ with bispectra therefore relies on the different redshift dependence of $C_{1\delta}$ and $C_2$ (eqs.~\eqref{eq:C1_delta} and~\eqref{eq:C2}).

\subsection{Skew-spectra}\label{sec:skew_spectra}

As mentioned in the introduction, the \fourxthree\ data vector is very large, since bispectra depend on three multipoles and must be computed for four probe combinations and all triplets of tomographic bins. This motivates the use of compressed statistics to reduce the size of the data vector and its covariance. To this end, we use skew-spectra, defined as the cross-power spectrum between the product of two fields and a third field. In data analyses, skew-spectra can be measured with the same tools as standard power spectra, while remaining sensitive to the bispectrum.

Skew-spectra have long been used to compress three-point information in cosmology, first in CMB analyses targeting primordial and secondary non-Gaussianity~\cite{Cooray:2001ps,MunshiHeavens2010,2011MNRAS.414.3173M,Munshi:2014tua}, and later for galaxy clustering and weak lensing~\cite{Pratten2011,Schmittfull:2014tca,Munshi:2020ofi,Munshi:2020nfh,MoradinezhadDizgah:2019xun,Schmittfull:2020hoi,Hou:2022rcd}, including in a cosmological analysis of BOSS galaxies~\cite{Hou:2024blc}. They have more recently been extended to cross-bispectra between different fields~\cite{Chakraborty:2022aok,Harscouet:2024hnq}, with measurements of CMB temperature--thermal Sunyaev--Zel'dovich skew-spectra~\cite{Timmons:2017xwv} and of CMB lensing--galaxy skew-spectra~\cite{kou:tel-04732792,Farren:2023yna,Harscouet:2025ksm}.

Following refs.~\cite{Cooray:2001ps,Harscouet:2024hnq}, which proposed filtering the fields entering skew-spectra, we smooth the fields $X$ and $Y$ with Gaussian filters of angular widths $\theta_X$ and $\theta_Y$, whose harmonic-space window is
\begin{equation}\label{eq:gaussian_filter}
    \mathcal{F}_\ell(\theta) = \exp\left[-\frac{\ell(\ell+1)\theta^2}{2}\right].
\end{equation}
We then multiply the smoothed maps and cross-correlate the product with the field $Z$. The resulting skew-spectrum, $S_\ell^{XYZ}(\theta_X,\theta_Y)$, is related to the bispectrum through
\begin{equation}\label{eq:skew_spectrum}
    S_{\ell}^{XYZ}(\theta_X,\theta_Y) = \sum_{\ell_1,\ell_2}\sqrt{\frac{(2\ell_1+1)(2\ell_2+1)}{4\pi(2\ell+1)}}
    \begin{pmatrix}
      \ell_1 & \ell_2 & \ell \\
      0 & 0 & 0
    \end{pmatrix}
    \mathcal{F}_{\ell_1}(\theta_X)\,\mathcal{F}_{\ell_2}(\theta_Y)\,B_{\ell_1\ell_2\ell}^{XYZ}.
\end{equation}
Each skew-spectrum multipole is therefore a weighted sum over all triangle configurations sharing one side $\ell$, so that skew-spectra are a linear compression of bispectra. The filters act as low-pass filters: $\mathcal{F}_{\ell_1}(\theta_X)$ retains the multipoles $\ell_1\lesssim1/\theta_X$ and suppresses higher ones, so that increasing $\theta_X$ shifts the weight towards configurations with smaller $\ell_1$, while $\theta_X=0$ corresponds to no smoothing. We describe how we choose the fields and filter widths in section~\ref{sec:skew_compression}. In practice, we restrict the sums over $\ell_1$ and $\ell_2$ in eq.~\eqref{eq:skew_spectrum} to $\ell_1,\ell_2\leq\ell_\mathrm{max}^{XYZ}$, and only use skew-spectrum multipoles $\ell\leq\ell_\mathrm{max}^{XYZ}$, where $\ell_\mathrm{max}^{XYZ}$ is the maximum multipole of the bispectrum $B^{XYZ}_{\ell_1\ell_2\ell_3}$ determined in section~\ref{sec:scale_cuts}. Only triangle configurations within our scale cuts therefore contribute to skew-spectra. In a data analysis, this corresponds to applying a top-hat filter in harmonic space to the fields $X$ and $Y$, in addition to the Gaussian filters, before multiplying them.

\section{Forecast setup}\label{sec:forecast_setup}

In this section, we describe the survey configuration, the parameters and their fiducial values, the covariances and the Fisher formalism used to forecast constraints from the combination of CMB, \threextwo\ and \fourxthree\ observables. The multipole ranges used for \threextwo\ and \fourxthree\ observables are determined in section~\ref{sec:scale_cuts}.

\subsection{Survey specifications}\label{sec:survey}

\begin{figure}[htbp]
\centering
\includegraphics{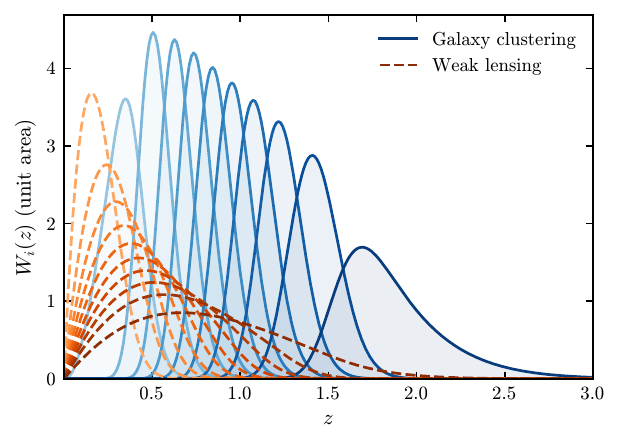}
\caption{Galaxy clustering (solid lines) and weak lensing (dashed lines) kernels of the ten tomographic bins of our \Euclid-like survey, for the fiducial cosmology. Each kernel is normalised to unit area for visualisation purposes.\label{fig:kernels}}
\end{figure}

We consider a photometric galaxy survey with the specifications of the \Euclid\ forecasts of ref.~\cite{Euclid:2019clj}, as an example of the photometric galaxy clustering and weak lensing analyses of upcoming stage-IV surveys, such as \Euclid~\cite{Euclid:2024yrr} and the Vera C. Rubin Observatory Legacy Survey of Space and Time (\LSST)~\cite{LSST:2008ijt}. The overall redshift distribution of the galaxies is
\begin{equation}\label{eq:nz}
    n(z) \propto \left(\frac{z}{z_0}\right)^2\exp\left[-\left(\frac{z}{z_0}\right)^{3/2}\right],
\end{equation}
with $z_0=0.9/\sqrt{2}$. We split this distribution into ten equipopulated tomographic bins spanning the range $0.001<z<2.5$, numbered from 1 to 10 by increasing redshift, and obtain the redshift distributions $n_i(z)$ by convolving each bin with the photometric redshift uncertainties described in ref.~\cite{Euclid:2019clj}. We evaluate these distributions, and the integrals of eqs.~\eqref{eq:kernel_gamma}, \eqref{eq:cl_limber} and~\eqref{eq:bispectrum_limber}, up to $z_\mathrm{max}=3$, beyond which they are negligible. The resulting galaxy clustering and weak lensing kernels are shown in figure~\ref{fig:kernels}. The weak lensing kernels are much broader than the galaxy clustering ones and, even for high-redshift sources, receive contributions from low redshifts, where the matter field is strongly non-linear.

Each tomographic bin has a galaxy surface density $\bar{n}=3\,\mathrm{arcmin}^{-2}$, and we assume an intrinsic ellipticity dispersion $\epsilon=0.3$ per ellipticity component. We assume that the survey covers a sky fraction $f_\mathrm{sky}=0.36$, corresponding to $15\,000\,\mathrm{deg}^2$. We use the same galaxy sample for galaxy clustering and weak lensing, so that both probes share the same tomographic bins.

\subsection{Parameters and fiducial values}\label{sec:parameters}

One of the main goals of upcoming galaxy surveys is to determine whether the dark energy equation of state evolves with time. We parametrise it with the Chevallier--Polarski--Linder (CPL) form~\cite{Chevallier:2000qy,Linder:2002et},
\begin{equation}\label{eq:cpl}
    w(a)=w_0+(1-a)\,w_a.
\end{equation}
Our cosmological parameters are the physical baryon and cold dark matter densities, $\Omega_\mathrm{b}h^2$ and $\Omega_\mathrm{c}h^2$, the Hubble constant $H_0$, the spectral index $n_\mathrm{s}$, the combination $10^9A_\mathrm{s}e^{-2\tau}$ of the amplitude of primordial fluctuations $A_\mathrm{s}$ and the optical depth to reionisation $\tau$, the optical depth itself, and the dark energy parameters $w_0$ and $w_a$. We fix the sum of neutrino masses to $0.06\,\mathrm{eV}$, assuming a single massive neutrino, and the effective number of relativistic species to $N_\mathrm{eff}=3.046$.
We adopt the \Planck\ 2018 best-fit cosmology~\cite{Planck:2018vyg} as our fiducial model, together with $w_0=-1$ and $w_a=0$. We also vary the baryonic feedback parameter $\log_{10} T_\mathrm{AGN}$, with $T_\mathrm{AGN}$ expressed in kelvin, which only enters the weak lensing power spectra in our model. We take a fiducial value of $7.8$ and a Gaussian prior of width $0.3$, covering the range of values calibrated on hydrodynamical simulations in ref.~\cite{Mead:2020vgs}. The fiducial values of the cosmological, intrinsic alignment and baryonic feedback parameters are listed in table~\ref{tab:constraints}.

For each tomographic bin, the fiducial linear bias is $b_1=\sqrt{1+\bar{z}_i}$, where $\bar{z}_i$ is the mean redshift of bin $i$, following ref.~\cite{Euclid:2019clj}. We set the fiducial values of the higher-order bias parameters with relations
obtained for dark matter haloes: $b_2$ follows the fit to simulations of
ref.~\cite{Lazeyras:2015lgp},
\begin{equation}
    b_2 = 0.412 - 2.143\,b_1 + 0.929\,b_1^2 + 0.008\,b_1^3 ,
\end{equation}
while $b_s = -4(b_1-1)/7$ and $b_{3\mathrm{nl}} = b_1-1$ follow the co-evolution
predictions~\cite{Chan:2012jj,Baldauf:2012hs,Saito:2014qha}.
Following~\cite{DES:2020yyz}, we take $b_{k^2}=0$ as a fiducial value. Finally, the fiducial values of the intrinsic alignment parameters are taken from the \Euclid\ analysis of ref.~\cite{Euclid:2026klv}.

\subsection{Covariances}\label{sec:covariances}

We model the covariance between two angular power spectra $C_{\ell}^{AB}$ and $C_{\ell'}^{CD}$ as Gaussian,
\begin{equation}\label{eq:covariance}
    \mathrm{Cov}\left(C_\ell^{AB},C_{\ell'}^{CD}\right) = \frac{\delta^\mathrm{K}_{\ell\ell'}}{(2\ell+1) f_\mathrm{sky}}
    \left(\tilde{C}_\ell^{AC}\tilde{C}_\ell^{BD}+\tilde{C}_\ell^{AD}\tilde{C}_\ell^{BC}\right),
\end{equation}
where $A$, $B$, $C$ and $D$ each denote a tracer in a given tomographic bin, and $\tilde{C}_\ell^{AB}=C_\ell^{AB}+N_\ell^{AB}$ includes the noise power spectrum $N_\ell^{AB}$. The noise only contributes to auto-correlations of a given tracer in a given bin: $N_\ell^{G_iG_j}=\delta^\mathrm{K}_{ij}/\bar{n}_i$ for galaxy clustering and $N_\ell^{L_iL_j}=\delta^\mathrm{K}_{ij}\,\epsilon^2/\bar{n}_i$ for weak lensing, while $N_\ell^{GL}=0$, where $\bar{n}_i$ is the angular number density of galaxies in bin $i$. We bin power spectra in linearly spaced multipole bins $b$ of width $\Delta\ell=30$, using the mode-weighted average
\begin{equation}\label{eq:binned_cl}
    C_b^{AB} = \frac{1}{N_b}\sum_{\ell\in b}(2\ell+1)\,C_\ell^{AB}, \qquad N_b=\sum_{\ell\in b}(2\ell+1),
\end{equation}
whose covariance follows exactly from eq.~\eqref{eq:covariance},
\begin{equation}\label{eq:binned_cl_covariance}
    \mathrm{Cov}\left(C_b^{AB},C_{b'}^{CD}\right) = \frac{\delta^\mathrm{K}_{bb'}}{N_b^2\,f_\mathrm{sky}}\sum_{\ell\in b}(2\ell+1)\left(\tilde{C}_\ell^{AC}\tilde{C}_\ell^{BD}+\tilde{C}_\ell^{AD}\tilde{C}_\ell^{BC}\right).
\end{equation}

Similarly, we model the covariance of bispectra as Gaussian,
\begin{equation}\label{eq:covariance_bispectra}
\begin{split}
    f_\mathrm{sky}\,\mathrm{Cov}\left(B_{\ell_1\ell_2\ell_3}^{XYZ},B_{\ell_1'\ell_2'\ell_3'}^{X'Y'Z'}\right) ={}&
      \delta^\mathrm{K}_{\ell_1\ell_1'}\tilde{C}_{\ell_1}^{XX'}\left(\delta^\mathrm{K}_{\ell_2\ell_2'}\delta^\mathrm{K}_{\ell_3\ell_3'}\tilde{C}_{\ell_2}^{YY'}\tilde{C}_{\ell_3}^{ZZ'}+\delta^\mathrm{K}_{\ell_2\ell_3'}\delta^\mathrm{K}_{\ell_3\ell_2'}\tilde{C}_{\ell_2}^{YZ'}\tilde{C}_{\ell_3}^{ZY'}\right) \\
    &+\delta^\mathrm{K}_{\ell_1\ell_2'}\tilde{C}_{\ell_1}^{XY'}\left(\delta^\mathrm{K}_{\ell_2\ell_1'}\delta^\mathrm{K}_{\ell_3\ell_3'}\tilde{C}_{\ell_2}^{YX'}\tilde{C}_{\ell_3}^{ZZ'}+\delta^\mathrm{K}_{\ell_2\ell_3'}\delta^\mathrm{K}_{\ell_3\ell_1'}\tilde{C}_{\ell_2}^{YZ'}\tilde{C}_{\ell_3}^{ZX'}\right) \\
    &+\delta^\mathrm{K}_{\ell_1\ell_3'}\tilde{C}_{\ell_1}^{XZ'}\left(\delta^\mathrm{K}_{\ell_2\ell_1'}\delta^\mathrm{K}_{\ell_3\ell_2'}\tilde{C}_{\ell_2}^{YX'}\tilde{C}_{\ell_3}^{ZY'}+\delta^\mathrm{K}_{\ell_2\ell_2'}\delta^\mathrm{K}_{\ell_3\ell_1'}\tilde{C}_{\ell_2}^{YY'}\tilde{C}_{\ell_3}^{ZX'}\right),
\end{split}
\end{equation}
where the six terms correspond to the possible pairings between the fields of the two bispectra. To keep the size of the data vector manageable, we also bin bispectra in linearly spaced multipole bins of width $\Delta\ell=30$. Following ref.~\cite{Bucher:2015ura}, we define the binned bispectrum in the multipole bins $I_1$, $I_2$ and $I_3$ as
\begin{equation}\label{eq:binned_bispectrum}
    B^{XYZ}_{I_1I_2I_3} = \frac{1}{\Xi_{I_1I_2I_3}}\sum_{\ell_1\in I_1}\sum_{\ell_2\in I_2}\sum_{\ell_3\in I_3} h^2_{\ell_1\ell_2\ell_3}\,b^{XYZ}_{\ell_1\ell_2\ell_3},
\end{equation}
where $\Xi_{I_1I_2I_3}$ is the number of multipole triplets in the bins that satisfy the triangle inequality and have an even sum $\ell_1+\ell_2+\ell_3$. This quantity is the average over the bin of $h^2_{\ell_1\ell_2\ell_3}b^{XYZ}_{\ell_1\ell_2\ell_3}=h_{\ell_1\ell_2\ell_3}B^{XYZ}_{\ell_1\ell_2\ell_3}$, and its covariance follows from eq.~\eqref{eq:covariance_bispectra} as
\begin{equation}\label{eq:binned_covariance}
    \mathrm{Cov}\left(B^{XYZ}_{I_1I_2I_3},B^{X'Y'Z'}_{I'_1I'_2I'_3}\right) = \frac{1}{\Xi_{I_1I_2I_3}\Xi_{I'_1I'_2I'_3}}\sum_{\ell_a\in I_a}\sum_{\ell'_a\in I'_a} h_{\ell_1\ell_2\ell_3}h_{\ell'_1\ell'_2\ell'_3}\,\mathrm{Cov}\left(B^{XYZ}_{\ell_1\ell_2\ell_3},B^{X'Y'Z'}_{\ell'_1\ell'_2\ell'_3}\right).
\end{equation}

Skew-spectra are linear combinations of bispectra (eq.~\eqref{eq:skew_spectrum}), so we obtain their covariance by applying the same linear transformation to eq.~\eqref{eq:covariance_bispectra}. Contrary to power spectra, this covariance is not diagonal in multipoles, since a given triangle configuration contributes to several skew-spectrum multipoles. We include the cross-covariance between all skew-spectra entering the data vector. For all statistics, our forecasts use the multipoles from $\ell_\mathrm{min}=10$ up to the scale cuts determined in section~\ref{sec:scale_cuts}.

Our covariances neglect non-Gaussian contributions, which arise from the connected higher-order correlation functions of the fields and from super-sample covariance~\cite{Takada:2013wfa,Lacasa:2016yva,Beauchamps:2021fhb,Euclid:2023ove}, and which grow on small scales. We also neglect the cross-covariance between power spectra and bispectra, as commonly done in forecasts combining them~\cite{Heinrich:2020xub,Frugte:2025afv,Euclid:2025hlc,Euclid:2026tee,Randrianjanahary:2026ozt}, and leave its inclusion for future work. Finally, shot noise only enters our covariances, whereas for Poisson-sampled galaxies it also contributes to the mean bispectra, through terms that depend on the cosmological and nuisance parameters via the power spectra. We show in section~\ref{sec:robustness} that including these contributions, as well as realistic deviations from Poisson statistics, has a negligible impact on our results.

\subsection{CMB data}\label{sec:cmb}

We combine galaxy clustering and weak lensing observables with the CMB temperature and polarisation power spectra $C_\ell^{TT}$, $C_\ell^{TE}$ and $C_\ell^{EE}$, which mainly constrain the parameters describing the early Universe. We consider a combination of \SO~\cite{SimonsObservatory:2018koc,SimonsObservatory:2025wwn} and \Planck\ data, and compute the spectra with \texttt{CAMB}~\cite{Lewis:1999bs,Howlett12,Lewis:2026mif}. Their covariance follows eq.~\eqref{eq:covariance}. For \SO, we use the publicly available foreground-cleaned noise curves of the baseline configuration. For \Planck, we model the noise power spectra as
\begin{equation}\label{eq:Nl}
    N_\ell^{XX} = \Delta_X^2\exp\left[\frac{\ell(\ell+1)\,\theta_\mathrm{FWHM}^2}{8\ln2}\right],
\end{equation}
where $X\in\{T,E\}$, with noise levels $\Delta_T=23\,\mu\mathrm{K}\,\mathrm{arcmin}$ and $\Delta_E=42\,\mu\mathrm{K}\,\mathrm{arcmin}$, and a beam of full width at half maximum $\theta_\mathrm{FWHM}=7\,\mathrm{arcmin}$.

For \SO, we use the multipole ranges $40\leq\ell\leq3000$ for $C_\ell^{TT}$ and $40\leq\ell\leq5000$ for $C_\ell^{TE}$ and $C_\ell^{EE}$, the more conservative range for temperature reflecting the larger foreground contamination on small scales. For \Planck, we use $40\leq\ell\leq2500$ for all spectra and replace the information from the lowest multipoles by a Gaussian prior on the optical depth to reionisation, $\sigma(\tau)=0.0073$. To avoid modelling the covariance between the two experiments where their footprints overlap, we use a sky fraction of $40\%$ for \SO\ and of $30\%$ for \Planck, the latter excluding the region observed by \SO.

\subsection{Fisher formalism}\label{sec:fisher}

We forecast parameter constraints with the Fisher formalism~\cite{Tegmark:1996bz}. For a data vector $\mathcal{D}$ with covariance $\mathrm{Cov}$, the Fisher matrix is
\begin{equation}\label{eq:fisher}
    F_{\alpha\beta} = \left(\frac{\partial \mathcal{D}}{\partial p_\alpha}\right)^\mathrm{T}\mathrm{Cov}^{-1}\,\frac{\partial \mathcal{D}}{\partial p_\beta},
\end{equation}
where $p_\alpha$ and $p_\beta$ are two of the parameters of the analysis, and the derivatives are evaluated at the fiducial model. The forecast marginalised $1\sigma$ uncertainty on parameter $p_\alpha$ is then
\begin{equation}\label{eq:sigma_x}
    \sigma(p_\alpha) = \sqrt{\left(F^{-1}\right)_{\alpha\alpha}}.
\end{equation}
We treat the CMB, \threextwo\ and \fourxthree\ observables as independent and combine them by summing their Fisher matrices. For \threextwo\ and \fourxthree, this follows from neglecting the cross-covariance between power spectra and bispectra. For the CMB, it relies on the result of ref.~\cite{Kou:2025hvg}, which found that the covariance between CMB and \threextwo\ observables has a limited impact on cosmological constraints. The Gaussian prior on $\log_{10}T_\mathrm{AGN}$ is added to the Fisher matrix in all cases.

Our baseline results combine galaxy clustering and weak lensing with the CMB. For ten tomographic bins, we also give constraints without the CMB. Since galaxy clustering and weak lensing depend on $A_\mathrm{s}$ and $\tau$ only through $A_\mathrm{s}$, we then fix $\tau$ and use $10^9A_\mathrm{s}$ instead of $10^9A_\mathrm{s}e^{-2\tau}$, and we add a Gaussian prior from big bang nucleosynthesis (BBN) on the baryon density, $\sigma(\Omega_\mathrm{b}h^2)=0.00036$~\cite{Mossa:2020gjc}, which is otherwise poorly constrained by these observables.

To quantify dark energy constraints, we use the figure of merit defined in refs.~\cite{Albrecht:2006um,Euclid:2019clj},
\begin{equation}\label{eq:fom}
    \mathrm{FoM}=\sqrt{\det\left(\tilde{F}_{w_0w_a}\right)},
\end{equation}
where $\tilde{F}_{w_0w_a}=\left[\left(F^{-1}\right)_{w_0w_a}\right]^{-1}$ is the Fisher matrix of $w_0$ and $w_a$ after marginalisation over all other parameters, obtained by inverting the $2\times2$ block of $F^{-1}$ corresponding to these two parameters. The figure of merit is a useful way of quantifying how tight the constraints on dark energy parameters $w_0$, $w_a$ are, taking into account the correlation between the two parameters.

\section{Scale cuts}\label{sec:scale_cuts}

Our modelling is restricted to mildly non-linear scales: galaxy clustering and intrinsic alignment power spectra are computed at 1-loop order, and all bispectra at tree-level. These predictions break down on small scales, and we need to estimate where this happens before performing forecasts. Since we aim to forecast the constraints obtainable from a given survey, we define the scale cuts as the scales beyond which the inaccuracy of our model would significantly bias the parameters it is used to constrain, given the precision of this survey. In this section, we compare our perturbative predictions to more accurate, non-perturbative predictions, and derive scale cuts adapted to our \Euclid-like survey. Our goal is to obtain physically motivated estimates of the scale cuts rather than choosing them arbitrarily, and we check in section~\ref{sec:robustness} that the bias they induce on cosmological parameters remains small.

\subsection{Reference predictions}\label{sec:reference_predictions}\label{sec:sim_measurements}

We compare our predictions to non-perturbative reference predictions, computed at the fiducial cosmology of our forecasts, so that any difference between the two is a model error. For power spectra, our reference is the non-linear matter power spectrum predicted by \texttt{HMcode}~\cite{Mead:2020vgs}. For bispectra, we use the \texttt{BiHalofit} fitting function~\cite{Takahashi:2019hth}, calibrated on $N$-body simulations, which reproduces the matter bispectrum of these simulations within about $10\%$ up to $k=3\,h\,\mathrm{Mpc}^{-1}$ for $0\leq z\leq3$ for a \Planck\ 2015 cosmology. These references only describe the matter field, and we therefore compute our predictions keeping only terms proportional to matter quantities, and using those as proxies to determine scales where modelling errors become significant. We hence compute both our predictions and the references with the fiducial values of the linear galaxy bias and intrinsic alignment parameters of section~\ref{sec:parameters}, i.e.\ $b_1$, $A_1$ and $\eta_1$, and set all higher-order parameters, $b_2$, $b_s$, $b_{3\mathrm{nl}}$, $b_{k^2}$, $A_2$ and $A_\delta$, to zero. The galaxy clustering field is then $b_1\delta$, the intrinsic shear follows the linear alignment term $C_1s_{ij}$ of eq.~\eqref{eq:tatt}, and the quadratic kernels of eqs.~\eqref{eq:K2_g}--\eqref{eq:K2_IA} reduce to those of matter, multiplied by the linear kernels. Table~\ref{tab:reference_predictions} summarises the resulting comparison for each observable. For the bispectra, we replace the matter part of the tree-level bispectrum, $2F_2(\mathbf{k}_1,\mathbf{k}_2)P_\mathrm{L}(k_1)P_\mathrm{L}(k_2)+\text{2 cyclic perms.}$, by the \texttt{BiHalofit} prediction, multiplied by the product of the linear kernels of the three fields; for weak lensing, this applies to both the cosmic shear and the intrinsic alignment components. Weak lensing power spectra do not appear in table~\ref{tab:reference_predictions}: since we model their cosmic shear contribution with \texttt{HMcode}, we derive their scale cuts with a different criterion, described in section~\ref{sec:wl_cuts}.

\begin{table}[htbp]
  \centering
  \caption{Predictions compared to derive the scale cuts, for linear galaxy bias and intrinsic alignments (all higher-order parameters set to zero). $\gamma$ and $I$ denote the cosmic shear and intrinsic alignment components of weak lensing.\label{tab:reference_predictions}}
  \begin{tabular}{lll}
    \toprule
    Observable & Model & Reference \\
    \midrule
    $C_\ell^{GG}$ & 1-loop: $b_1^2P_{\delta\delta}$ (eq.~\eqref{eq:pk_1loop}) & $b_1^2P_\mathrm{NL}$ (\texttt{HMcode}) \\
    $B^{GGG}$ & tree-level & $b_1^3B_\mathrm{NL}$ (\texttt{BiHalofit}) \\
    $B^{LLL}$ & tree-level & $\gamma$ and $I$: $B_\mathrm{NL}$ (\texttt{BiHalofit}) \\
    \bottomrule
  \end{tabular}
\end{table}

This comparison validates our modelling of the non-linear evolution of the matter field, but not that of the higher-order galaxy bias and intrinsic alignment terms. These terms are described by the same perturbative expansion as the matter field, and we assume that they remain accurate on the scales where the matter predictions are. Testing this assumption explicitly would require galaxy mocks including intrinsic alignments, which is beyond the scope of this work.

\subsection{Figure of bias criterion}\label{sec:recipe}

For a given observable, we order the binned data points by multipole, and bispectrum configurations by the largest multipole bin of each triangle, so that all multipoles of all triangles retained satisfy $\ell\leq\ell_\mathrm{cut}$. We denote by $\mathcal{D}_i$ the model prediction for data point $i$, by
\begin{equation}\label{eq:residual}
    \Delta_i = \mathcal{D}^\mathrm{ref}_i-\mathcal{D}_i
\end{equation}
its difference with the reference, and by $\sigma_i$ its uncertainty for our \Euclid-like survey, given by the Gaussian covariance of section~\ref{sec:covariances}, which is diagonal for the auto-spectra and auto-bispectra that we consider.

As a simple test of where our modelling starts to fail, we use the bias that the residuals induce on an amplitude parameter $A$, defined by $\mathcal{D}_i\to(1+A)\,\mathcal{D}_i$, whose fiducial value is $A=0$. Fitting the residuals on the multipoles $\ell_\mathrm{min}\leq\ell\leq\ell_\mathrm{cut}$ biases this amplitude by
\begin{equation}\label{eq:amplitude_bias}
    \delta A(\ell_\mathrm{cut}) = \frac{\sum_{i\leq\ell_\mathrm{cut}}\mathcal{D}_i\Delta_i/\sigma_i^2}{\sum_{i\leq\ell_\mathrm{cut}}\mathcal{D}_i^2/\sigma_i^2},
    \qquad
    \sigma_A(\ell_\mathrm{cut}) = \left(\sum_{i\leq\ell_\mathrm{cut}}\frac{\mathcal{D}_i^2}{\sigma_i^2}\right)^{-1/2},
\end{equation}
where $\sigma_A$ is the uncertainty on $A$ obtained from this observable alone, with all other parameters fixed. We define the figure of bias as
\begin{equation}\label{eq:fob}
    \mathrm{FoB}(\ell_\mathrm{cut}) = \frac{\left|\delta A(\ell_\mathrm{cut})\right|}{\sigma_A(\ell_\mathrm{cut})},
\end{equation}
and the scale cut as the largest $\ell_\mathrm{cut}$ such that $\mathrm{FoB}(\ell')\leq\mathrm{FoB}_\mathrm{max}$ for all $\ell'\leq\ell_\mathrm{cut}$, choosing $\mathrm{FoB}_\mathrm{max}=0.3$.  The figure of bias only accounts for the part of the residual that is degenerate with the amplitude, and the remaining part could bias other parameters. We check in section~\ref{sec:robustness} that the resulting bias on cosmological parameters remains small.

\subsection{Weak lensing power spectra}\label{sec:wl_cuts}

Our model of the cosmic shear power spectrum already relies on \texttt{HMcode}, so that the comparison above would only test the intrinsic alignment terms. The accuracy of weak lensing power spectra on small scales is instead limited by baryonic feedback, by the accuracy of the non-linear matter power spectrum, and by the non-Gaussian contributions to the covariance that we neglect. For these spectra, we follow refs.~\cite{Doux:2020duk,DES:2022qpf,Euclid:2026klv} and define a maximum wavenumber $k_\mathrm{max}$, converted into a maximum multipole in each tomographic bin by requiring that wavenumbers $k>k_\mathrm{max}$ contribute at most $5\%$ of the power spectrum,
\begin{equation}\label{eq:kmax_criterion}
    \int_{-\infty}^{\ln k_\mathrm{max}}\mathrm{d}\ln k\,\frac{\mathrm{d}C_\ell^{LL}}{\mathrm{d}\ln k}\geq\alpha\,C_\ell^{LL}, \qquad \alpha=0.95.
\end{equation}
We evaluate this criterion for the full weak lensing auto-spectra of eq.~\eqref{eq:cl_LL}, including intrinsic alignments, and use $k_\mathrm{max}=3\,h\,\mathrm{Mpc}^{-1}$, the less conservative of the two values considered in ref.~\cite{Euclid:2026klv}. The criterion itself is nevertheless conservative: it excludes all multipoles receiving more than $5\%$ of their signal from $k>k_\mathrm{max}$, including many for which most of the signal still comes from $k<k_\mathrm{max}$.

\subsection{Adopted scale cuts}\label{sec:final_cuts}

We apply the procedure above to the auto-spectra and auto-bispectra of galaxy clustering and weak lensing in each tomographic bin, testing each observable separately. Testing them separately is optimistic: when several observables are combined, their biases can add up, which we assess in section~\ref{sec:robustness}. We then extend the cuts to the other statistics as follows. For statistics combining galaxy clustering and weak lensing, we use the cuts of the corresponding galaxy clustering statistics. At a given multipole, the wavenumbers probed, $k=(\ell+1/2)/\chi(z)$, and the degree of non-linearity of the fields at these wavenumbers both depend on the redshifts contributing to the Limber integral. Since the galaxy clustering kernels are much narrower than the weak lensing ones, the product of the kernels restricts these redshifts to nearly those of the galaxy clustering statistic, so that both statistics probe similar scales with similar levels of non-linearity. For statistics involving several tomographic bins, we adopt the most conservative of the cuts of the corresponding auto-statistics.

\begin{figure}[htbp]
\centering
\includegraphics{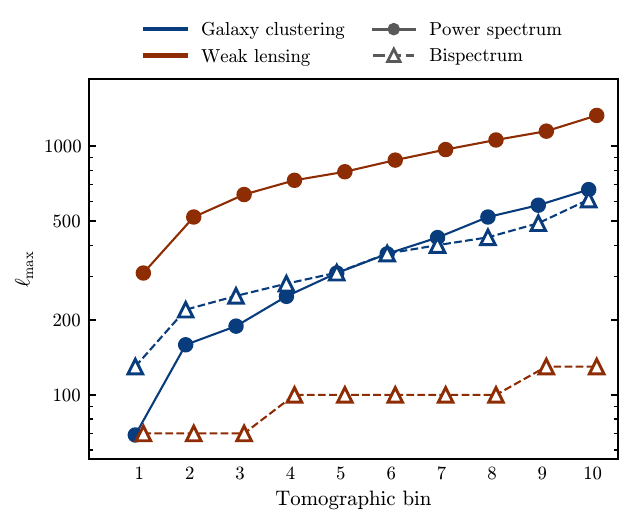}
\caption{Maximum multipole $\ell_\mathrm{max}$ used for galaxy clustering (blue) and weak lensing (red) auto-statistics in each tomographic bin, for power spectra (circles, solid lines) and bispectra (triangles, dashed lines).\label{fig:scale_cuts}}
\end{figure}

The resulting scale cuts are shown in figure~\ref{fig:scale_cuts}. As expected, $\ell_\mathrm{max}$ increases with redshift, for two reasons. First, a given wavenumber $k$ corresponds to a higher multipole at larger comoving distance. Second, the fields become more linear at higher redshift, so that a larger $k_\mathrm{max}$ can be reached for the same modelling error.

Interestingly, the galaxy clustering cuts are rather similar for power spectra and bispectra, even though the former are modelled at 1-loop and the latter at tree-level. This is because our cuts are not set by the absolute accuracy of the models, but by the bias that the model error induces relative to the precision of each statistic. Since bispectra are measured with a much lower signal-to-noise ratio than power spectra, they can tolerate a larger relative model error.

Overall, our scale cuts are more conservative than those typically adopted in forecasts, especially for galaxy clustering power spectra. This is partly because many analyses use an effective approach, in which galaxy power spectra are modelled from non-perturbative predictions of the matter power spectrum, calibrated on simulations, whereas we use instead a perturbative 1-loop model and only keep the scales where it is accurate.

The cuts are also low for weak lensing bispectra. This is because weak lensing kernels receive significant contributions from low redshifts (figure~\ref{fig:kernels}), where a given multipole corresponds to small physical scales on which the matter field is strongly non-linear. Using weak lensing bispectra on smaller scales would require modelling beyond tree-level, for instance with the \texttt{BiHalofit} approach used to validate our scale cuts. However, \texttt{BiHalofit} has not been validated on CPL cosmologies and shows larger errors for $w$CDM than for \lcdm. We therefore keep the more conservative tree-level modelling of weak lensing auto-bispectra, and leave their extension to smaller scales for future work.

\section{Results with three tomographic bins}\label{sec:results_3bins}

In this section, we restrict the analysis to the three lowest redshift tomographic bins, for which we can compute the full \fourxthree\ data vector and its covariance. The \threextwo\ data vector contains all auto- and cross-spectra between these bins, and the \fourxthree\ data vector contains the binned bispectra $B^{GGG}$, $B^{GGL}$, $B^{GLL}$ and $B^{LLL}$ of all combinations of tomographic bins, within the scale cuts of section~\ref{sec:scale_cuts}. We compare the constraints obtained from the combination of CMB and \threextwo\ observables to those obtained when adding bispectra. We also performed the full bispectrum analysis with four and five tomographic bins, which we use in sections~\ref{sec:origin_improvement} and~\ref{sec:skew_validation} to study how the gain evolves with the number of bins. The results obtained when replacing bispectra by skew-spectra, which are also shown in the figures and tables of this section, are discussed in section~\ref{sec:skew_compression}.

\subsection{Constraints}\label{sec:constraints_3bins}

\begin{figure}[htbp]
\centering
\includegraphics{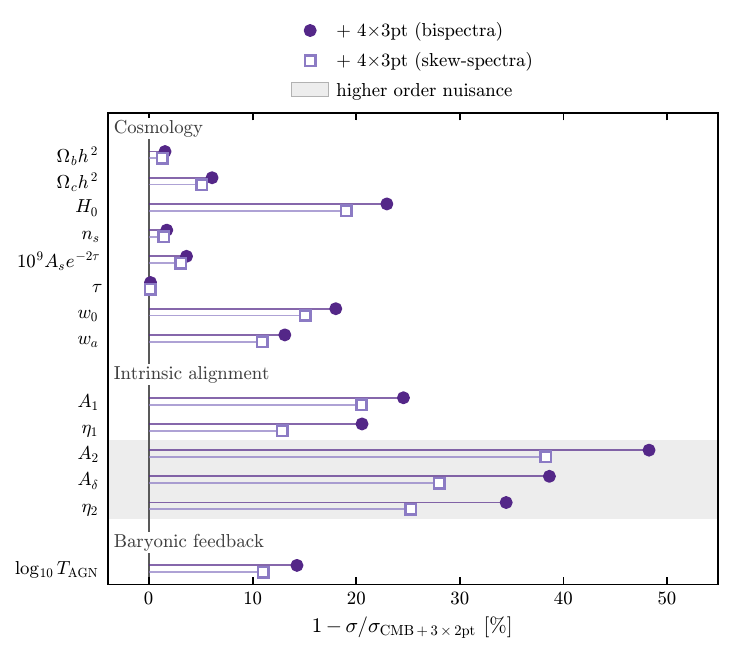}
\caption{Relative improvement of the marginalised $1\sigma$ uncertainties on cosmological, intrinsic alignment and baryonic feedback parameters when adding \fourxthree\ observables to CMB and \threextwo\ observables, for the three lowest redshift tomographic bins. Filled circles correspond to the full set of bispectra, and open squares to the set of skew-spectra selected in section~\ref{sec:skew_compression}. The shaded area highlights higher-order nuisance parameters.\label{fig:params_improvement_3bins}}
\end{figure}

\begin{figure}[htbp]
\centering
\includegraphics{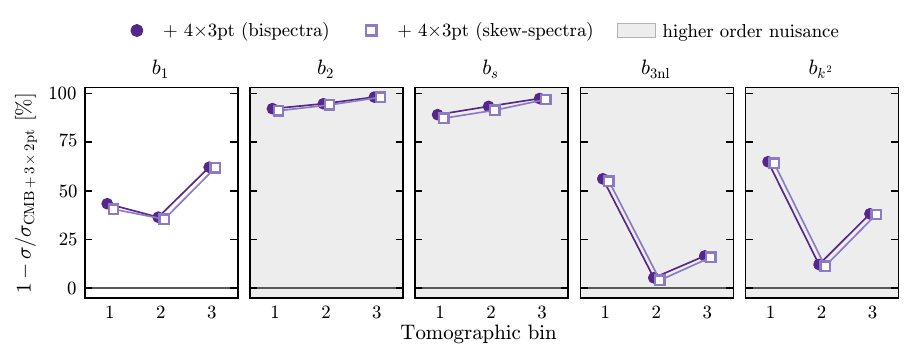}
\caption{Same as figure~\ref{fig:params_improvement_3bins} for the galaxy bias parameters of each tomographic bin.\label{fig:bias_improvement_3bins}}
\end{figure}

\begin{table}[htbp]
  \centering
  \caption{Fiducial values and marginalised $1\sigma$ uncertainties on cosmological, intrinsic alignment and baryonic feedback parameters for the three lowest redshift tomographic bins, together with the dark energy figure of merit. We give the uncertainties obtained from CMB and \threextwo\ observables, and when adding the \fourxthree\ in the form of bispectra or of the skew-spectra selected in section~\ref{sec:skew_compression}. The gain is the relative reduction of the uncertainty with respect to CMB and \threextwo, except for the figure of merit, for which we give the ratio. $R$ is the fraction of the bispectrum gain recovered by skew-spectra (eq.~\eqref{eq:recovery}), only shown when the bispectrum gain exceeds $10\%$.\label{tab:constraints}}
  \begin{tabular}{lrrrrrrr}
\toprule
 &  & \multicolumn{1}{c}{CMB +} & \multicolumn{2}{c}{+ $4{\times}3$pt} & \multicolumn{2}{c}{+ $4{\times}3$pt} &  \\
 &  & \multicolumn{1}{c}{$3{\times}2$pt} & \multicolumn{2}{c}{(bispectra)} & \multicolumn{2}{c}{(skew-spectra)} &  \\
\cmidrule(lr){3-3}\cmidrule(lr){4-5}\cmidrule(lr){6-7}
Parameter & Fiducial & $\sigma$ & $\sigma$ & Gain & $\sigma$ & Gain & $R$ \\
\midrule
\multicolumn{8}{l}{\textit{Cosmology}} \\
$\Omega_b h^2$ & $0.02237$ & $7.29\times10^{-5}$ & $7.18\times10^{-5}$ & $2\%$ & $7.20\times10^{-5}$ & $1\%$ & -- \\
$\Omega_c h^2$ & $0.1200$ & $6.98\times10^{-4}$ & $6.56\times10^{-4}$ & $6\%$ & $6.63\times10^{-4}$ & $5\%$ & -- \\
$H_0$ & $67.36$ & $3.24$ & $2.50$ & $23\%$ & $2.62$ & $19\%$ & $84\%$ \\
$n_s$ & $0.9649$ & $0.00232$ & $0.00228$ & $2\%$ & $0.00229$ & $1\%$ & -- \\
$10^{9}A_s e^{-2\tau}$ & $1.883$ & $0.00423$ & $0.00408$ & $4\%$ & $0.00410$ & $3\%$ & -- \\
$\tau$ & $0.05440$ & $0.00624$ & $0.00623$ & $0\%$ & $0.00624$ & $0\%$ & -- \\
$w_0$ & $-1$ & $0.309$ & $0.253$ & $18\%$ & $0.262$ & $15\%$ & $84\%$ \\
$w_a$ & $0$ & $0.794$ & $0.690$ & $13\%$ & $0.707$ & $11\%$ & $84\%$ \\
\addlinespace
\multicolumn{8}{l}{\textit{Intrinsic alignment}} \\
$A_1$ & $0.9200$ & $0.153$ & $0.115$ & $25\%$ & $0.121$ & $21\%$ & $84\%$ \\
$\eta_1$ & $2.260$ & $1.06$ & $0.840$ & $21\%$ & $0.921$ & $13\%$ & $63\%$ \\
$A_2$ & $0.4000$ & $0.402$ & $0.208$ & $48\%$ & $0.248$ & $38\%$ & $79\%$ \\
$A_\delta$ & $-0.8300$ & $0.475$ & $0.291$ & $39\%$ & $0.342$ & $28\%$ & $72\%$ \\
$\eta_2$ & $2.690$ & $7.35$ & $4.82$ & $34\%$ & $5.49$ & $25\%$ & $74\%$ \\
\addlinespace
\multicolumn{8}{l}{\textit{Baryonic feedback}} \\
$\log_{10}T_{\rm AGN}$ & $7.800$ & $0.223$ & $0.191$ & $14\%$ & $0.198$ & $11\%$ & $78\%$ \\
\midrule
FoM & -- & $16.4$ & $21.8$ & $\times1.33$ & $20.6$ & $\times1.26$ & $78\%$ \\
\bottomrule
\end{tabular}

\end{table}

\begin{table}[htbp]
  \centering
  \caption{Same as table~\ref{tab:constraints} for the galaxy bias parameters of each tomographic bin, numbered from the lowest to the highest redshift.\label{tab:constraints_bias}}
  \begin{tabular}{lrrrrrrr}
\toprule
 &  & \multicolumn{1}{c}{CMB +} & \multicolumn{2}{c}{+ $4{\times}3$pt} & \multicolumn{2}{c}{+ $4{\times}3$pt} &  \\
 &  & \multicolumn{1}{c}{$3{\times}2$pt} & \multicolumn{2}{c}{(bispectra)} & \multicolumn{2}{c}{(skew-spectra)} &  \\
\cmidrule(lr){3-3}\cmidrule(lr){4-5}\cmidrule(lr){6-7}
Parameter & Fiducial & $\sigma$ & $\sigma$ & Gain & $\sigma$ & Gain & $R$ \\
\midrule
\multicolumn{8}{l}{\textit{$b_1$}} \\
$b_1^{(1)}$ & $1.156$ & $0.0737$ & $0.0417$ & $43\%$ & $0.0438$ & $41\%$ & $93\%$ \\
$b_1^{(2)}$ & $1.233$ & $0.0603$ & $0.0383$ & $36\%$ & $0.0389$ & $35\%$ & $97\%$ \\
$b_1^{(3)}$ & $1.282$ & $0.104$ & $0.0392$ & $62\%$ & $0.0395$ & $62\%$ & $100\%$ \\
\addlinespace
\multicolumn{8}{l}{\textit{$b_2$}} \\
$b_2^{(1)}$ & $-0.8115$ & $1.76$ & $0.138$ & $92\%$ & $0.156$ & $91\%$ & $99\%$ \\
$b_2^{(2)}$ & $-0.8029$ & $1.55$ & $0.0824$ & $95\%$ & $0.0919$ & $94\%$ & $99\%$ \\
$b_2^{(3)}$ & $-0.7916$ & $4.84$ & $0.0908$ & $98\%$ & $0.0952$ & $98\%$ & $100\%$ \\
\addlinespace
\multicolumn{8}{l}{\textit{$b_s$}} \\
$b_s^{(1)}$ & $-0.08902$ & $3.67$ & $0.401$ & $89\%$ & $0.467$ & $87\%$ & $98\%$ \\
$b_s^{(2)}$ & $-0.1333$ & $4.12$ & $0.274$ & $93\%$ & $0.356$ & $91\%$ & $98\%$ \\
$b_s^{(3)}$ & $-0.1612$ & $11.6$ & $0.305$ & $97\%$ & $0.358$ & $97\%$ & $100\%$ \\
\addlinespace
\multicolumn{8}{l}{\textit{$b_{3\mathrm{nl}}$}} \\
$b_{3\mathrm{nl}}^{(1)}$ & $0.1558$ & $13.0$ & $5.72$ & $56\%$ & $5.88$ & $55\%$ & $98\%$ \\
$b_{3\mathrm{nl}}^{(2)}$ & $0.2334$ & $6.29$ & $5.96$ & $5\%$ & $6.04$ & $4\%$ & -- \\
$b_{3\mathrm{nl}}^{(3)}$ & $0.2820$ & $9.06$ & $7.56$ & $17\%$ & $7.62$ & $16\%$ & $96\%$ \\
\addlinespace
\multicolumn{8}{l}{\textit{$b_{k^2}$}} \\
$b_{k^2}^{(1)}$ & $0$ & $183$ & $64.3$ & $65\%$ & $65.6$ & $64\%$ & $99\%$ \\
$b_{k^2}^{(2)}$ & $0$ & $60.9$ & $53.5$ & $12\%$ & $54.1$ & $11\%$ & $92\%$ \\
$b_{k^2}^{(3)}$ & $0$ & $98.9$ & $61.2$ & $38\%$ & $61.5$ & $38\%$ & $99\%$ \\
\bottomrule
\end{tabular}

\end{table}

Figures~\ref{fig:params_improvement_3bins} and~\ref{fig:bias_improvement_3bins} show the relative improvement of the marginalised uncertainties when adding bispectra to CMB and \threextwo\ observables, and table~\ref{tab:constraints} gives the corresponding constraints. Among cosmological parameters, the improvement is concentrated on $H_0$, $w_0$ and $w_a$, whose uncertainties decrease by $23\%$, $18\%$ and $13\%$ respectively, while the other cosmological parameters improve by at most $6\%$. The dark energy figure of merit increases by a factor $1.33$, from $16.4$ to $21.8$. These values remain much lower than those obtained in \Euclid\ forecasts with ten tomographic bins~\cite{Euclid:2019clj}, since we only use three low redshift bins here, with conservative scale cuts and a larger number of nuisance parameters.

The improvement is much larger for nuisance parameters, in particular for second-order ones. The uncertainties on the second-order intrinsic alignment parameters $A_2$, $A_\delta$ and $\eta_2$ decrease by $48\%$, $39\%$ and $35\%$, and those on the linear parameters $A_1$ and $\eta_1$ by $25\%$ and $21\%$. The second-order galaxy bias parameters $b_2$ and $b_s$ are the most improved parameters (figure~\ref{fig:bias_improvement_3bins} and table~\ref{tab:constraints_bias}), with gains ranging from $89\%$ to $98\%$. The bias parameters $b_{3\mathrm{nl}}$ and $b_{k^2}$ do not enter our tree-level bispectrum model, and their improvement only results from the better determination of the parameters with which they are degenerate in the 1-loop power spectra. Similarly, baryonic feedback does not enter our bispectrum model, but the uncertainty on $\log_{10} T_\mathrm{AGN}$ decreases by $14\%$, a degeneracy-breaking effect that we discuss in section~\ref{sec:origin_improvement}.

\subsection{Origin of the improvement}\label{sec:origin_improvement}

\begin{figure}[htbp]
\centering
\includegraphics[width=\textwidth]{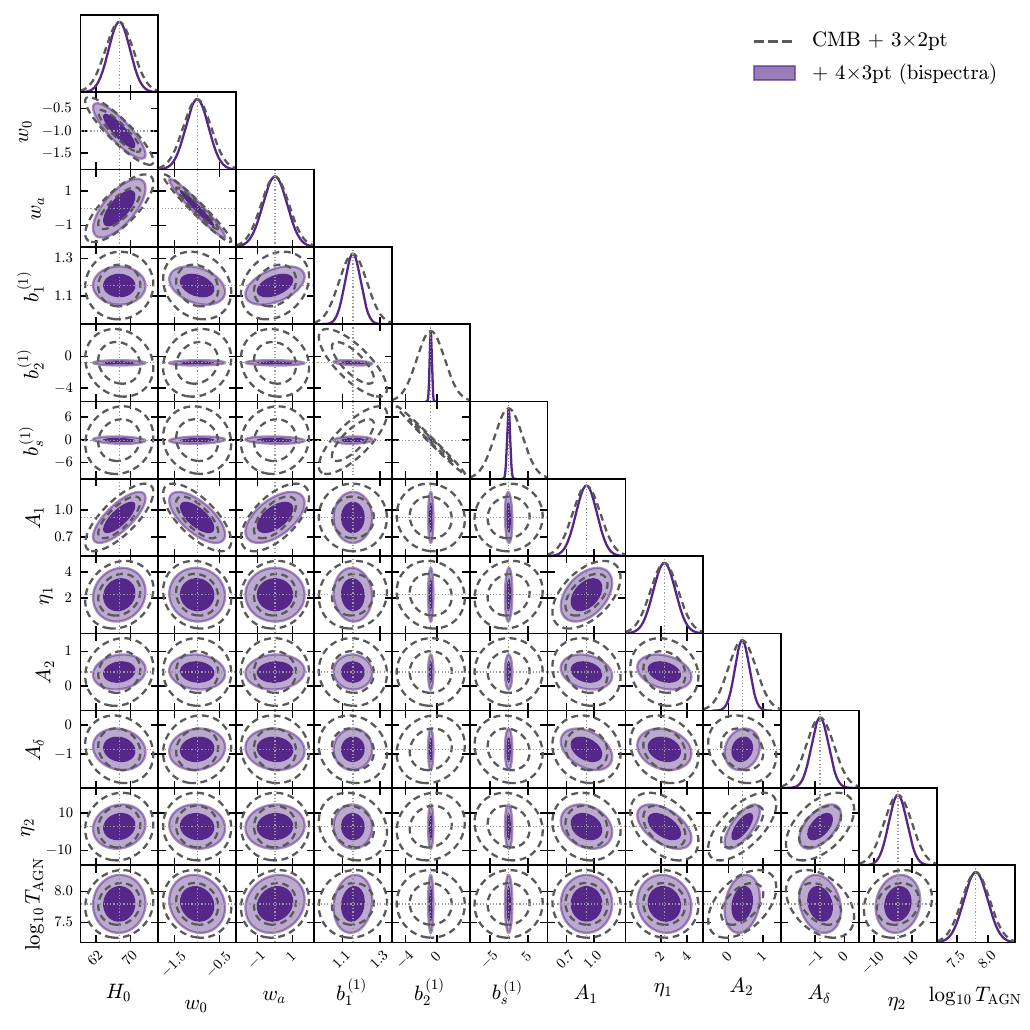}
\caption{Marginalised $68\%$ and $95\%$ confidence contours obtained from CMB and \threextwo\ observables (dashed lines), and when adding \fourxthree\ bispectra (filled contours), for the three lowest redshift tomographic bins. We show the cosmological parameters most affected by the \fourxthree, the bias parameters of the first tomographic bin, the intrinsic alignment parameters, and the baryonic feedback parameter $\log_{10}(T_\mathrm{AGN})$. Dotted lines indicate the fiducial values.\label{fig:triangle_plot}}
\end{figure}

The improvements described above follow from the fact that second-order nuisance parameters enter power spectra only through loop corrections, whereas they contribute to bispectra already at tree-level. Figure~\ref{fig:triangle_plot} illustrates the resulting mechanism. In the CMB and \threextwo\ analysis, $b_2$ and $b_s$ are strongly degenerate with each other and with $b_1$. Adding bispectra breaks the $b_2$--$b_s$ degeneracy, which in turn tightens the constraint on $b_1$, and this calibration propagates to the cosmological parameters correlated with $b_1$. A similar mechanism operates for intrinsic alignments, for which bispectra especially improve constraints in the plane $\eta_1$--$\eta_2$.

\begin{table}[htbp]
  \centering
  \caption{Relative improvement from adding \fourxthree\ bispectra to CMB and \threextwo\ observables, for the three lowest redshift tomographic bins, when all nuisance parameters are marginalised over (as in table~\ref{tab:constraints}), and when some higher-order nuisance parameters are fixed to their fiducial values in both analyses: all of them ($b_2$, $b_s$, $b_{3\mathrm{nl}}$, $b_{k^2}$, $A_2$, $A_\delta$ and $\eta_2$), only the higher-order galaxy bias parameters ($b_2$, $b_s$, $b_{3\mathrm{nl}}$ and $b_{k^2}$), or only the higher-order intrinsic alignment parameters ($A_2$, $A_\delta$ and $\eta_2$). For the figures of merit, we give that obtained from CMB and \threextwo\ observables, and report the improvement obtained when adding the \fourxthree\ bispectra.\label{tab:constraints_fixed}}
  \begin{tabular}{lrrrr}
    \toprule
    & & \multicolumn{3}{c}{Fixed higher-order parameters} \\
    \cmidrule(lr){3-5}
    Parameter & Marginalised & All & Galaxy bias & Intrinsic alignment \\
    \midrule
    $H_0$ & $23\%$ & $4\%$ & $6\%$ & $23\%$ \\
    $w_0$ & $18\%$ & $3\%$ & $3\%$ & $18\%$ \\
    $w_a$ & $13\%$ & $3\%$ & $2\%$ & $13\%$ \\
    $A_1$ & $25\%$ & $3\%$ & $7\%$ & $29\%$ \\
    $\eta_1$ & $21\%$ & $3\%$ & $14\%$ & $22\%$ \\
    $\log_{10} T_\mathrm{AGN}$ & $14\%$ & $1\%$ & $14\%$ & $1\%$ \\
    \midrule
    FoM (CMB + \threextwo) & $16.4$ & $28.0$ & $26.9$ & $17.2$ \\
    FoM (+ \fourxthree)& $\times1.33$ & $\times1.07$ & $\times1.09$ & $\times1.31$ \\
    \bottomrule
  \end{tabular}
\end{table}

To identify which parameters drive the improvement, we repeat the forecast with some of the higher-order nuisance parameters fixed to their fiducial values in both the CMB and \threextwo\ analysis and the analysis including bispectra, and report the results in table~\ref{tab:constraints_fixed}. When all higher-order nuisance parameters are fixed, the CMB and \threextwo\ constraints tighten considerably, with the figure of merit increasing from $16.4$ to $28.0$, and adding bispectra then improves the constraints on cosmological parameters by at most $4\%$ and the figure of merit by a factor $1.07$ only. Fixing only the higher-order galaxy bias parameters gives almost the same result, with a figure of merit increased by a factor $1.09$, whereas fixing only the higher-order intrinsic alignment parameters leaves the improvement on cosmological parameters almost unchanged. Bispectra therefore improve the cosmological constraints mostly by calibrating the higher-order galaxy bias parameters, which limit the cosmological information that can be extracted from the 1-loop galaxy clustering power spectra. The small remaining gain when these parameters are fixed comes from the direct sensitivity of bispectra to cosmology and to linear nuisance parameters.

The baryonic feedback parameter behaves differently. Its improvement is unaffected when the higher-order galaxy bias parameters are fixed, but vanishes when the higher-order intrinsic alignment parameters are fixed. In our model, baryonic feedback and higher-order intrinsic alignments therefore both modify the small-scale weak lensing power spectra in a partly degenerate way: with CMB and \threextwo\ observables, the correlation coefficients between $\log_{10}T_\mathrm{AGN}$ and $A_2$ and $A_\delta$ are $0.48$ and $-0.39$, but reduce to $0.21$ and $-0.21$ when the \fourxthree\ bispectra are added, which can be observed on figure~\ref{fig:triangle_plot}. Bispectra therefore seem to be able to partially break the degeneracy that we observe between baryonic feedback and intrinsic alignment parameters, an interesting finding that we return to in section~\ref{sec:tagn_ia}.

\begin{table}[htbp]
  \centering
  \setlength{\tabcolsep}{5pt}
  \caption{Relative improvement from adding subsets of the \fourxthree\ bispectra to CMB and \threextwo\ observables. The improvement on the parameters is given for the three lowest redshift tomographic bins, and the ratio of the dark energy figures of merit for three, four and five tomographic bins. $T_\mathrm{AGN}$ stands for $\log_{10} T_\mathrm{AGN}$\label{tab:bispectrum_types}}
  \begin{tabular}{lrrrrrrrrr}
    \toprule
 & \multicolumn{6}{c}{Gain, $N_\mathrm{bin}=3$}
 & \multicolumn{3}{c}{FoM ratio} \\
    \cmidrule(lr){8-10}
    Bispectra & $H_0$ & $w_0$ & $w_a$ & $A_2$ & $A_\delta$ & $T_\mathrm{AGN}$ & $N_\mathrm{bin}=3$ & $4$ & $5$ \\
    \midrule
    $GGG$ & $16\%$ & $13\%$ & $10\%$ & $13\%$ & $5\%$ & $1\%$ & $\times1.22$ & $\times1.20$ & $\times1.18$ \\
    $GGG+LLL$ & $17\%$ & $14\%$ & $10\%$ & $13\%$ & $6\%$ & $1\%$ & $\times1.22$ & $\times1.20$ & $\times1.18$ \\
    $GGG+GGL$ & $21\%$ & $16\%$ & $12\%$ & $48\%$ & $38\%$ & $14\%$ & $\times1.30$ & $\times1.28$ & $\times1.26$ \\
    $GGG+GLL$ & $19\%$ & $15\%$ & $11\%$ & $17\%$ & $22\%$ & $5\%$ & $\times1.26$ & $\times1.25$ & $\times1.22$ \\
    $GGG+GGL+GLL$ & $21\%$ & $17\%$ & $12\%$ & $48\%$ & $39\%$ & $14\%$ & $\times1.31$ & $\times1.29$ & $\times1.27$ \\
    $GGG+GGL+LLL$ & $22\%$ & $17\%$ & $12\%$ & $48\%$ & $38\%$ & $14\%$ & $\times1.31$ & $\times1.29$ & $\times1.27$ \\
    All & $23\%$ & $18\%$ & $13\%$ & $48\%$ & $39\%$ & $14\%$ & $\times1.33$ & $\times1.32$ & $\times1.32$ \\
    \bottomrule
  \end{tabular}
\end{table}

To identify which bispectra drive these improvements, we repeat the forecast including only some of the four types of bispectra, and report the results in table~\ref{tab:bispectrum_types}. For three tomographic bins, the galaxy clustering bispectra $B^{GGG}$ alone provide about two thirds of the improvement of the figure of merit, consistently with the role of galaxy bias identified above, and adding the cross-bispectra $B^{GGL}$ brings this fraction to $90\%$. The cross-bispectra $B^{GGL}$ are also responsible for almost all of the improvement on the second-order intrinsic alignment parameters, and for that on baryonic feedback. Adding either $B^{GLL}$ or $B^{LLL}$ to $B^{GGG}$ and $B^{GGL}$ changes the improvement on cosmological parameters by at most one percentage point, but adding both provides the remaining $10\%$ of the improvement of the figure of merit, so that these two types of bispectra mostly contribute in combination. With four and five tomographic bins, we find the same hierarchy, with a growing contribution of $B^{GLL}$ and $B^{LLL}$: for three, four and five bins, the fraction of the improvement of the figure of merit provided by $B^{GGG}$ decreases from $65\%$ to $62\%$ and $56\%$, and that provided by $B^{GGG}$ and $B^{GGL}$ together from $90\%$ to $86\%$ and $81\%$. These findings suggest that high redshift bins may benefit more significantly from the inclusion of $B^{GLL}$ and $B^{LLL}$ than the low redshift bins.

\subsection{Robustness tests}\label{sec:robustness}

We test the robustness of these results to three aspects of our analysis: the accuracy of our model within the scale cuts, the Poisson nature of the galaxy shot noise, and the absence of priors on higher-order galaxy bias parameters.

The scale cuts of section~\ref{sec:scale_cuts} limit the bias that the error of our model induces on the amplitude of each observable taken separately. To check that the resulting bias on cosmological parameters remains small when all observables are combined, we compute, at linear order in the residuals $\Delta$ between the reference predictions of section~\ref{sec:reference_predictions} and our model, the bias $\delta p=F^{-1}\,(\partial\mathcal{D}/\partial p)^\mathrm{T}\,\mathrm{Cov}^{-1}\,\Delta$ on all parameters, for the full data vector of power spectra and bispectra. We performed this test for three, four and five tomographic bins. In all cases, the bias on $w_0$ and $w_a$ remains below $0.5\sigma$. Since we find it to increase with the number of bins, we repeat this test for ten tomographic bins in section~\ref{sec:constraints_10bins}.

In our baseline analysis, the galaxy shot noise only enters the covariances. For a Poisson-sampled field, however, shot noise also contributes to the mean power spectra and bispectra. In the galaxy clustering auto-power spectra, it adds a constant $1/\bar{n}_i$, where $\bar{n}_i$ is the angular number density of galaxies in bin $i$. In the bispectra, each pair of fields describing the same galaxies contributes the power spectrum between this pair and the third field, evaluated at the multipole of the third field~\cite{Matarrese:1997sk,Oddo:2019run}. These contributions depend on cosmological and nuisance parameters through the power spectra. To assess their impact, together with that of deviations from Poisson statistics, we write the shot-noise contributions as
\begin{equation}
C_\ell^{G_iG_i} \supset \frac{1+\alpha_G^{(i)}}{\bar{n}_i}
\label{eq:stoch_P}
\end{equation}
for the power spectra, and, for two galaxy clustering fields in the same bin $i$ and a third field $X$,
\begin{equation}
b_{\ell_1\ell_2\ell_3}^{G_iG_iX} \supset \frac{1+\alpha_{GX}^{(i)}}{\bar{n}_i}\,C_{\ell_3}^{G_iX}
\label{eq:stoch_B1}
\end{equation}
for the reduced bispectra, with one amplitude, $\alpha_{GG}^{(i)}$, for $X=G_k$ and another, $\alpha_{GL}^{(i)}$, for $X=L_k$. The $G_iG_iG_i$ bispectra contain three such pairs, together with a term involving the three identical fields:
\begin{equation}
b_{\ell_1\ell_2\ell_3}^{G_iG_iG_i} \supset \frac{1+\alpha_{GG}^{(i)}}{\bar{n}_i}\left(C_{\ell_1}^{G_iG_i}+C_{\ell_2}^{G_iG_i}+C_{\ell_3}^{G_iG_i}\right) + \frac{1+\alpha_{GGG}^{(i)}}{\bar{n}_i^2}\,.
\label{eq:stoch_B2}
\end{equation}
Poisson statistics corresponds to $\alpha_G^{(i)}=\alpha_{GG}^{(i)}=\alpha_{GL}^{(i)}=\alpha_{GGG}^{(i)}=0$, which we take as fiducial values. We neglect the analogous terms arising when shapes are measured on the same galaxies, i.e.\ from pairs formed by the position and the shape of the same galaxies (as in $G_iG_jL_i$ or $G_iL_iL_k$), or by the shapes of the same galaxies (as in $G_kL_iL_i$ or $L_iL_iL_k$). Like the terms above, they scale as $1/\bar{n}_i$, and they involve weak lensing power spectra, so that we expect their impact to be at most comparable. We first include the Poisson contributions, with all amplitudes fixed to zero, together with their dependence on the cosmological and nuisance parameters. Their derivatives are much smaller than those of the bispectra within our scale cuts, so that the constraints change by less than $0.1\%$ and bispectra still improve the figure of merit by a factor $1.33$. We then vary the amplitudes $\alpha$ in each tomographic bin, with a Gaussian prior of unit width, which allows deviations of $100\%$ from Poisson statistics. The improvement of the figure of merit is again unchanged, with a factor $1.33$, and so is the improvement on baryonic feedback. Shot noise, whether Poissonian or with realistic deviations from Poisson statistics, therefore does not affect our conclusions.

Finally, some higher-order galaxy bias parameters are almost unconstrained by the CMB and \threextwo\ observables (table~\ref{tab:constraints_bias}). In this regime, the Fisher matrix is close to singular, and the marginalised uncertainties on these parameters are sensitive to small numerical errors in the derivatives, which could in turn affect the gain from bispectra. To test the robustness of our results to these nearly unconstrained directions, we repeated the analysis adding Gaussian priors on $b_2$, $b_s$, $b_{3\mathrm{nl}}$ and $b_{k^2}$, which regularise the Fisher matrix. For prior widths larger than about unity, the gain in figure of merit from bispectra is only marginally affected, for three, four and five tomographic bins. Our results are therefore robust to the treatment of these nearly unconstrained directions.

\section{Compression with skew-spectra}\label{sec:skew_compression}

Extending the full \fourxthree\ analysis to ten tomographic bins is very challenging computationally. In this section, we therefore use the three-bin configuration of section~\ref{sec:results_3bins} to select a small set of skew-spectra recovering most of the information of the full bispectrum analysis, and we propose a way to generalise this set to an arbitrary number of tomographic bins.

\subsection{Selection of skew-spectra}\label{sec:skew_selection}

A skew-spectrum $S_\ell^{XYZ}(\theta_X,\theta_Y)$ is defined by the three fields $X$, $Y$ and $Z$, each corresponding to galaxy clustering or weak lensing in a given tomographic bin, and by the widths of the filters applied to $X$ and $Y$. We set these widths relative to the maximum multipoles determined in section~\ref{sec:scale_cuts},
\begin{equation}\label{eq:filter_width}
   \theta_X=\frac{\lambda_X}{\ell_\mathrm{max}^{XYZ}},
\end{equation}
where $\ell_\mathrm{max}^{XYZ}$ is the maximum multipole used for the bispectrum $B_{\ell_1\ell_2\ell_3}^{XYZ}$ (section~\ref{sec:final_cuts}), and the dimensionless factor $\lambda_X$ takes values in a set $\Lambda$. With this choice, the filters select multipoles relative to the range on which our model is valid for each combination of fields, rather than fixed multipoles.

To quantify the information recovered by a set of skew-spectra, we compare its improvement of the dark energy figure of merit with that of the full set of bispectra, and define
\begin{equation}\label{eq:recovery}
    R\left(\mathrm{FoM}\right) = \frac{\mathrm{FoM}^{S}-\mathrm{FoM}^{\mathrm{ref}}}{\mathrm{FoM}^{B}-\mathrm{FoM}^{\mathrm{ref}}},
\end{equation}
where $\mathrm{FoM}^{\mathrm{ref}}$ is the figure of merit obtained from CMB and \threextwo\ observables, and $\mathrm{FoM}^{B}$ and $\mathrm{FoM}^{S}$ are those obtained when adding bispectra and skew-spectra respectively. We similarly define, for any parameter $p$, the fraction $R(p)$ of the reduction of its uncertainty recovered by skew-spectra, by replacing the figures of merit in eq.~\eqref{eq:recovery} by the marginalised uncertainties $\sigma(p)$.

We select skew-spectra with a greedy algorithm. Starting from CMB and \threextwo\ observables, we compute $R(\mathrm{FoM})$ obtained when adding each candidate skew-spectrum, where the candidates are all combinations of three fields within the three tomographic bins and of filter factors in $\Lambda$. We retain the skew-spectrum maximising $R(\mathrm{FoM})$, and repeat the procedure, each time adding the candidates to the skew-spectra already retained, until $R(\mathrm{FoM})$ reaches a target value. We tested several sets of filter factors, $\Lambda=\{0\}$, $\{0,1\}$, $\{0,3\}$, $\{0,5\}$, $\{0,1,3\}$, $\{0,1,5\}$ and $\{0,3,5\}$, and obtained similar results in most cases apart from the unfiltered case $\Lambda=\{0\}$ where only part of the information contained within bispectra could be recovered. We use $\Lambda=\{0,3\}$, as this set gave slightly better results among the sets containing one filter (on top of $0$).

\begin{figure}[htbp]
\centering
\includegraphics{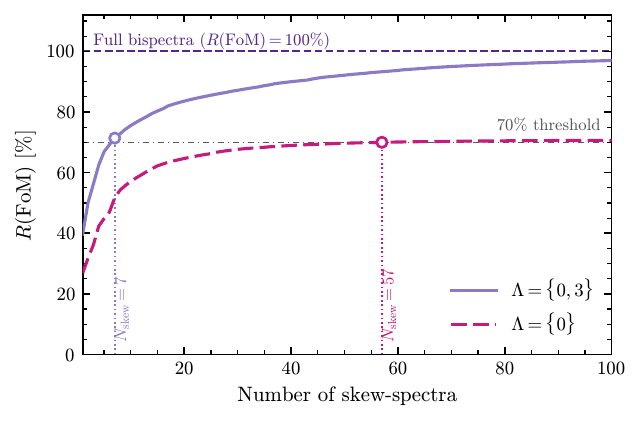}
\caption{Fraction $R(\mathrm{FoM})$ of the gain in dark energy figure of merit from bispectra that is recovered by skew-spectra, as a function of the number of skew-spectra selected by the greedy algorithm, for the three lowest redshift tomographic bins. The dashed line corresponds to the full set of bispectra, and the dash-dotted line to our target of $70\%$, reached with seven skew-spectra when using filters among $\Lambda=\{0,3\}$, against fifty seven when using no filter ($\Lambda=\{0\}$).\label{fig:R_vs_N}}
\end{figure}

Figure~\ref{fig:R_vs_N} shows $R(\mathrm{FoM})$ as a function of the number of selected skew-spectra. With $\Lambda=\{0,3\}$, most of the information is contained in the first few skew-spectra, and $R(\mathrm{FoM})$ approaches unity when about a hundred skew-spectra are included. Filtering is essential to reach this efficiency: without filters ($\Lambda=\{0\}$), $R(\mathrm{FoM})$ only barely exceeds $70\%$, after fifty seven skew-spectra. That a small number of filtered skew-spectra suffices is consistent with the mechanism identified in section~\ref{sec:origin_improvement}: bispectra improve dark energy constraints mostly by calibrating a few second-order galaxy bias parameters, so that the relevant information is concentrated in a small number of combinations of bispectra. We therefore set a target of $70\%$, reached with the seven skew-spectra listed in table~\ref{tab:selected_skew_spectra}, for which $R(\mathrm{FoM})=71.4\%$. Higher targets would require many more skew-spectra, and their number grows further once the selection is generalised to more tomographic bins (section~\ref{sec:skew_generalisation}).

\begin{table}[htbp]
  \centering
  \caption{Skew-spectra selected by the greedy algorithm for the three lowest redshift tomographic bins, in the order of selection. $G_i$ and $L_i$ denote galaxy clustering and weak lensing in tomographic bin $i$, $\lambda_X$ and $\lambda_Y$ are the dimensionless filter factors applied to the fields $X$ and $Y$ (eq.~\eqref{eq:filter_width}), and the last column gives the fraction $R(\mathrm{FoM})$ recovered by the skew-spectra selected up to each step.\label{tab:selected_skew_spectra}}
  \begin{tabular}{lrrr}
    \toprule
    $(X, Y, Z)$ & $\lambda_X$ & $\lambda_Y$ & $R(\mathrm{FoM})$ \\
    \midrule
    $(G_2, G_2, G_2)$ & $3$ & $3$ & $39.8\%$ \\
    $(G_2, G_3, G_2)$ & $0$ & $0$ & $50.1\%$ \\
    $(G_1, G_1, G_1)$ & $0$ & $0$ & $56.2\%$ \\
    $(G_1, G_1, L_3)$ & $0$ & $0$ & $62.5\%$ \\
    $(G_2, L_3, G_2)$ & $3$ & $0$ & $66.9\%$ \\
    $(G_2, G_3, G_3)$ & $0$ & $0$ & $69.2\%$ \\
    $(G_3, G_3, L_3)$ & $3$ & $3$ & $71.4\%$ \\
    \bottomrule
  \end{tabular}
\end{table}

All selected skew-spectra involve at least two galaxy clustering fields: four are of type $GGG$ and three of type $GGL$. In our setup, the information relevant to dark energy is therefore mostly contained in configurations dominated by galaxy clustering, consistently with the role of galaxy bias identified in section~\ref{sec:origin_improvement}, and with the fact that $B^{GGG}$ and $B^{GGL}$ provide $90\%$ of the gain in figure of merit from bispectra (table~\ref{tab:bispectrum_types}). Weak lensing auto-bispectra are also penalised by the very restrictive scale cuts we use (section~\ref{sec:final_cuts}), and modelling them beyond tree-level could make them more informative. Moreover, $B^{GLL}$ and $B^{LLL}$ mostly contribute in combination as we showed in section~\ref{sec:origin_improvement}, which a greedy algorithm, adding one skew-spectrum at a time, is less likely to capture.

\subsection{Generalisation to more tomographic bins}\label{sec:skew_generalisation}

The selection above is specific to the three-bin configuration, and we need a rule to define the corresponding set of skew-spectra for a larger number of bins. We generalise each selected skew-spectrum to all skew-spectra with the same field types, the same filter factors and the same arrangement of tomographic bins. To define this arrangement, we associate to each triplet of tomographic indices $(i,j,k)$ its order structure $\pi(i,j,k)$, obtained by replacing each index by its rank among the distinct values of the triplet. For instance, $\pi(1,1,2)=\pi(2,2,3)=(1,1,2)$, $\pi(3,2,3)=\pi(4,1,4)=(2,1,2)$ and $\pi(1,2,4)=\pi(3,6,8)=(1,2,3)$. The order structure encodes which fields share a tomographic bin and how the distinct bins are ordered in redshift, but not their absolute redshifts nor their separation.

A skew-spectrum of fields $X_i$, $Y_j$ and $Z_k$ is then generalised to all skew-spectra of fields $X_l$, $Y_m$ and $Z_n$, with $1\leq l,m,n\leq N_\mathrm{bin}$, which have the same field types and filter factors and satisfy $\pi(l,m,n)=\pi(i,j,k)$. For example, the skew-spectrum $(G_2,L_3,G_2)$ is generalised to all skew-spectra $(G_i,L_j,G_i)$ with $i<j$. Since we keep the factors $\lambda_X$ and $\lambda_Y$ fixed rather than the filter widths, the filters automatically adapt to the scale cuts of each bin through eq.~\eqref{eq:filter_width}. The union of the skew-spectra generalised from all selected ones, without duplicates, defines our data vector for $N_\mathrm{bin}$ bins.

\begin{table}[htbp]
  \centering
  \caption{Classes of skew-spectra obtained by generalising the skew-spectra of table~\ref{tab:selected_skew_spectra} to $N_\mathrm{bin}$ tomographic bins. The last column gives the tomographic indices included in each class.\label{tab:skew_spectra_generalisation}}
  \begin{tabular}{lrrl}
    \toprule
    $(X, Y, Z)$ & $\lambda_X$ & $\lambda_Y$ & Indices\\
    \midrule
    $(G_i, G_i, G_i)$ & $3$ & $3$ & $1\leq i\leq N_\mathrm{bin}$\\
    $(G_i, G_j, G_i)$ & $0$ & $0$ & $1\leq i<j\leq N_\mathrm{bin}$\\
    $(G_i, G_i, G_i)$ & $0$ & $0$ & $1\leq i\leq N_\mathrm{bin}$\\
    $(G_i, G_i, L_j)$ & $0$ & $0$ & $1\leq i<j\leq N_\mathrm{bin}$\\
    $(G_i, L_j, G_i)$ & $3$ & $0$ & $1\leq i<j\leq N_\mathrm{bin}$\\
    $(G_i, G_j, G_j)$ & $0$ & $0$ & $1\leq i<j\leq N_\mathrm{bin}$\\
    $(G_i, G_i, L_i)$ & $3$ & $3$ & $1\leq i\leq N_\mathrm{bin}$\\
    \bottomrule
  \end{tabular}
\end{table}

Table~\ref{tab:skew_spectra_generalisation} lists the resulting classes, one per selected skew-spectrum. Three classes contain one skew-spectrum per bin and four contain one per pair of bins, so that the data vector contains $3N_\mathrm{bin}+2N_\mathrm{bin}(N_\mathrm{bin}-1)$ skew-spectra, i.e.\ $21$, $36$, $55$ and $210$ skew-spectra for three, four, five and ten bins respectively.

\subsection{Performance and validation}\label{sec:skew_validation}

For three bins, the generalised set of $21$ skew-spectra gives $R(\mathrm{FoM})=78\%$, compared to $71.4\%$ for the seven skew-spectra selected by the greedy algorithm. Figures~\ref{fig:params_improvement_3bins} and~\ref{fig:bias_improvement_3bins} and table~\ref{tab:constraints} compare the constraints obtained with this set to those of the full bispectrum analysis. Skew-spectra recover $84\%$ of the gain from bispectra on $H_0$, $w_0$ and $w_a$, and almost all of it for the higher-order galaxy bias parameters, consistently with ref.~\cite{MoradinezhadDizgah:2019xun}, which showed that weighted skew-spectra can recover the information on galaxy bias contained in the bispectrum. The recovery is lower for intrinsic alignment parameters, between $63\%$ and $84\%$, and is $78\%$ for baryonic feedback.

\begin{figure}[htbp]
\centering
\includegraphics{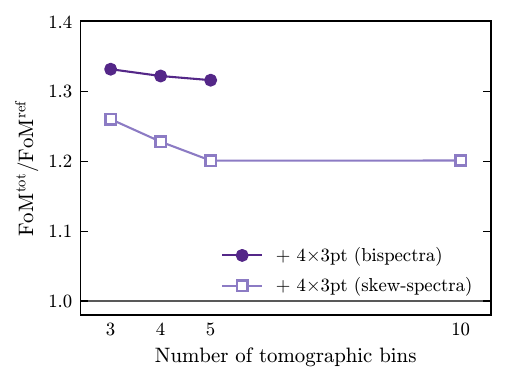}
\caption{Ratio of the dark energy figure of merit obtained when adding \fourxthree\ bispectra (filled circles) or skew-spectra (open squares) to CMB and \threextwo\ observables, to that obtained in the reference case (CMB and \threextwo\ observables), as a function of the number of tomographic bins.\label{fig:fom_vs_nbins}}
\end{figure}

To test the generalisation on configurations not used in the selection, we perform the full bispectrum analysis for four and five tomographic bins, and compare it to the analyses with the corresponding $36$ and $55$ skew-spectra. Figure~\ref{fig:fom_vs_nbins} shows the resulting gains on the figure of merit. With bispectra, the figure of merit increases by a factor $1.33$, $1.32$ and $1.32$ for three, four and five bins: this gain is remarkably stable with the number of bins. The gain obtained with skew-spectra decreases slightly, from a factor $1.26$ to $1.23$ and $1.20$, so that $R(\mathrm{FoM})$ decreases from $78\%$ to $71\%$ and $64\%$, reflecting the fact that the selection was not optimised for these configurations. This decrease is consistent with the growing contribution of $B^{GLL}$ and $B^{LLL}$ with the number of bins found in section~\ref{sec:origin_improvement}: our skew-spectra only involve the field types of $B^{GGG}$ and $B^{GGL}$, whose share of the improvement of the figure of merit decreases from $90\%$ to $86\%$ and $81\%$ for three, four and five bins. For ten bins, skew-spectra increase the figure of merit by a factor $1.20$ (section~\ref{sec:constraints_10bins}), as for five bins, which suggests that the loss of information due to the compression may stabilise as the number of bins increases. 

\section{Results with ten tomographic bins and discussion}\label{sec:results_10bins}

\subsection{Constraints}\label{sec:constraints_10bins}

\begin{figure}[htbp]
\centering
\includegraphics{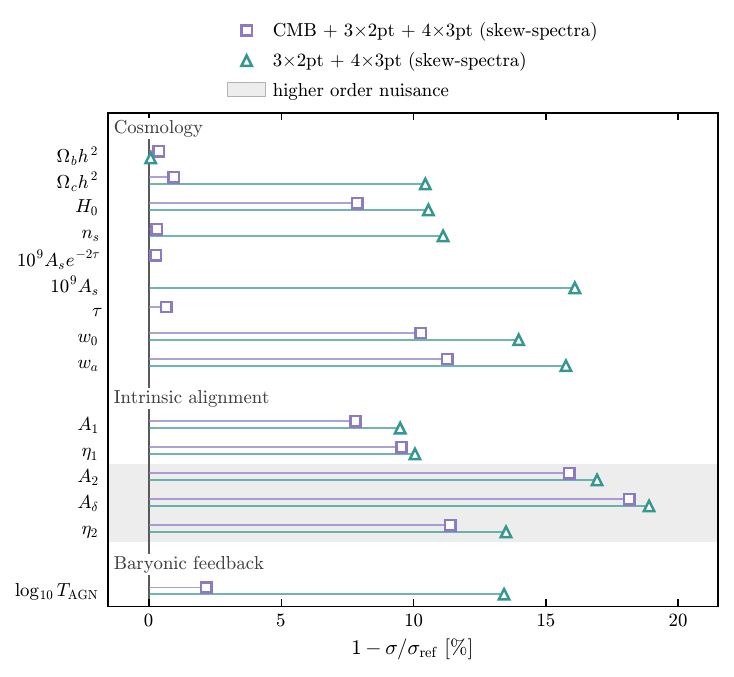}
\caption{Relative improvement of the marginalised $1\sigma$ uncertainties on cosmological, intrinsic alignment and baryonic feedback parameters when adding the $210$ skew-spectra of section~\ref{sec:skew_generalisation} to CMB and \threextwo\ observables (open squares) and to \threextwo\ observables alone (open triangles) with a prior on $\Omega_bh^2$ coming from BBN, for the ten tomographic bins of our \Euclid-like survey. The shaded area highlights higher-order nuisance parameters.\label{fig:constraints_10bins}}
\end{figure}

We now apply the skew-spectrum compression of section~\ref{sec:skew_compression} to the full set of ten tomographic bins, which yields $210$ skew-spectra. Figure~\ref{fig:constraints_10bins} shows how much these skew-spectra improve the constraints from CMB and \threextwo\ observables, and from \threextwo\ observables alone, and tables~\ref{tab:constraints_10bins} and~\ref{tab:constraints_10bins_nocmb} give the corresponding constraints.

\begin{table}[htbp]
  \centering
  \caption{Fiducial values and marginalised $1\sigma$ uncertainties on cosmological, intrinsic alignment and baryonic feedback parameters for the ten tomographic bins, together with the dark energy figure of merit. We give the uncertainties obtained from CMB and \threextwo\ observables and when adding the $210$ skew-spectra of section~\ref{sec:skew_generalisation}. The gain is the relative reduction of the uncertainty, except for the figure of merit, for which we give the ratio.\label{tab:constraints_10bins}}
  \begin{tabular}{lrrrr}
\toprule
 &  & \multicolumn{1}{c}{CMB +} & \multicolumn{2}{c}{+ $4{\times}3$pt} \\
 &  & \multicolumn{1}{c}{$3{\times}2$pt} & \multicolumn{2}{c}{(skew-spectra)} \\
\cmidrule(lr){3-3}\cmidrule(lr){4-5}
Parameter & Fiducial & $\sigma$ & $\sigma$ & Gain \\
\midrule
\multicolumn{5}{l}{\textit{Cosmology}} \\
$\Omega_b h^2$ & $0.02237$ & $6.65\times10^{-5}$ & $6.63\times10^{-5}$ & $0\%$ \\
$\Omega_c h^2$ & $0.1200$ & $4.45\times10^{-4}$ & $4.41\times10^{-4}$ & $1\%$ \\
$H_0$ & $67.36$ & $1.37$ & $1.26$ & $8\%$ \\
$n_s$ & $0.9649$ & $0.00210$ & $0.00209$ & $0\%$ \\
$10^{9}A_s e^{-2\tau}$ & $1.883$ & $0.00345$ & $0.00344$ & $0\%$ \\
$\tau$ & $0.05440$ & $0.00606$ & $0.00602$ & $1\%$ \\
$w_0$ & $-1$ & $0.142$ & $0.127$ & $11\%$ \\
$w_a$ & $0$ & $0.369$ & $0.327$ & $11\%$ \\
\addlinespace
\multicolumn{5}{l}{\textit{Intrinsic alignment}} \\
$A_1$ & $0.9200$ & $0.0507$ & $0.0467$ & $8\%$ \\
$\eta_1$ & $2.260$ & $0.0860$ & $0.0778$ & $10\%$ \\
$A_2$ & $0.4000$ & $0.0626$ & $0.0527$ & $16\%$ \\
$A_\delta$ & $-0.8300$ & $0.105$ & $0.0861$ & $18\%$ \\
$\eta_2$ & $2.690$ & $0.393$ & $0.348$ & $11\%$ \\
\addlinespace
\multicolumn{5}{l}{\textit{Baryonic feedback}} \\
$\log_{10}T_{\rm AGN}$ & $7.800$ & $0.0686$ & $0.0671$ & $2\%$ \\
\midrule
FoM & -- & $115$ & $138$ & $\times1.20$ \\
\bottomrule
\end{tabular}

\end{table}

\begin{table}[htbp]
  \centering
  \caption{Same as table~\ref{tab:constraints_10bins} without the CMB, with $\tau$ fixed and a BBN prior on $\Omega_\mathrm{b}h^2$ (section~\ref{sec:fisher}).\label{tab:constraints_10bins_nocmb}}
  \begin{tabular}{lrrrr}
\toprule
 &  &  & \multicolumn{2}{c}{+ $4{\times}3$pt} \\
 &  & \multicolumn{1}{c}{$3{\times}2$pt} & \multicolumn{2}{c}{(skew-spectra)} \\
\cmidrule(lr){3-3}\cmidrule(lr){4-5}
Parameter & Fiducial & $\sigma$ & $\sigma$ & Gain \\
\midrule
\multicolumn{5}{l}{\textit{Cosmology}} \\
$\Omega_b h^2$ & $0.02237$ & $3.59\times10^{-4}$ & $3.59\times10^{-4}$ & $0\%$ \\
$\Omega_c h^2$ & $0.1200$ & $0.00479$ & $0.00429$ & $10\%$ \\
$H_0$ & $67.36$ & $1.94$ & $1.74$ & $10\%$ \\
$n_s$ & $0.9649$ & $0.0239$ & $0.0212$ & $11\%$ \\
$10^{9}A_s$ & $2.099$ & $0.186$ & $0.156$ & $16\%$ \\
$w_0$ & $-1$ & $0.183$ & $0.158$ & $14\%$ \\
$w_a$ & $0$ & $0.605$ & $0.510$ & $16\%$ \\
\addlinespace
\multicolumn{5}{l}{\textit{Intrinsic alignment}} \\
$A_1$ & $0.9200$ & $0.0578$ & $0.0523$ & $10\%$ \\
$\eta_1$ & $2.260$ & $0.0874$ & $0.0786$ & $10\%$ \\
$A_2$ & $0.4000$ & $0.0638$ & $0.0530$ & $17\%$ \\
$A_\delta$ & $-0.8300$ & $0.121$ & $0.0978$ & $19\%$ \\
$\eta_2$ & $2.690$ & $0.416$ & $0.359$ & $14\%$ \\
\addlinespace
\multicolumn{5}{l}{\textit{Baryonic feedback}} \\
$\log_{10}T_{\rm AGN}$ & $7.800$ & $0.173$ & $0.150$ & $13\%$ \\
\midrule
FoM & -- & $29.5$ & $40.4$ & $\times1.37$ \\
\bottomrule
\end{tabular}

\end{table}

With the CMB, the cosmological parameters benefiting most from the \fourxthree\ are again $H_0$, $w_0$ and $w_a$, whose uncertainties all decrease by about $10\%$. The dark energy figure of merit increases by a factor $1.20$, from $115$ to $138$. Among nuisance parameters, the second-order intrinsic alignment parameters $A_2$ and $A_\delta$ still improve substantially, by $16\%$ and $18\%$. The figures of merit we obtain are lower than those of the \Euclid\ forecasts of ref.~\cite{Euclid:2019clj}, owing to our conservative scale cuts and to the larger number of nuisance parameters in our galaxy bias, intrinsic alignment and baryonic feedback models.

Without the CMB, skew-spectra improve all cosmological parameters (except $\Omega_bh^2$) by $10$ to $16\%$, and the figure of merit by a factor $1.37$, from $29.5$ to $40.4$. The improvement is larger than with the CMB, since the CMB no longer constrains the parameters describing the early Universe, so that the calibration of nuisance parameters by the \fourxthree\ benefits all cosmological parameters. Without the CMB, the uncertainty on baryonic feedback also decreases by $14\%$, compared to $2\%$ with the CMB.

The relative improvement on individual parameters is smaller than for three bins. This trend is already present with the full bispectra: from three to five bins, the improvement on $H_0$ decreases from $23\%$ to $19\%$, and those on $w_0$ and $w_a$ from $18\%$ and $13\%$ to $12\%$ and $9\%$. This decrease is, however, much smaller once correlations between parameters are taken into account: the gain in figure of merit remains stable, and the ratio of the volumes of the Fisher ellipsoids in the space of the eight cosmological parameters, marginalised over nuisance parameters, only decreases from $1.50$ to $1.40$. This residual decrease can be explained by the better calibration of nuisance parameters by the \threextwo\ alone as bins are added. First, each lens bin has more weak lensing bins behind it, which improves the calibration of galaxy bias. Second, intrinsic alignment parameters are shared by all tomographic bins, so that they are constrained by more weak lensing bins, leaving less room for three-point statistics to improve them. Analyses assigning independent intrinsic alignment parameters to each tomographic bin would not benefit from this sharing, and three-point statistics could then improve them more.

For ten bins, the compression adds a further loss of information: the fraction of the bispectrum gain recovered by skew-spectra decreases as more bins are included (section~\ref{sec:skew_validation}), since our selection, optimised for three bins, contains no skew-spectra of types $GLL$ and $LLL$, whose bispectra contribute more as the number of bins increases (section~\ref{sec:origin_improvement}). The calibration of intrinsic alignments nevertheless plays a larger role with ten bins than with three: when the higher-order intrinsic alignment parameters are fixed, the gain in figure of merit from skew-spectra decreases from a factor $1.20$ to $1.15$ for ten bins, against $1.26$ to $1.25$ for three bins.

Finally, we repeat the test of section~\ref{sec:robustness} for this configuration, computing the bias induced on all parameters by the model errors on small scales, estimated from the reference predictions of section~\ref{sec:reference_predictions}. The bias on $w_0$ and $w_a$ remains below $0.3\sigma$, and the largest bias on cosmological parameters is $0.38\sigma$, on $H_0$, which indicates that our scale cuts remain adequate for ten tomographic bins. This bias is slightly smaller than that obtained with the full \fourxthree\ bispectra for five bins, which could be explained by the fact that the skew-spectra do not extract all the information contained in bispectra, and may therefore be less sensitive to model errors.

\subsection{Baryonic feedback and intrinsic alignments}\label{sec:tagn_ia}

In section~\ref{sec:origin_improvement}, we found that baryonic feedback is partly degenerate with the higher-order intrinsic alignment parameters in the CMB and \threextwo\ analysis. With ten tomographic bins, this degeneracy remains, but mostly with $A_\delta$: the correlation coefficient between $\log_{10}T_\mathrm{AGN}$ and $A_\delta$ is $-0.29$, whereas that with $A_2$ drops from $0.48$ for three bins to $0.09$. In contrast, ref.~\cite{Euclid:2026klv} did not find any degeneracy between intrinsic alignment parameters and $T_\mathrm{AGN}$ in their analysis of \Euclid\ \threextwo\ observables. To understand this difference, we study the degeneracy in the \threextwo\ analysis without the CMB.

\begin{figure}[htbp]
\centering
\includegraphics{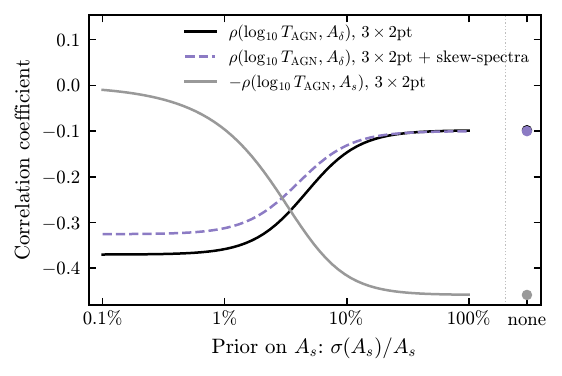}
\caption{Correlation coefficients of $\log_{10}T_\mathrm{AGN}$ with $A_\delta$, from \threextwo\ observables (black solid line) and when adding the $210$ skew-spectra of section~\ref{sec:skew_generalisation} (dashed line), and with $A_s$, from \threextwo\ observables (grey line, plotted with a minus sign), as a function of the width of a Gaussian prior on $A_s$, for ten tomographic bins without the CMB. As in ref.~\cite{Euclid:2026klv}, $\Omega_\mathrm{b}h^2$ and $n_s$ are fixed. The points on the right show the results without prior on $A_s$.\label{fig:tagn_as_prior}}
\end{figure}

Without the CMB, the correlation coefficient between $\log_{10}T_\mathrm{AGN}$ and $A_\delta$ drops to $-0.17$, while $\log_{10}T_\mathrm{AGN}$ becomes strongly degenerate with the amplitude of the primordial power spectrum, with a correlation coefficient of $0.86$ with $A_s$. When $A_s$ is free, the suppression of the small-scale weak lensing power spectra caused by baryonic feedback can thus be compensated by a change of the overall amplitude of the matter power spectrum, so that the degeneracy with intrinsic alignments does not appear. To compare with ref.~\cite{Euclid:2026klv}, we fix $\Omega_\mathrm{b}h^2$ and $n_s$ as in their analysis, and add a Gaussian prior on $A_s$ of varying width. Figure~\ref{fig:tagn_as_prior} shows how the degeneracy emerges as $A_s$ gets constrained. Without prior, the correlation coefficient between $\log_{10}T_\mathrm{AGN}$ and $A_\delta$ is only $-0.10$, consistently with ref.~\cite{Euclid:2026klv}. As the prior tightens, the correlation with $A_s$ vanishes, while that with $A_\delta$ increases, reaching $-0.36$ for a $1\%$ prior. A constraint on $A_s$ at the percent level, as provided by CMB experiments, is therefore enough for the degeneracy between baryonic feedback and intrinsic alignments to appear.

Three-point statistics could help break this degeneracy, by calibrating the higher-order intrinsic alignment parameters: for three bins, bispectra recover about $70\%$ of the information on $\log_{10}T_\mathrm{AGN}$ lost by marginalising over these parameters, almost entirely through the cross-bispectra $B^{GGL}$ (section~\ref{sec:origin_improvement}). With ten bins, our skew-spectra, which were selected to maximise the dark energy figure of merit, do not significantly reduce this degeneracy, and skew-spectra optimised for this purpose, for instance including more configurations of type $GGL$, could perform better. This would be even more valuable for intrinsic alignments stronger than in our fiducial model, or for more flexible baryonic feedback models with several free parameters, which could open further degeneracies with intrinsic alignment parameters.

\section{Conclusion}\label{sec:conclusion}

In this work, we investigated the combination of galaxy clustering and weak lensing bispectra, which we refer to as the \fourxthree, with the standard \threextwo. Only a handful of works had previously considered this combination, and to our knowledge this is the first forecast of the improvement of dark energy constraints for stage-IV galaxy surveys obtained by combining 1-loop power spectra with all tree-level auto- and cross-bispectra of galaxy clustering and weak lensing. For a \Euclid-like survey with very conservative scale cuts, we find that, in combination with CMB data, the \fourxthree\ improves the dark energy figure of merit by about $30\%$, a gain that remains stable for three, four and five tomographic bins.

Extending the \fourxthree\ to ten tomographic bins requires compressing the bispectra, whose data vector and covariance would otherwise become too large. We proposed a compression based on skew-spectra, and identified seven classes of skew-spectra retaining most of the information. Although this compression is suboptimal, it improves the figure of merit by $20\%$ when combined with \threextwo\ and CMB data, and by $37\%$ without the CMB.

We showed that the improvement mainly comes from breaking the degeneracies between linear and second-order galaxy bias parameters: all of them contribute to bispectra at tree-level, whereas the higher-order parameters only enter power spectra through loop corrections. Accordingly, the galaxy clustering bispectra alone provide between $56\%$ and $65\%$ of the improvement of the figure of merit for three to five tomographic bins. Bispectra also substantially improve the constraints on the second-order intrinsic alignment parameters of the TATT model, although, for three tomographic bins, this barely translates into better cosmological constraints. It does, however, partly break a degeneracy between second-order intrinsic alignments and baryonic feedback, which appears once the CMB constrains the amplitude of the matter power spectrum. Both effects come almost entirely from the cross-bispectra between two galaxy clustering fields and one weak lensing field, which demonstrates the value of including cross-bispectra in such analyses. This could become more valuable in future analyses allowing more freedom in intrinsic alignment and baryonic feedback modelling, for instance through redshift-dependent intrinsic alignment parameters or more flexible baryonic feedback models.

Several limitations remain to be addressed before applying this framework to data. These include the cross-covariance between power spectra and bispectra, non-Gaussian and super-sample covariance contributions, photometric redshift uncertainties, and magnification bias, which should be modelled consistently in power spectra and bispectra. In addition, our scale cuts are validated against matter predictions only, and should be tested with galaxy mocks including intrinsic alignments. Conversely, these scale cuts rely on perturbation theory alone and are deliberately conservative: hybrid approaches combining perturbation theory with corrections calibrated on simulations could extend the analysis to smaller scales.

Despite these limitations, our results demonstrate the potential of combining galaxy clustering and weak lensing bispectra with standard two-point analyses to improve cosmological constraints and calibrate astrophysical nuisance parameters. They motivate further work on more complete covariance modelling and improved descriptions of non-linear scales, with the aim of extracting more information from the higher-order statistics of future galaxy surveys.

\acknowledgments
This work has been supported by UK STFC grant ST/X001040/1. We used Claude Opus 5.5 (Anthropic) to improve the writing of this paper and to perform cross-checks of the analysis. The authors reviewed all suggestions and take full responsibility for the content.

\bibliographystyle{JHEP}
\bibliography{biblio}

@article{Heinrich:2020xub,
    author = "Heinrich, Chen and Dor{\'e}, Olivier",
    title = "{Cross-bispectra constraints on modified gravity theories from the Nancy Grace Roman Space Telescope and the Rubin Observatory Legacy Survey of Space and Time}",
    eprint = "2006.03138",
    archivePrefix = "arXiv",
    primaryClass = "astro-ph.CO",
    doi = "10.1103/PhysRevD.102.123549",
    journal = "Phys. Rev. D",
    volume = "102",
    number = "12",
    pages = "123549",
    year = "2020"
}

@article{Euclid:2019clj,
    author = "Blanchard, A. and others",
    collaboration = "Euclid",
    title = "{Euclid preparation. VII. Forecast validation for Euclid cosmological probes}",
    eprint = "1910.09273",
    archivePrefix = "arXiv",
    primaryClass = "astro-ph.CO",
    doi = "10.1051/0004-6361/202038071",
    journal = "Astron. Astrophys.",
    volume = "642",
    pages = "A191",
    year = "2020"
}

@article{McDonald:2009dh,
    author = "McDonald, Patrick and Roy, Arabindo",
    title = "{Clustering of dark matter tracers: generalizing bias for the coming era of precision LSS}",
    eprint = "0902.0991",
    archivePrefix = "arXiv",
    primaryClass = "astro-ph.CO",
    doi = "10.1088/1475-7516/2009/08/020",
    journal = "JCAP",
    volume = "08",
    pages = "020",
    year = "2009"
}

@article{Desjacques:2016bnm,
    author = "Desjacques, Vincent and Jeong, Donghui and Schmidt, Fabian",
    title = "{Large-Scale Galaxy Bias}",
    eprint = "1611.09787",
    archivePrefix = "arXiv",
    primaryClass = "astro-ph.CO",
    doi = "10.1016/j.physrep.2017.12.002",
    journal = "Phys. Rept.",
    volume = "733",
    pages = "1--193",
    year = "2018"
}

@article{Blazek:2017wbz,
    author = "Blazek, Jonathan and MacCrann, Niall and Troxel, M. A. and Fang, Xiao",
    title = "{Beyond linear galaxy alignments}",
    eprint = "1708.09247",
    archivePrefix = "arXiv",
    primaryClass = "astro-ph.CO",
    doi = "10.1103/PhysRevD.100.103506",
    journal = "Phys. Rev. D",
    volume = "100",
    number = "10",
    pages = "103506",
    year = "2019"
}

@article{Hirata:2007np,
    author = "Hirata, Christopher M. and Mandelbaum, Rachel and Ishak, Mustapha and Seljak, Uros and Nichol, Robert and Pimbblet, Kevin A. and Ross, Nicholas P. and Wake, David",
    title = "{Intrinsic galaxy alignments from the 2SLAQ and SDSS surveys: Luminosity and redshift scalings and implications for weak lensing surveys}",
    eprint = "astro-ph/0701671",
    archivePrefix = "arXiv",
    doi = "10.1111/j.1365-2966.2007.12312.x",
    journal = "Mon. Not. Roy. Astron. Soc.",
    volume = "381",
    pages = "1197--1218",
    year = "2007"
}

@article{Bridle:2007ft,
    author = "Bridle, Sarah and King, Lindsay",
    title = "{Dark energy constraints from cosmic shear power spectra: impact of intrinsic alignments on photometric redshift requirements}",
    eprint = "0705.0166",
    archivePrefix = "arXiv",
    primaryClass = "astro-ph",
    doi = "10.1088/1367-2630/9/12/444",
    journal = "New J. Phys.",
    volume = "9",
    pages = "444",
    year = "2007"
}

@article{Euclid:2026klv,
    author = "Navarro-Giron{\'e}s, D. and others",
    collaboration = "Euclid",
    title = "{Euclid preparation. Impact of galaxy intrinsic alignment modelling choices on Euclid 3x2pt cosmology}",
    eprint = "2602.16448",
    archivePrefix = "arXiv",
    primaryClass = "astro-ph.CO",
    month = "2",
    year = "2026"
}

@article{Catelan:2000vm,
    author = "Catelan, Paolo and Kamionkowski, Marc and Blandford, Roger D.",
    title = "{Intrinsic and extrinsic galaxy alignment}",
    eprint = "astro-ph/0005470",
    archivePrefix = "arXiv",
    doi = "10.1046/j.1365-8711.2001.04105.x",
    journal = "Mon. Not. Roy. Astron. Soc.",
    volume = "320",
    pages = "L7--L13",
    year = "2001"
}

@article{Hirata:2004gc,
    author = "Hirata, Christopher M. and Seljak, Uros",
    title = "{Intrinsic alignment-lensing interference as a contaminant of cosmic shear}",
    eprint = "astro-ph/0406275",
    archivePrefix = "arXiv",
    doi = "10.1103/PhysRevD.82.049901",
    journal = "Phys. Rev. D",
    volume = "70",
    pages = "063526",
    year = "2004",
    note = "[Erratum: Phys.Rev.D 82, 049901 (2010)]"
}

@article{Krause:2015jqa,
    author = "Krause, Elisabeth and Eifler, Tim and Blazek, Jonathan",
    title = "{The impact of intrinsic alignment on current and future cosmic shear surveys}",
    eprint = "1506.08730",
    archivePrefix = "arXiv",
    primaryClass = "astro-ph.CO",
    doi = "10.1093/mnras/stv2615",
    journal = "Mon. Not. Roy. Astron. Soc.",
    volume = "456",
    number = "1",
    pages = "207--222",
    year = "2016"
}

@article{Brown:2000gt,
    author = "Brown, M. L. and Taylor, A. N. and Hambly, N. C. and Dye, S.",
    title = "{Measurement of intrinsic alignments in galaxy ellipticities}",
    eprint = "astro-ph/0009499",
    archivePrefix = "arXiv",
    doi = "10.1046/j.1365-8711.2002.05354.x",
    journal = "Mon. Not. Roy. Astron. Soc.",
    volume = "333",
    pages = "501",
    year = "2002"
}

@article{Hambly:2001yi,
    author = "Hambly, Nigel and Irwin, Mike and MacGillivray, Harvey",
    title = "{The supercosmos sky survey. paper. 2. Image detection, parameterisation, classification and photometry}",
    eprint = "astro-ph/0108290",
    archivePrefix = "arXiv",
    doi = "10.1111/j.1365-2966.2001.04661.x",
    journal = "Mon. Not. Roy. Astron. Soc.",
    volume = "326",
    pages = "1295",
    year = "2001"
}

@article{DiDio:2018unb,
    author = "Di Dio, Enea and Durrer, Ruth and Maartens, Roy and Montanari, Francesco and Umeh, Obinna",
    title = "{The Full-Sky Angular Bispectrum in Redshift Space}",
    eprint = "1812.09297",
    archivePrefix = "arXiv",
    primaryClass = "astro-ph.CO",
    reportNumber = "IFT-UAM/CSIC-18-134, HIP-2018-38/TH",
    doi = "10.1088/1475-7516/2019/04/053",
    journal = "JCAP",
    volume = "04",
    pages = "053",
    year = "2019"
}

@article{Montandon:2025uul,
    author = "Montandon, Thomas and Di Dio, Enea and Rampf, Cornelius and Adamek, Julian",
    title = "{Angular bispectrum of matter number counts in cosmic structures}",
    eprint = "2501.05422",
    archivePrefix = "arXiv",
    primaryClass = "astro-ph.CO",
    doi = "10.33232/001c.155624",
    journal = "Open J. Astrophys.",
    volume = "9",
    pages = "155624",
    year = "2026"
}

@article{Farren:2023yna,
    author = "Farren, Gerrit S. and Sherwin, Blake D. and Bolliet, Boris and Namikawa, Toshiya and Ferraro, Simone and Krolewski, Alex",
    title = "{Detection of the CMB lensing -- galaxy bispectrum}",
    eprint = "2311.04213",
    archivePrefix = "arXiv",
    primaryClass = "astro-ph.CO",
    month = "11",
    year = "2023"
}

@article{Bucher:2015ura,
    author = "Bucher, Martin and Racine, Benjamin and van Tent, Bartjan",
    title = "{The binned bispectrum estimator: template-based and non-parametric CMB non-Gaussianity searches}",
    eprint = "1509.08107",
    archivePrefix = "arXiv",
    primaryClass = "astro-ph.CO",
    reportNumber = "LPT-ORSAY-15-67",
    doi = "10.1088/1475-7516/2016/05/055",
    journal = "JCAP",
    volume = "05",
    pages = "055",
    year = "2016"
}

@article{Verdiani:2025jgz,
    author = "Verdiani, Francesco and Harscouet, Lea and Zennaro, Matteo and Alonso, David and Hadzhiyska, Boryana",
    title = "{Cosmological constraints from the angular power spectrum and bispectrum of luminous red galaxies and CMB lensing}",
    eprint = "2510.17796",
    archivePrefix = "arXiv",
    primaryClass = "astro-ph.CO",
    month = "10",
    year = "2025"
}

@article{Frugte:2025afv,
    author = "Frugte, Jonas and Meerburg, P. Daniel",
    title = "{Future Parameter Constraints from Weak Lensing CMB and Galaxy Lensing Power- and Bispectra}",
    eprint = "2507.20262",
    archivePrefix = "arXiv",
    primaryClass = "astro-ph.CO",
    month = "7",
    year = "2025"
}

@article{Harscouet:2025ksm,
    author = "Harscouet, Lea and Alonso, David and Nicola, Andrina and Slosar, An{\v{z}}e",
    title = "{Constraints from CMB lensing tomography with projected bispectra}",
    eprint = "2507.07968",
    archivePrefix = "arXiv",
    primaryClass = "astro-ph.CO",
    doi = "10.33232/001c.155341",
    month = "7",
    year = "2025"
}

@article{Karacayli:2024udo,
    author = {Kara{\c{c}}ayl{\i}, Naim G{\"o}ksel and others},
    title = "{CMB lensing and Ly{\ensuremath{\alpha}} forest cross bispectrum from DESI{\textquoteright}s first-year quasar sample}",
    eprint = "2405.14988",
    archivePrefix = "arXiv",
    primaryClass = "astro-ph.CO",
    doi = "10.1103/PhysRevD.110.063505",
    journal = "Phys. Rev. D",
    volume = "110",
    number = "6",
    pages = "063505",
    year = "2024"
}

@article{CHIME:2026rbk,
    author = "Chakraborty, Arnab and others",
    collaboration = "CHIME",
    title = "{The Squeezed Bispectrum from CHIME HI Emission and Planck CMB Lensing: Current Sensitivity and Forecasts}",
    eprint = "2601.03240",
    archivePrefix = "arXiv",
    primaryClass = "astro-ph.CO",
    month = "1",
    year = "2026"
}

@article{DES:2025llp,
    author = "Gomes, R. C. H. and others",
    collaboration = "DES",
    title = "{Dark Energy Survey Year 3 Results: Cosmological constraints from second- and third-order shear statistics}",
    eprint = "2508.14018",
    archivePrefix = "arXiv",
    primaryClass = "astro-ph.CO",
    reportNumber = "FERMILAB-PUB-25-0572-PPD",
    doi = "10.1103/sxlz-t9gb",
    journal = "Phys. Rev. D",
    volume = "112",
    number = "12",
    pages = "123515",
    year = "2025"
}

@article{Sugiyama:2025dfc,
    author = "Sugiyama, Sunao and Gomes, Rafael C. H. and Jain, Bhuvnesh",
    title = "{Cosmology from a joint analysis of second- and third-order shear statistics with Subaru Hyper Suprime-Cam Year 3 data}",
    eprint = "2508.14019",
    archivePrefix = "arXiv",
    primaryClass = "astro-ph.CO",
    doi = "10.1103/6wk9-q5gh",
    journal = "Phys. Rev. D",
    volume = "112",
    number = "12",
    pages = "123545",
    year = "2025"
}

@article{Halder:2023kfy,
    author = "Halder, Anik and Gong, Zhengyangguang and Barreira, Alexandre and Friedrich, Oliver and Seitz, Stella and Gruen, Daniel",
    title = "{Beyond 3{\texttimes}2-point cosmology: the integrated shear and galaxy 3-point correlation functions}",
    eprint = "2305.17132",
    archivePrefix = "arXiv",
    primaryClass = "astro-ph.CO",
    doi = "10.1088/1475-7516/2023/10/028",
    journal = "JCAP",
    volume = "10",
    pages = "028",
    year = "2023"
}

@article{Lu:2025gki,
    author = "Lu, Zhiyu and Simon, Th{\'e}o and Zhang, Pierre",
    title = "{Preference for evolving dark energy in light of the galaxy bispectrum}",
    eprint = "2503.04602",
    archivePrefix = "arXiv",
    primaryClass = "astro-ph.CO",
    month = "3",
    year = "2025"
}

@article{DESI:2026haa,
    author = "Bansal, P. and others",
    collaboration = "DESI",
    title = "{FolpsD: combining EFT and phenomenological approaches for joint power spectrum and bispectrum analyses}",
    eprint = "2604.08895",
    archivePrefix = "arXiv",
    primaryClass = "astro-ph.CO",
    reportNumber = "FERMILAB-PUB-26-0266-PPD",
    doi = "10.5281/zenodo.19302052",
    month = "4",
    year = "2026"
}

@article{Porth:2023dzx,
    author = "Porth, Lucas and Heydenreich, Sven and Burger, Pierre and Linke, Laila and Schneider, Peter",
    title = "{A road map to cosmological parameter analysis with third-order shear statistics - III. Efficient estimation of third-order shear correlation functions and an application to the KiDS-1000 data}",
    eprint = "2309.08601",
    archivePrefix = "arXiv",
    primaryClass = "astro-ph.CO",
    doi = "10.1051/0004-6361/202347987",
    journal = "Astron. Astrophys.",
    volume = "689",
    pages = "A227",
    year = "2024"
}

@article{Oel:2026eni,
    author = "Oel, Jonathan and Porth, Lucas and Schneider, Peter and Silvestre-Rosello, Elena",
    title = "{Fourth-order galaxy-galaxy-lensing: Theoretical framework and direct estimation}",
    eprint = "2604.18378",
    archivePrefix = "arXiv",
    primaryClass = "astro-ph.CO",
    month = "4",
    year = "2026"
}

@article{Burger:2023qef,
    author = "Burger, Pierre A. and others",
    title = "{KiDS-1000 cosmology: Combined second- and third-order shear statistics}",
    eprint = "2309.08602",
    archivePrefix = "arXiv",
    primaryClass = "astro-ph.CO",
    doi = "10.1051/0004-6361/202347986",
    journal = "Astron. Astrophys.",
    volume = "683",
    pages = "A103",
    year = "2024"
}

@article{Silvestre-Rosello:2025rmt,
    author = "Silvestre-Rosello, Elena and Porth, Lucas and Schneider, Peter and Linke, Laila and Krueger, Jonas and Grandis, Sebastian and Oel, Jonathan",
    title = "{Towards an application of fourth-order shear statistics I. The information content of $\langle M_\mathrm{ap}^4 \rangle $}",
    eprint = "2509.07973",
    archivePrefix = "arXiv",
    primaryClass = "astro-ph.CO",
    doi = "10.1051/0004-6361/202557147",
    journal = "Astron. Astrophys.",
    volume = "708",
    pages = "A243",
    year = "2026"
}

@article{Porth:2025pem,
    author = "Porth, Lucas and Silvestre-Rosello, Elena and Schneider, Peter and Larma, Martina",
    title = "{Towards an application of fourth-order shear statistics II: Efficient estimation of fourth-order shear correlation functions and an application to the DES Y3 data}",
    eprint = "2509.07974",
    archivePrefix = "arXiv",
    primaryClass = "astro-ph.CO",
    month = "9",
    year = "2025"
}

@article{Sugiyama:2024uqo,
    author = "Sugiyama, Sunao and Gomes, Rafael C. H. and Jarvis, Mike",
    title = "{Fast modeling of the shear three-point correlation function}",
    eprint = "2407.01798",
    archivePrefix = "arXiv",
    primaryClass = "astro-ph.CO",
    month = "7",
    year = "2024"
}

@article{Pyne:2020ijd,
    author = "Pyne, Susan and Joachimi, Benjamin",
    title = "{Self-calibration of weak lensing systematic effects using combined two- and three-point statistics}",
    eprint = "2010.00614",
    archivePrefix = "arXiv",
    primaryClass = "astro-ph.CO",
    doi = "10.1093/mnras/stab413",
    journal = "Mon. Not. Roy. Astron. Soc.",
    volume = "503",
    number = "2",
    pages = "2300--2317",
    year = "2021"
}

@article{DES:2022oqz,
    author = "Secco, L. F. and others",
    collaboration = "DES",
    title = "{Dark Energy Survey Year 3 Results: Three-point shear correlations and mass aperture moments}",
    eprint = "2201.05227",
    archivePrefix = "arXiv",
    primaryClass = "astro-ph.CO",
    reportNumber = "DES-2021-0681, FERMILAB-PUB-22-001-PPD-SCD",
    doi = "10.1103/PhysRevD.105.103537",
    journal = "Phys. Rev. D",
    volume = "105",
    number = "10",
    pages = "103537",
    year = "2022"
}

@article{Jung:2021weh,
    author = "Jung, Gabriel and Namikawa, Toshiya and Liguori, Michele and Munshi, Dipak and Heavens, Alan",
    title = "{The integrated angular bispectrum of weak lensing}",
    eprint = "2102.05521",
    archivePrefix = "arXiv",
    primaryClass = "astro-ph.CO",
    doi = "10.1088/1475-7516/2021/06/055",
    journal = "JCAP",
    volume = "06",
    pages = "055",
    year = "2021"
}

@article{Munshi:2021kth,
    author = "Munshi, D. and Jung, G. and Kitching, T. D. and McEwen, J. and Liguori, M. and Namikawa, T. and Heavens, A.",
    title = "{Position-dependent correlation function of weak-lensing convergence}",
    eprint = "2104.01185",
    archivePrefix = "arXiv",
    primaryClass = "astro-ph.CO",
    doi = "10.1103/PhysRevD.107.043516",
    journal = "Phys. Rev. D",
    volume = "107",
    number = "4",
    pages = "043516",
    year = "2023"
}

@article{Halder:2021itp,
    author = "Halder, Anik and Friedrich, Oliver and Seitz, Stella and Varga, Tamas N.",
    title = "{The integrated three-point correlation function of cosmic shear}",
    eprint = "2102.10177",
    archivePrefix = "arXiv",
    primaryClass = "astro-ph.CO",
    doi = "10.1093/mnras/stab1801",
    journal = "Mon. Not. Roy. Astron. Soc.",
    volume = "506",
    number = "2",
    pages = "2780--2803",
    year = "2021"
}

@article{Samario-Nava:2025auc,
    author = "Samario-Nava, Sofia and Aviles, Alejandro and Hidalgo, Juan Carlos",
    title = "{Non-Gaussian statistics in galaxy weak lensing: compressed three-point correlations and cosmological forecasts}",
    eprint = "2506.19811",
    archivePrefix = "arXiv",
    primaryClass = "astro-ph.CO",
    doi = "10.1088/1475-7516/2025/10/006",
    journal = "JCAP",
    volume = "10",
    pages = "006",
    year = "2025"
}

@article{Spaar:2023his,
    author = "Spaar, Simon and Zhang, Pierre",
    title = "{Cosmological constraints from combined probes with the three-point statistics of galaxies at one-loop precision}",
    eprint = "2312.15164",
    archivePrefix = "arXiv",
    primaryClass = "astro-ph.CO",
    doi = "10.1103/PhysRevD.111.023529",
    journal = "Phys. Rev. D",
    volume = "111",
    number = "2",
    pages = "023529",
    year = "2025"
}

@article{Bakx:2025pop,
    author = "Bakx, Thomas and Ivanov, Mikhail M. and Philcox, Oliver H. E. and Vlah, Zvonimir",
    title = "{One-Loop Galaxy Bispectrum: Consistent Theory, Efficient Analysis with COBRA, and Implications for Cosmological Parameters}",
    eprint = "2507.22110",
    archivePrefix = "arXiv",
    primaryClass = "astro-ph.CO",
    reportNumber = "MIT-CTP/5891, RBI-ThPhys-2025-23",
    month = "7",
    year = "2025"
}

@article{Euclid:2026glh,
    author = "Pardede, K. and others",
    collaboration = "Euclid",
    title = "{Euclid preparation. Galaxy power spectrum and bispectrum modelling}",
    eprint = "2603.27966",
    archivePrefix = "arXiv",
    primaryClass = "astro-ph.CO",
    month = "3",
    year = "2026"
}

@article{Chudaykin:2025aux,
    author = "Chudaykin, Anton and Ivanov, Mikhail M. and Philcox, Oliver H. E.",
    title = "{Reanalyzing DESI DR1. I. {\ensuremath{\Lambda}}CDM constraints from the power spectrum and bispectrum}",
    eprint = "2507.13433",
    archivePrefix = "arXiv",
    primaryClass = "astro-ph.CO",
    reportNumber = "MIT-CPT/5890",
    doi = "10.1103/qsnt-dppc",
    journal = "Phys. Rev. D",
    volume = "113",
    number = "6",
    pages = "063502",
    year = "2026"
}

@article{Novell-Masot:2026sgg,
    author = "Novell-Masot, S. and others",
    title = "{Cosmological constraints from the DESI DR1 joint power spectrum and bispectrum analysis}",
    eprint = "2603.19356",
    archivePrefix = "arXiv",
    primaryClass = "astro-ph.CO",
    reportNumber = "FERMILAB-PUB-26-0197-PPD",
    doi = "10.5281/zenodo.19099550",
    month = "3",
    year = "2026"
}

@article{Gil-Marin:2016wya,
    author = "Gil-Mar{\'\i}n, H{\'e}ctor and Percival, Will J. and Verde, Licia and Brownstein, Joel R. and Chuang, Chia-Hsun and Kitaura, Francisco-Shu and Rodr{\'\i}guez-Torres, Sergio A. and Olmstead, Matthew D.",
    title = "{The clustering of galaxies in the SDSS-III Baryon Oscillation Spectroscopic Survey: RSD measurement from the power spectrum and bispectrum of the DR12 BOSS galaxies}",
    eprint = "1606.00439",
    archivePrefix = "arXiv",
    primaryClass = "astro-ph.CO",
    doi = "10.1093/mnras/stw2679",
    journal = "Mon. Not. Roy. Astron. Soc.",
    volume = "465",
    number = "2",
    pages = "1757--1788",
    year = "2017"
}

@article{NovellMasot:2025fju,
    author = "Novell Masot, S. and others",
    title = "{Full-Shape analysis of the power spectrum and bispectrum of DESI DR1 LRG and QSO samples}",
    eprint = "2503.09714",
    archivePrefix = "arXiv",
    primaryClass = "astro-ph.CO",
    reportNumber = "FERMILAB-PUB-25-0173-PPD",
    doi = "10.1088/1475-7516/2025/06/005",
    journal = "JCAP",
    volume = "06",
    pages = "005",
    year = "2025"
}

@article{Gil-Marin:2014baa,
    author = "Gil-Mar{\'\i}n, H{\'e}ctor and Verde, Licia and Nore{\~n}a, Jorge and Cuesta, Antonio J. and Samushia, Lado and Percival, Will J. and Wagner, Christian and Manera, Marc and Schneider, Donald P.",
    title = "{The power spectrum and bispectrum of SDSS DR11 BOSS galaxies {\textendash} II. Cosmological interpretation}",
    eprint = "1408.0027",
    archivePrefix = "arXiv",
    primaryClass = "astro-ph.CO",
    doi = "10.1093/mnras/stv1359",
    journal = "Mon. Not. Roy. Astron. Soc.",
    volume = "452",
    number = "2",
    pages = "1914--1921",
    year = "2015"
}

@article{Philcox:2021kcw,
    author = "Philcox, Oliver H. E. and Ivanov, Mikhail M.",
    title = "{BOSS DR12 full-shape cosmology: {\ensuremath{\Lambda}}CDM constraints from the large-scale galaxy power spectrum and bispectrum monopole}",
    eprint = "2112.04515",
    archivePrefix = "arXiv",
    primaryClass = "astro-ph.CO",
    doi = "10.1103/PhysRevD.105.043517",
    journal = "Phys. Rev. D",
    volume = "105",
    number = "4",
    pages = "043517",
    year = "2022"
}

@article{DAmico:2022osl,
    author = "D'Amico, Guido and Donath, Yaniv and Lewandowski, Matthew and Senatore, Leonardo and Zhang, Pierre",
    title = "{The BOSS bispectrum analysis at one loop from the Effective Field Theory of Large-Scale Structure}",
    eprint = "2206.08327",
    archivePrefix = "arXiv",
    primaryClass = "astro-ph.CO",
    reportNumber = "NUHEP-TH/22-05",
    doi = "10.1088/1475-7516/2024/05/059",
    journal = "JCAP",
    volume = "05",
    pages = "059",
    year = "2024"
}

@article{Bera:2026ykq,
    author = "Bera, Avijit and Harnois-D{\'e}raps, Joachim and Mena-Fern{\'a}ndez, Juan and Jarvis, Mike and Doux, Cyrille and Heitmann, Katrin and Ishak, Mustapha",
    collaboration = "LSST Dark Energy Science",
    title = "{$4\times3$ Point Correlation Functions in Galaxy Surveys: Impact of Baryonic Feedback}",
    eprint = "2605.00818",
    archivePrefix = "arXiv",
    primaryClass = "astro-ph.CO",
    month = "5",
    year = "2026"
}

@article{Schneider:2005cs,
    author = "Schneider, Peter and Watts, Peter",
    title = "{Galaxy-galaxy-galaxy lensing: Third-order correlations between the galaxy and mass distributions in the Universe}",
    eprint = "astro-ph/0502552",
    archivePrefix = "arXiv",
    doi = "10.1051/0004-6361:20041923",
    journal = "Astron. Astrophys.",
    volume = "432",
    pages = "783",
    year = "2005"
}

@article{Simon:2007yg,
    author = "Simon, P. and Watts, P. and Schneider, P. and Hoekstra, H. and Gladders, M. D. and Yee, H. K. C. and Hsieh, B. C. and Lin, H.",
    title = "{First detection of galaxy-galaxy-galaxy lensing in RCS -- a new tool for studying the matter environment of galaxy pairs}",
    eprint = "0707.0066",
    archivePrefix = "arXiv",
    primaryClass = "astro-ph",
    reportNumber = "FERMILAB-PUB-08-009-CD",
    doi = "10.1051/0004-6361:20078197",
    journal = "Astron. Astrophys.",
    volume = "479",
    pages = "655--667",
    year = "2008"
}

@article{Simon:2013ps,
    author = "Simon, P. and others",
    title = "{CFHTLenS: Higher-order galaxy-mass correlations probed by galaxy-galaxy-galaxy lensing}",
    eprint = "1301.1863",
    archivePrefix = "arXiv",
    primaryClass = "astro-ph.CO",
    doi = "10.1093/mnras/stt069",
    journal = "Mon. Not. Roy. Astron. Soc.",
    volume = "430",
    pages = "2476",
    year = "2013"
}

@article{Chen:2021vba,
    author = "Chen, Shu-Fan and Lee, Hayden and Dvorkin, Cora",
    title = "{Precise and accurate cosmology with CMB{\texttimes}LSS power spectra and bispectra}",
    eprint = "2103.01229",
    archivePrefix = "arXiv",
    primaryClass = "astro-ph.CO",
    doi = "10.1088/1475-7516/2021/05/030",
    journal = "JCAP",
    volume = "05",
    pages = "030",
    year = "2021"
}

@article{Euclid:2024yrr,
    author = "Mellier, Y. and others",
    collaboration = "Euclid",
    title = "{Euclid. I. Overview of the Euclid mission}",
    eprint = "2405.13491",
    archivePrefix = "arXiv",
    primaryClass = "astro-ph.CO",
    doi = "10.1051/0004-6361/202450810",
    journal = "Astron. Astrophys.",
    volume = "697",
    pages = "A1",
    year = "2025"
}

@article{LSST:2008ijt,
    author = "Ivezi{\'c}, {\v{Z}}eljko and others",
    collaboration = "LSST",
    title = "{LSST: from Science Drivers to Reference Design and Anticipated Data Products}",
    eprint = "0805.2366",
    archivePrefix = "arXiv",
    primaryClass = "astro-ph",
    reportNumber = "SLAC-PUB-16076",
    doi = "10.3847/1538-4357/ab042c",
    journal = "Astrophys. J.",
    volume = "873",
    number = "2",
    pages = "111",
    year = "2019"
}

@article{Chevallier:2000qy,
    author = "Chevallier, Michel and Polarski, David",
    title = "{Accelerating universes with scaling dark matter}",
    eprint = "gr-qc/0009008",
    archivePrefix = "arXiv",
    doi = "10.1142/S0218271801000822",
    journal = "Int. J. Mod. Phys. D",
    volume = "10",
    pages = "213--224",
    year = "2001"
}

@article{Linder:2002et,
    author = "Linder, Eric V.",
    title = "{Exploring the expansion history of the universe}",
    eprint = "astro-ph/0208512",
    archivePrefix = "arXiv",
    doi = "10.1103/PhysRevLett.90.091301",
    journal = "Phys. Rev. Lett.",
    volume = "90",
    pages = "091301",
    year = "2003"
}

@article{Limber:1954zz,
    author = "Limber, D. Nelson",
    title = "{The Analysis of Counts of the Extragalactic Nebulae in Terms of a Fluctuating Density Field. II}",
    doi = "10.1086/145870",
    journal = "Astrophys. J.",
    volume = "119",
    pages = "655",
    year = "1954"
}

@article{LSSTDarkEnergyScience:2018yem,
    author = "Chisari, Nora Elisa and others",
    collaboration = "LSST Dark Energy Science",
    title = "{Core Cosmology Library: Precision Cosmological Predictions for LSST}",
    eprint = "1812.05995",
    archivePrefix = "arXiv",
    primaryClass = "astro-ph.CO",
    doi = "10.3847/1538-4365/ab1658",
    journal = "Astrophys. J. Suppl.",
    volume = "242",
    number = "1",
    pages = "2",
    year = "2019"
}

@article{McEwen:2016fjn,
    author = "McEwen, Joseph E. and Fang, Xiao and Hirata, Christopher M. and Blazek, Jonathan A.",
    title = "{FAST-PT: a novel algorithm to calculate convolution integrals in cosmological perturbation theory}",
    eprint = "1603.04826",
    archivePrefix = "arXiv",
    primaryClass = "astro-ph.CO",
    doi = "10.1088/1475-7516/2016/09/015",
    journal = "JCAP",
    volume = "09",
    pages = "015",
    year = "2016"
}

@article{Fang:2016wcf,
    author = "Fang, Xiao and Blazek, Jonathan A. and McEwen, Joseph E. and Hirata, Christopher M.",
    title = "{FAST-PT II: an algorithm to calculate convolution integrals of general tensor quantities in cosmological perturbation theory}",
    eprint = "1609.05978",
    archivePrefix = "arXiv",
    primaryClass = "astro-ph.CO",
    doi = "10.1088/1475-7516/2017/02/030",
    journal = "JCAP",
    volume = "02",
    pages = "030",
    year = "2017"
}

@article{Schmitz:2018rfw,
    author = "Schmitz, Denise M. and Hirata, Christopher M. and Blazek, Jonathan and Krause, Elisabeth",
    title = "{Time evolution of intrinsic alignments of galaxies}",
    eprint = "1805.02649",
    archivePrefix = "arXiv",
    primaryClass = "astro-ph.CO",
    doi = "10.1088/1475-7516/2018/07/030",
    journal = "JCAP",
    volume = "07",
    pages = "030",
    year = "2018"
}

@article{Bakx:2023mld,
    author = "Bakx, Thomas and Kurita, Toshiki and Chisari, Nora Elisa and Vlah, Zvonimir and Schmidt, Fabian",
    title = "{Effective field theory of intrinsic alignments at one loop order: a comparison to dark matter simulations}",
    eprint = "2303.15565",
    archivePrefix = "arXiv",
    primaryClass = "astro-ph.CO",
    doi = "10.1088/1475-7516/2023/10/005",
    journal = "JCAP",
    volume = "10",
    pages = "005",
    year = "2023"
}

@article{Bakx:2025uvf,
    author = "Bakx, Thomas and Kurita, Toshiki and Eggemeier, Alexander and Chisari, Nora Elisa and Vlah, Zvonimir",
    title = "{The Bispectrum of Intrinsic Alignments: II. Precision Comparison Against Dark Matter Simulations}",
    eprint = "2507.06818",
    archivePrefix = "arXiv",
    primaryClass = "astro-ph.CO",
    doi = "10.33232/001c.156361",
    month = "7",
    year = "2025"
}

@article{Gomes:2026wki,
    author = "Gomes, Rafael C. H. and Miller, Kyle and Sugiyama, Sunao and Blazek, Jonathan and Bakx, Thomas and Jain, Bhuvnesh",
    title = "{Tidal alignment and tidal torquing modeling for the cosmic shear three-point correlation function and mass aperture skewness}",
    eprint = "2601.09133",
    archivePrefix = "arXiv",
    primaryClass = "astro-ph.CO",
    month = "1",
    year = "2026"
}

@article{Mead:2020vgs,
    author = {Mead, Alexander and Brieden, Samuel and Tr{\"o}ster, Tilman and Heymans, Catherine},
    title = "{hmcode-2020: improved modelling of non-linear cosmological power spectra with baryonic feedback}",
    eprint = "2009.01858",
    archivePrefix = "arXiv",
    primaryClass = "astro-ph.CO",
    doi = "10.1093/mnras/stab082",
    journal = "Mon. Not. Roy. Astron. Soc.",
    volume = "502",
    number = "1",
    pages = "1401--1422",
    year = "2021"
}

@article{Cooray:2001ps,
    author = "Cooray, Asantha",
    title = "{Squared temperature-temperature power spectrum as a probe of the CMB bispectrum}",
    eprint = "astro-ph/0105415",
    archivePrefix = "arXiv",
    doi = "10.1103/PhysRevD.64.043516",
    journal = "Phys. Rev. D",
    volume = "64",
    pages = "043516",
    year = "2001"
}

@ARTICLE{MunshiHeavens2010,
       author = {{Munshi}, Dipak and {Heavens}, Alan},
        title = "{A new approach to probing primordial non-Gaussianity}",
      journal = {\mnras},
         year = 2010,
        month = feb,
       volume = {401},
       number = {4},
        pages = {2406-2418},
          doi = {10.1111/j.1365-2966.2009.15820.x},
archivePrefix = {arXiv},
       eprint = {0904.4478},
 primaryClass = {astro-ph.CO},
       adsurl = {https://ui.adsabs.harvard.edu/abs/2010MNRAS.401.2406M}
}

@ARTICLE{Pratten2011,
       author = {{Pratten}, Geraint and {Munshi}, Dipak},
        title = "{Non-Gaussianity in large-scale structure and Minkowski functionals}",
      journal = {\mnras},
         year = 2012,
        month = jul,
       volume = {423},
       number = {4},
        pages = {3209-3226},
          doi = {10.1111/j.1365-2966.2012.21103.x},
archivePrefix = {arXiv},
       eprint = {1108.1985},
 primaryClass = {astro-ph.CO},
       adsurl = {https://ui.adsabs.harvard.edu/abs/2012MNRAS.423.3209P}
}

@ARTICLE{2011MNRAS.414.3173M,
       author = {{Munshi}, Dipak and {Heavens}, Alan and {Cooray}, Asantha and {Valageas}, Patrick},
        title = "{Secondary non-Gaussianity and cross-correlation analysis}",
      journal = {\mnras},
         year = 2011,
        month = jul,
       volume = {414},
       number = {4},
        pages = {3173-3197},
          doi = {10.1111/j.1365-2966.2011.18616.x},
archivePrefix = {arXiv},
       eprint = {0907.3229},
 primaryClass = {astro-ph.CO},
       adsurl = {https://ui.adsabs.harvard.edu/abs/2011MNRAS.414.3173M}
}

@article{Schmittfull:2014tca,
    author = "Schmittfull, Marcel and Baldauf, Tobias and Seljak, Uro{\v{s}}",
    title = "{Near optimal bispectrum estimators for large-scale structure}",
    eprint = "1411.6595",
    archivePrefix = "arXiv",
    primaryClass = "astro-ph.CO",
    doi = "10.1103/PhysRevD.91.043530",
    journal = "Phys. Rev. D",
    volume = "91",
    number = "4",
    pages = "043530",
    year = "2015"
}

@article{Schmittfull:2020hoi,
    author = "Schmittfull, Marcel and Moradinezhad Dizgah, Azadeh",
    title = "{Galaxy skew-spectra in redshift-space}",
    eprint = "2010.14267",
    archivePrefix = "arXiv",
    primaryClass = "astro-ph.CO",
    doi = "10.1088/1475-7516/2021/03/020",
    journal = "JCAP",
    volume = "03",
    pages = "020",
    year = "2021"
}

@article{Hou:2022rcd,
    author = "Hou, Jiamin and Moradinezhad Dizgah, Azadeh and Hahn, ChangHoon and Massara, Elena",
    title = "{Cosmological information in skew spectra of biased tracers in redshift space}",
    eprint = "2210.12743",
    archivePrefix = "arXiv",
    primaryClass = "astro-ph.CO",
    doi = "10.1088/1475-7516/2023/03/045",
    journal = "JCAP",
    volume = "03",
    pages = "045",
    year = "2023"
}

@article{Hou:2024blc,
    author = "Hou, Jiamin and Moradinezhad Dizgah, Azadeh and Hahn, ChangHoon and Eickenberg, Michael and Ho, Shirley and Lemos, Pablo and Massara, Elena and Modi, Chirag and Parker, Liam and Blancard, Bruno R{\'e}galdo-Saint",
    title = "{Cosmological constraints from the redshift-space galaxy skew spectra}",
    eprint = "2401.15074",
    archivePrefix = "arXiv",
    primaryClass = "astro-ph.CO",
    doi = "10.1103/PhysRevD.109.103528",
    journal = "Phys. Rev. D",
    volume = "109",
    number = "10",
    pages = "103528",
    year = "2024"
}

@article{Munshi:2020ofi,
    author = "Munshi, D. and Namikawa, T. and Kitching, T. D. and McEwen, J. D. and Bouchet, F. R.",
    title = "{Weak Lensing Skew-Spectrum}",
    eprint = "2006.12832",
    archivePrefix = "arXiv",
    primaryClass = "astro-ph.CO",
    doi = "10.1093/mnras/staa2769",
    journal = "Mon. Not. Roy. Astron. Soc.",
    volume = "498",
    number = "4",
    pages = "6057--6068",
    year = "2020"
}

@article{Munshi:2020nfh,
    author = "Munshi, D. and McEwen, J. D.",
    title = "{Higher order spectra of weak lensing convergence maps in parametrized theories of modified gravity}",
    eprint = "2004.07021",
    archivePrefix = "arXiv",
    primaryClass = "astro-ph.CO",
    doi = "10.1093/mnras/staa2706",
    journal = "Mon. Not. Roy. Astron. Soc.",
    volume = "498",
    number = "4",
    pages = "5299--5316",
    year = "2020"
}

@article{MoradinezhadDizgah:2019xun,
    author = "Moradinezhad Dizgah, Azadeh and Lee, Hayden and Schmittfull, Marcel and Dvorkin, Cora",
    title = "{Capturing non-Gaussianity of the large-scale structure with weighted skew-spectra}",
    eprint = "1911.05763",
    archivePrefix = "arXiv",
    primaryClass = "astro-ph.CO",
    doi = "10.1088/1475-7516/2020/04/011",
    journal = "JCAP",
    volume = "04",
    pages = "011",
    year = "2020"
}

@article{Munshi:2014tua,
    author = "Munshi, D. and Hu, B. and Renzi, A. and Heavens, A. and Coles, P.",
    title = "{Probing Modified Gravity Theories with ISW and CMB Lensing}",
    eprint = "1403.0852",
    archivePrefix = "arXiv",
    primaryClass = "astro-ph.CO",
    doi = "10.1093/mnras/stu916",
    journal = "Mon. Not. Roy. Astron. Soc.",
    volume = "442",
    number = "1",
    pages = "821--837",
    year = "2014"
}

@article{Chakraborty:2022aok,
    author = "Chakraborty, Priyesh and Chen, Shu-Fan and Dvorkin, Cora",
    title = "{Skewing the CMB{\texttimes}LSS: a fast method for bispectrum analysis}",
    eprint = "2202.11724",
    archivePrefix = "arXiv",
    primaryClass = "astro-ph.CO",
    doi = "10.1088/1475-7516/2022/07/038",
    journal = "JCAP",
    volume = "07",
    number = "07",
    pages = "038",
    year = "2022"
}

@article{Timmons:2017xwv,
    author = "Timmons, Nicholas and Cooray, Asantha and Feng, Chang and Keating, Brian",
    title = "{Halo Pressure Profile through the Skew Cross-power Spectrum of the Sunyaev{\textendash}Zel{\textquoteright}dovich Effect and CMB Lensing in $\textit{Planck}$}",
    eprint = "1707.02298",
    archivePrefix = "arXiv",
    primaryClass = "astro-ph.CO",
    doi = "10.3847/2041-8213/aa89e8",
    journal = "Astrophys. J. Lett.",
    volume = "849",
    number = "1",
    pages = "L6",
    year = "2017"
}

@phdthesis{kou:tel-04732792,
  author = {Kou, Rapha{\"e}l},
  title  = {{Cosmic microwave background observations for large-scale structure studies}},
  school = {Universit{\'e} Paris Cit{\'e}, \href{https://theses.hal.science/tel-04732792}{tel-04732792}},
  year   = {2023}
}

@article{Harscouet:2024hnq,
    author = "Harscouet, Lea and Cowell, Jessica A. and Ereza, Julia and Alonso, David and Camacho, Hugo and Nicola, Andrina and Slosar, An{\v{z}}e",
    title = "{Fast Projected Bispectra: the filter-square approach}",
    eprint = "2409.07980",
    archivePrefix = "arXiv",
    primaryClass = "astro-ph.CO",
    doi = "10.33232/001c.128309",
    month = "9",
    year = "2024"
}

@article{Kou:2025hvg,
    author = {Kou, Rapha{\"e}l and Lewis, Antony},
    title = "{Improving CMB constraints on early Universe physics with LSS: a multi-probe forecast including cross-covariance}",
    eprint = "2504.13913",
    archivePrefix = "arXiv",
    primaryClass = "astro-ph.CO",
    doi = "10.1088/1475-7516/2025/08/031",
    journal = "JCAP",
    volume = "08",
    pages = "031",
    year = "2025"
}

@article{SimonsObservatory:2018koc,
    author = "Ade, Peter and others",
    collaboration = "Simons Observatory",
    title = "{The Simons Observatory: Science goals and forecasts}",
    eprint = "1808.07445",
    archivePrefix = "arXiv",
    primaryClass = "astro-ph.CO",
    doi = "10.1088/1475-7516/2019/02/056",
    journal = "JCAP",
    volume = "02",
    pages = "056",
    year = "2019"
}

@article{SimonsObservatory:2025wwn,
    author = "Abitbol, M. and others",
    collaboration = "Simons Observatory",
    title = "{The Simons Observatory: science goals and forecasts for the enhanced Large Aperture Telescope}",
    eprint = "2503.00636",
    archivePrefix = "arXiv",
    primaryClass = "astro-ph.IM",
    reportNumber = "FERMILAB-PUB-25-0188-PPD",
    doi = "10.1088/1475-7516/2025/08/034",
    journal = "JCAP",
    volume = "08",
    pages = "034",
    year = "2025"
}

@article{Lewis:1999bs,
    author = "Lewis, Antony and Challinor, Anthony and Lasenby, Anthony",
    title = "{Efficient computation of CMB anisotropies in closed FRW models}",
    eprint = "astro-ph/9911177",
    archivePrefix = "arXiv",
    doi = "10.1086/309179",
    journal = "Astrophys. J.",
    volume = "538",
    pages = "473--476",
    year = "2000"
}

@ARTICLE{Howlett12,
       author = {{Howlett}, Cullan and {Lewis}, Antony and {Hall}, Alex and {Challinor}, Anthony},
        title = "{CMB power spectrum parameter degeneracies in the era of precision cosmology}",
      journal = {\jcap},
         year = 2012,
        month = apr,
       volume = {2012},
       number = {4},
          eid = {027},
        pages = {027},
          doi = {10.1088/1475-7516/2012/04/027},
archivePrefix = {arXiv},
       eprint = {1201.3654},
 primaryClass = {astro-ph.CO},
       adsurl = {https://ui.adsabs.harvard.edu/abs/2012JCAP...04..027H}
}

@article{Lewis:2026mif,
    author = "Lewis, Antony",
    title = "{CAMB v2: cosmological power spectra for high-precision surveys}",
    eprint = "2607.14854",
    archivePrefix = "arXiv",
    primaryClass = "astro-ph.CO",
    month = "7",
    year = "2026"
}

@article{Albrecht:2006um,
    author = "Albrecht, Andreas and others",
    title = "{Report of the Dark Energy Task Force}",
    eprint = "astro-ph/0609591",
    archivePrefix = "arXiv",
    reportNumber = "FERMILAB-FN-0793-A",
    month = "9",
    year = "2006"
}

@article{Eggemeier:2018qae,
    author = "Eggemeier, Alexander and Scoccimarro, Roman and Smith, Robert E.",
    title = "{Bias Loop Corrections to the Galaxy Bispectrum}",
    eprint = "1812.03208",
    archivePrefix = "arXiv",
    primaryClass = "astro-ph.CO",
    doi = "10.1103/PhysRevD.99.123514",
    journal = "Phys. Rev. D",
    volume = "99",
    number = "12",
    pages = "123514",
    year = "2019"
}

@article{Lazeyras:2015lgp,
    author = "Lazeyras, Titouan and Wagner, Christian and Baldauf, Tobias and Schmidt, Fabian",
    title = "{Precision measurement of the local bias of dark matter halos}",
    eprint = "1511.01096",
    archivePrefix = "arXiv",
    primaryClass = "astro-ph.CO",
    doi = "10.1088/1475-7516/2016/02/018",
    journal = "JCAP",
    volume = "02",
    pages = "018",
    year = "2016"
}

@article{Baldauf:2012hs,
    author = "Baldauf, Tobias and Seljak, Uros and Desjacques, Vincent and McDonald, Patrick",
    title = "{Evidence for Quadratic Tidal Tensor Bias from the Halo Bispectrum}",
    eprint = "1201.4827",
    archivePrefix = "arXiv",
    primaryClass = "astro-ph.CO",
    doi = "10.1103/PhysRevD.86.083540",
    journal = "Phys. Rev. D",
    volume = "86",
    pages = "083540",
    year = "2012"
}

@article{Chan:2012jj,
    author = "Chan, Kwan Chuen and Scoccimarro, Roman and Sheth, Ravi K.",
    title = "{Gravity and Large-Scale Non-local Bias}",
    eprint = "1201.3614",
    archivePrefix = "arXiv",
    primaryClass = "astro-ph.CO",
    doi = "10.1103/PhysRevD.85.083509",
    journal = "Phys. Rev. D",
    volume = "85",
    pages = "083509",
    year = "2012"
}

@article{Saito:2014qha,
    author = "Saito, Shun and Baldauf, Tobias and Vlah, Zvonimir and Seljak, Uro{\v{s}} and Okumura, Teppei and McDonald, Patrick",
    title = "{Understanding higher-order nonlocal halo bias at large scales by combining the power spectrum with the bispectrum}",
    eprint = "1405.1447",
    archivePrefix = "arXiv",
    primaryClass = "astro-ph.CO",
    reportNumber = "IPMU14-0105",
    doi = "10.1103/PhysRevD.90.123522",
    journal = "Phys. Rev. D",
    volume = "90",
    number = "12",
    pages = "123522",
    year = "2014"
}

@article{DES:2020yyz,
    author = "Pandey, S. and others",
    collaboration = "DES",
    title = "{Perturbation theory for modeling galaxy bias: validation with simulations of the Dark Energy Survey}",
    eprint = "2008.05991",
    archivePrefix = "arXiv",
    primaryClass = "astro-ph.CO",
    reportNumber = "DES-2019-0458, FERMILAB-PUB-20-421-AE",
    doi = "10.1103/PhysRevD.102.123522",
    journal = "Phys. Rev. D",
    volume = "102",
    pages = "123522",
    year = "2020"
}

@article{Takada:2013wfa,
    author = "Takada, Masahiro and Hu, Wayne",
    title = "{Power Spectrum Super-Sample Covariance}",
    eprint = "1302.6994",
    archivePrefix = "arXiv",
    primaryClass = "astro-ph.CO",
    doi = "10.1103/PhysRevD.87.123504",
    journal = "Phys. Rev. D",
    volume = "87",
    number = "12",
    pages = "123504",
    year = "2013"
}

@article{Lacasa:2016yva,
    author = "Lacasa, Fabien and Lima, Marcos and Aguena, Michel",
    title = "{Super-sample covariance approximations and partial sky coverage}",
    eprint = "1612.05958",
    archivePrefix = "arXiv",
    primaryClass = "astro-ph.CO",
    doi = "10.1051/0004-6361/201630281",
    journal = "Astron. Astrophys.",
    volume = "611",
    pages = "A83",
    year = "2018"
}

@article{Beauchamps:2021fhb,
    author = "Beauchamps, S. Gouyou and Lacasa, F. and Tutusaus, I. and Aubert, M. and Baratta, P. and Gorce, A. and Sakr, Z.",
    title = "{Impact of survey geometry and super-sample covariance on future photometric galaxy surveys}",
    eprint = "2109.02308",
    archivePrefix = "arXiv",
    primaryClass = "astro-ph.CO",
    doi = "10.1051/0004-6361/202142052",
    journal = "Astron. Astrophys.",
    volume = "659",
    pages = "A128",
    year = "2022"
}

@article{Euclid:2023ove,
    author = "Sciotti, D. and others",
    collaboration = "Euclid",
    title = "{Euclid preparation - LII. Forecast impact of super-sample covariance on 3\texttimes{}2pt analysis with Euclid}",
    eprint = "2310.15731",
    archivePrefix = "arXiv",
    primaryClass = "astro-ph.CO",
    doi = "10.1051/0004-6361/202348389",
    journal = "Astron. Astrophys.",
    volume = "691",
    pages = "A318",
    year = "2024"
}

@article{Euclid:2025hlc,
    author = "Finelli, F. and others",
    collaboration = "Euclid",
    title = "{Euclid preparation - XCIII. Expected constraints on initial conditions}",
    eprint = "2507.15819",
    archivePrefix = "arXiv",
    primaryClass = "astro-ph.CO",
    doi = "10.1051/0004-6361/202556600",
    journal = "Astron. Astrophys.",
    volume = "712",
    pages = "A166",
    year = "2026"
}

@article{Euclid:2026tee,
    author = "Linde, D. and others",
    collaboration = "Euclid",
    title = "{Euclid preparation: Testing multi-field inflation with galaxy power spectrum and bispectrum}",
    eprint = "2605.21436",
    archivePrefix = "arXiv",
    primaryClass = "astro-ph.CO",
    month = "5",
    year = "2026"
}

@article{Randrianjanahary:2026ozt,
    author = "Randrianjanahary, Liantsoa F. and Dash, Chandrachud B. V.",
    title = "{Exploring Beyond {\ensuremath{\Lambda}}CDM with the Weak Lensing Power Spectrum and Bispectrum}",
    eprint = "2603.12504",
    archivePrefix = "arXiv",
    primaryClass = "astro-ph.CO",
    month = "3",
    year = "2026"
}

@article{Tegmark:1996bz,
    author = "Tegmark, Max and Taylor, Andy and Heavens, Alan",
    title = "{Karhunen-Loeve eigenvalue problems in cosmology: How should we tackle large data sets?}",
    eprint = "astro-ph/9603021",
    archivePrefix = "arXiv",
    doi = "10.1086/303939",
    journal = "Astrophys. J.",
    volume = "480",
    pages = "22",
    year = "1997"
}

@article{Chudaykin:2019ock,
    author = "Chudaykin, Anton and Ivanov, Mikhail M.",
    title = "{Measuring neutrino masses with large-scale structure: Euclid forecast with controlled theoretical error}",
    eprint = "1907.06666",
    archivePrefix = "arXiv",
    primaryClass = "astro-ph.CO",
    reportNumber = "INR-TH-2019-014",
    doi = "10.1088/1475-7516/2019/11/034",
    journal = "JCAP",
    volume = "11",
    pages = "034",
    year = "2019"
}

@article{Agarwal:2020lov,
    author = "Agarwal, Nishant and Desjacques, Vincent and Jeong, Donghui and Schmidt, Fabian",
    title = "{Information content in the redshift-space galaxy power spectrum and bispectrum}",
    eprint = "2007.04340",
    archivePrefix = "arXiv",
    primaryClass = "astro-ph.CO",
    doi = "10.1088/1475-7516/2021/03/021",
    journal = "JCAP",
    volume = "03",
    pages = "021",
    year = "2021"
}

@article{Ivanov:2021kcd,
    author = "Ivanov, Mikhail M. and Philcox, Oliver H. E. and Nishimichi, Takahiro and Simonovi{\'c}, Marko and Takada, Masahiro and Zaldarriaga, Matias",
    title = "{Precision analysis of the redshift-space galaxy bispectrum}",
    eprint = "2110.10161",
    archivePrefix = "arXiv",
    primaryClass = "astro-ph.CO",
    reportNumber = "YITP-21-120, CERN-TH-2021-155",
    doi = "10.1103/PhysRevD.105.063512",
    journal = "Phys. Rev. D",
    volume = "105",
    number = "6",
    pages = "063512",
    year = "2022"
}

@article{Gil-Marin:2014sta,
    author = "Gil-Mar{\'\i}n, H{\'e}ctor and Nore{\~n}a, Jorge and Verde, Licia and Percival, Will J. and Wagner, Christian and Manera, Marc and Schneider, Donald P.",
    title = "{The power spectrum and bispectrum of SDSS DR11 BOSS galaxies {\textendash} I. Bias and gravity}",
    eprint = "1407.5668",
    archivePrefix = "arXiv",
    primaryClass = "astro-ph.CO",
    doi = "10.1093/mnras/stv961",
    journal = "Mon. Not. Roy. Astron. Soc.",
    volume = "451",
    number = "1",
    pages = "539--580",
    year = "2015"
}

@article{Linke:2020ssu,
    author = "Linke, Laila and others",
    title = "{KiDS+VIKING+GAMA: Testing semi-analytic models of galaxy evolution with galaxy{\textendash}galaxy{\textendash}galaxy lensing}",
    eprint = "2005.02419",
    archivePrefix = "arXiv",
    primaryClass = "astro-ph.CO",
    doi = "10.1051/0004-6361/202038355",
    journal = "Astron. Astrophys.",
    volume = "640",
    pages = "A59",
    year = "2020"
}

@article{Shi:2010yu,
    author = "Shi, X. and Joachimi, B. and Schneider, P.",
    title = "{Controlling intrinsic-shear alignment in three-point weak lensing statistics}",
    eprint = "1002.0693",
    archivePrefix = "arXiv",
    primaryClass = "astro-ph.CO",
    doi = "10.1051/0004-6361/201014191",
    journal = "Astron. Astrophys.",
    volume = "523",
    pages = "A60",
    year = "2010"
}

@article{Troxel:2011za,
    author = "Troxel, M. A. and Ishak, Mustapha",
    title = "{Self-Calibration Technique for 3-point Intrinsic Alignment Correlations in Weak Lensing Surveys}",
    eprint = "1109.4896",
    archivePrefix = "arXiv",
    primaryClass = "astro-ph.CO",
    doi = "10.1111/j.1365-2966.2011.20205.x",
    journal = "Mon. Not. Roy. Astron. Soc.",
    volume = "419",
    pages = "1804--1823",
    year = "2012"
}

@article{Troxel:2012mx,
    author = "Troxel, M. A. and Ishak, Mustapha",
    title = "{Self-Calibration for 3-point Intrinsic Alignment Auto-Correlations in Weak Lensing Surveys}",
    eprint = "1203.2138",
    archivePrefix = "arXiv",
    primaryClass = "astro-ph.CO",
    doi = "10.1111/j.1365-2966.2012.20987.x",
    journal = "Mon. Not. Roy. Astron. Soc.",
    volume = "423",
    pages = "1663--1673",
    year = "2012"
}

@article{Troxel:2012pu,
    author = "Troxel, M. A. and Ishak, Mustapha",
    title = "{Self-calibrating the gravitational shear-intrinsic ellipticity-intrinsic ellipticity (GII) cross-correlation}",
    eprint = "1205.1547",
    archivePrefix = "arXiv",
    primaryClass = "astro-ph.CO",
    doi = "10.1111/j.1365-2966.2012.21912.x",
    journal = "Mon. Not. Roy. Astron. Soc.",
    volume = "427",
    pages = "442--457",
    year = "2012"
}

@article{Semboloni:2008xa,
    author = "Semboloni, Elisabetta and Heymans, Catherine and van Waerbeke, Ludovic and Schneider, Peter",
    title = "{Sources of contamination to weak lensing three-point statistics: constraints from N-body simulations}",
    eprint = "0802.3978",
    archivePrefix = "arXiv",
    primaryClass = "astro-ph",
    doi = "10.1111/j.1365-2966.2008.13478.x",
    journal = "Mon. Not. Roy. Astron. Soc.",
    volume = "388",
    pages = "991",
    year = "2008"
}

@article{Merkel:2014wma,
    author = {Merkel, Philipp M. and Sch{\"a}fer, Bj{\"o}rn Malte},
    title = "{A theoretical estimate of intrinsic ellipticity bispectra induced by angular momenta alignments}",
    eprint = "1409.5197",
    archivePrefix = "arXiv",
    primaryClass = "astro-ph.CO",
    doi = "10.1093/mnras/stu1945",
    journal = "Mon. Not. Roy. Astron. Soc.",
    volume = "445",
    number = "3",
    pages = "2918--2929",
    year = "2014"
}

@article{Takahashi:2019hth,
    author = "Takahashi, Ryuichi and Nishimichi, Takahiro and Namikawa, Toshiya and Taruya, Atsushi and Kayo, Issha and Osato, Ken and Kobayashi, Yosuke and Shirasaki, Masato",
    title = "{Fitting the nonlinear matter bispectrum by the Halofit approach}",
    eprint = "1911.07886",
    archivePrefix = "arXiv",
    primaryClass = "astro-ph.CO",
    doi = "10.3847/1538-4357/ab908d",
    journal = "Astrophys. J.",
    volume = "895",
    number = "2",
    pages = "113",
    year = "2020"
}

@article{Doux:2020duk,
    author = "Doux, C. and others",
    collaboration = "DES",
    title = "{Consistency of cosmic shear analyses in harmonic and real space}",
    eprint = "2011.06469",
    archivePrefix = "arXiv",
    primaryClass = "astro-ph.CO",
    doi = "10.1093/mnras/stab661",
    journal = "Mon. Not. Roy. Astron. Soc.",
    volume = "503",
    number = "3",
    pages = "3796--3817",
    year = "2021"
}

@article{DES:2022qpf,
    author = "Doux, C. and others",
    collaboration = "DES",
    title = "{Dark Energy Survey Year 3 results: cosmological constraints from the analysis of cosmic shear in harmonic space}",
    eprint = "2203.07128",
    archivePrefix = "arXiv",
    primaryClass = "astro-ph.CO",
    doi = "10.1093/mnras/stac1826",
    journal = "Mon. Not. Roy. Astron. Soc.",
    year = "2022"
}

@article{Planck:2018vyg,
    author = "Aghanim, N. and others",
    collaboration = "Planck",
    title = "{Planck 2018 results. VI. Cosmological parameters}",
    eprint = "1807.06209",
    archivePrefix = "arXiv",
    primaryClass = "astro-ph.CO",
    doi = "10.1051/0004-6361/201833910",
    journal = "Astron. Astrophys.",
    volume = "641",
    pages = "A6",
    year = "2020",
    note = "[Erratum: Astron.Astrophys. 652, C4 (2021)]"
}

@article{Mossa:2020gjc,
    author = "Mossa, V. and others",
    title = "{The baryon density of the Universe from an improved rate of deuterium burning}",
    doi = "10.1038/s41586-020-2878-4",
    journal = "Nature",
    volume = "587",
    number = "7833",
    pages = "210--213",
    year = "2020"
}

@article{Matarrese:1997sk,
    author = "Matarrese, Sabino and Verde, Licia and Heavens, Alan F.",
    title = "{Large-scale bias in the universe: bispectrum method}",
    eprint = "astro-ph/9706059",
    archivePrefix = "arXiv",
    doi = "10.1093/mnras/290.4.651",
    journal = "Mon. Not. Roy. Astron. Soc.",
    volume = "290",
    pages = "651--662",
    year = "1997"
}

@article{Oddo:2019run,
    author = "Oddo, Andrea and Sefusatti, Emiliano and Porciani, Cristiano and Monaco, Pierluigi and S\'anchez, Ariel G.",
    title = "{Toward a robust inference method for the galaxy bispectrum: likelihood function and model selection}",
    eprint = "1908.01774",
    archivePrefix = "arXiv",
    primaryClass = "astro-ph.CO",
    journal = "JCAP",
    volume = "03",
    pages = "056",
    year = "2020"
}

@article{Vlah:2019byq,
    author = "Vlah, Zvonimir and Chisari, Nora Elisa and Schmidt, Fabian",
    title = "{An EFT description of galaxy intrinsic alignments}",
    eprint = "1910.08085",
    archivePrefix = "arXiv",
    primaryClass = "astro-ph.CO",
    doi = "10.1088/1475-7516/2020/01/025",
    journal = "JCAP",
    volume = "01",
    pages = "025",
    year = "2020"
}

@article{Vlah:2020ovg,
    author = "Vlah, Zvonimir and Chisari, Nora Elisa and Schmidt, Fabian",
    title = "{Galaxy shape statistics in the effective field theory}",
    eprint = "2012.04114",
    archivePrefix = "arXiv",
    primaryClass = "astro-ph.CO",
    doi = "10.1088/1475-7516/2021/05/061",
    journal = "JCAP",
    volume = "05",
    pages = "061",
    year = "2021"
}

@article{Bakx:2025siw,
    author = "Bakx, Thomas and Eggemeier, Alexander and Kurita, Toshiki and Chisari, Nora Elisa and Vlah, Zvonimir",
    title = "{The Bispectrum of Intrinsic Alignments: Theory Modelling and Forecasts for Stage IV Galaxy Surveys}",
    eprint = "2504.10009",
    archivePrefix = "arXiv",
    primaryClass = "astro-ph.CO",
    doi = "10.33232/001c.144388",
    month = "4",
    year = "2025"
}

@article{LoVerde:2008re,
    author = "LoVerde, Marilena and Afshordi, Niayesh",
    title = "{Extended Limber Approximation}",
    eprint = "0809.5112",
    archivePrefix = "arXiv",
    primaryClass = "astro-ph",
    doi = "10.1103/PhysRevD.78.123506",
    journal = "Phys. Rev. D",
    volume = "78",
    pages = "123506",
    year = "2008"
}

\appendix
\section{Intrinsic alignment kernels in the TATT model}
\label{appendix:IA_kernels}

In this appendix, we derive the linear and quadratic intrinsic-alignment kernels entering the tree-level bispectrum within the tidal alignment and tidal torquing (TATT) model. We first derive the E-mode intrinsic shear field from the three-dimensional intrinsic-shape tensor. We then compute separately the contributions proportional to the coefficients $C_1$, $C_{1\delta}$ and $C_2$ to the bispectrum $B_{\delta\delta\gamma_E}$, before combining them into the final quadratic kernel quoted in section~\ref{sec:bispectra}.
The TATT model describes the intrinsic shape tensor as
\begin{align}
    \gamma^I_{ij}(\mathbf{x})
    =
    C_1 s_{ij}
    + C_{1\delta}\,\delta s_{ij}
    + C_2
    \left(
        s_{ik}s_{kj}
        -\frac13\delta^\mathrm{K}_{ij}s^2
    \right),
\end{align}
where $s_{ij}$ denotes the tidal tensor. The observable quantity is obtained by projecting this three-dimensional tensor onto the plane of the sky. Introducing the usual $+$ and $\times$ shear components,
\begin{align}
    \gamma_+(\mathbf{k})
    &= \gamma^I_{xx}(\mathbf{k})-\gamma^I_{yy}(\mathbf{k}),\\
    \gamma_\times(\mathbf{k})
    &=2\gamma^I_{xy}(\mathbf{k}),
\end{align}
the corresponding E- and B-mode fields are given by~\cite{Blazek:2017wbz}
\begin{align}
    \gamma^I_E(\mathbf{k})
    &=
    p^{-1}(\hat{\mathbf{k}})
    \left[
        f_E(\hat{\mathbf{k}})
        \gamma_+(\mathbf{k})
        +
        f_B(\hat{\mathbf{k}})
        \gamma_\times(\mathbf{k})
    \right],\\
    \gamma^I_B(\mathbf{k})
    &=
    p^{-1}(\hat{\mathbf{k}})
    \left[
        -f_B(\hat{\mathbf{k}})
        \gamma_+(\mathbf{k})
        +
        f_E(\hat{\mathbf{k}})
        \gamma_\times(\mathbf{k})
    \right],
\end{align}
where
\begin{align}
    f_E(\hat{\mathbf{k}})
    &=\hat{k}_x^2-\hat{k}_y^2,\\
    f_B(\hat{\mathbf{k}})
    &=2\hat{k}_x\hat{k}_y,\\
    p(\hat{\mathbf{k}})
    &=1-\hat{k}_z^2.
\end{align}
Throughout this appendix we restrict ourselves to the E-mode component. Under the Limber approximation, only transverse Fourier modes contribute, implying $\hat{k}_z=0$ and therefore $p(\hat{\mathbf{k}})=1$. The observable intrinsic shear then reduces to
\begin{align}
    \gamma^I_E(\mathbf{k})
    =
    f_E(\hat{\mathbf{k}})
    \gamma_+(\mathbf{k})
    +
    f_B(\hat{\mathbf{k}})
    \gamma_\times(\mathbf{k}).
    \label{eq:E_modes_IA}
\end{align}
We first derive the bispectrum involving two matter overdensity fields and one intrinsic-alignment field. We consider successively the three operators entering the TATT model.

\subsection{Linear alignment contribution}

The linear alignment model corresponds to
\begin{align}
    \gamma_{ij}^I(\mathbf{x}) = C_1 s_{ij}(\mathbf{x}),
\end{align}
which, using the Fourier-space expression  of $s_{ij}(\mathbf{k})$ gives
\begin{align}
    \gamma_{ij}^I(\mathbf{k})
    =
    C_1
    \left(
        \hat{k}_i\hat{k}_j
        -\frac13\delta^\mathrm{K}_{ij}
    \right)
    \delta(\mathbf{k}).
\end{align}
Using eq.~\eqref{eq:E_modes_IA}, the corresponding E-mode field is
\begin{align}
    \gamma^I_E(\mathbf{k})
    &=
    C_1
    \left[
        (\hat{k}_x^2-\hat{k}_y^2)^2
        +
        4\hat{k}_x^2\hat{k}_y^2
    \right]
    \delta(\mathbf{k}) \nonumber\\
    &=
    C_1
    (\hat{k}_x^2+\hat{k}_y^2)^2
    \delta(\mathbf{k})
    =
    C_1\delta(\mathbf{k}),
\end{align}
where we used $\hat{k}_x^2+\hat{k}_y^2=1$ for transverse Fourier modes. The intrinsic-alignment field is therefore simply proportional to the matter overdensity,
\begin{align}
    \gamma^I_E(\mathbf{k})
    =
    C_1\delta(\mathbf{k}),
\end{align}
from which it immediately follows that
\begin{align}
    \left<
        \delta(\mathbf{k}_1)
        \delta(\mathbf{k}_2)
        \gamma^I_E(\mathbf{k}_3)
    \right>
    &=
    C_1
    \left<
        \delta(\mathbf{k}_1)
        \delta(\mathbf{k}_2)
        \delta(\mathbf{k}_3)
    \right>.
\end{align}
Hence, the corresponding bispectrum is simply
\begin{align}
    B_{\delta\delta\gamma^I_E}
    (\mathbf{k}_1,\mathbf{k}_2,\mathbf{k}_3)
    =
    C_1
    B_{\delta\delta\delta}
    (\mathbf{k}_1,\mathbf{k}_2,\mathbf{k}_3).
\end{align}

\subsection{Density-weighted alignment contribution}

We now consider the second operator of the TATT model,
\begin{align}
    \gamma_{ij}^I(\mathbf{x})
    =
    C_{1\delta}\,
    \delta(\mathbf{x})\,s_{ij}(\mathbf{x}).
\end{align}
Since this operator is quadratic in the density field, its Fourier transform is given by the convolution
\begin{align}
    \gamma_{ij}^I(\mathbf{k})
    =
    C_{1\delta}
    \int\frac{d^3q}{(2\pi)^3}
    \delta(\mathbf{k}-\mathbf{q})
    \left(
        \hat{q}_i\hat{q}_j
        -\frac13\delta^\mathrm{K}_{ij}
    \right)
    \delta(\mathbf{q}).
\end{align}
Projecting onto the observable E-mode using eq.~\eqref{eq:E_modes_IA} gives
\begin{align}
    \gamma^I_E(\mathbf{k})
    =
    C_{1\delta}
    \int\frac{d^3q}{(2\pi)^3}
    \delta(\mathbf{k}-\mathbf{q})
    \delta(\mathbf{q})
    \left[
        (\hat q_x^2-\hat q_y^2)(\hat k_x^2-\hat k_y^2)
        +
        4\hat q_x\hat q_y\hat k_x\hat k_y
    \right].
\end{align}
The bispectrum then follows from
\begin{multline}
    \left<
        \delta(\mathbf{k}_1)
        \delta(\mathbf{k}_2)
        \gamma^I_E(\mathbf{k}_3)
    \right>
    =
    C_{1\delta}
    \int\frac{d^3q}{(2\pi)^3}
    \Big[
        (\hat q_x^2-\hat q_y^2)(\hat k_{3x}^2-\hat k_{3y}^2)\\
        +4\hat q_x\hat q_y\hat k_{3x}\hat k_{3y}
    \Big]
    \left<
        \delta(\mathbf{k}_1)
        \delta(\mathbf{k}_2)
        \delta(\mathbf{k}_3-\mathbf{q})
        \delta(\mathbf{q})
    \right>.
\end{multline}
At tree-level, Wick's theorem gives two connected contractions, $\mathbf{q}=-\mathbf{k}_1$ and $\mathbf{q}=-\mathbf{k}_2$, each contributing a factor $P_L(k_1)P_L(k_2)$. We hence get,
\begin{multline}
    B_{\delta\delta\gamma^I_E}
    (\mathbf{k}_1,\mathbf{k}_2,\mathbf{k}_3)
    =
    C_{1\delta}
    P_L(k_1)
    P_L(k_2)
    \Big[
        (\hat{k}_{1x}^2-\hat{k}_{1y}^2
        +
        \hat{k}_{2x}^2-\hat{k}_{2y}^2)
        (\hat{k}_{3x}^2-\hat{k}_{3y}^2)\\
        +
        4(\hat{k}_{1x}\hat{k}_{1y}
        +
        \hat{k}_{2x}\hat{k}_{2y})
        \hat{k}_{3x}\hat{k}_{3y}
    \Big].
\end{multline}

\subsection{Quadratic tidal-torquing contribution}

We finally turn to the third operator entering the TATT model,
\begin{align}
    \gamma_{ij}^I(\mathbf{x})
    =
    C_2
    \left(
        s_{ik}(\mathbf{x})s_{kj}(\mathbf{x})
        -\frac13\delta^\mathrm{K}_{ij}s^2(\mathbf{x})
    \right).
\end{align}
Since this operator is quadratic in the tidal field, its Fourier transform is also a convolution,
\begin{align}
    \gamma_{ij}^I(\mathbf{k})
    =
    C_2
    \int\frac{d^3q}{(2\pi)^3}
    \left[
        s_{ik}(\mathbf{q})s_{kj}(\mathbf{p})
        -\frac13\delta^\mathrm{K}_{ij}
        s_{kl}(\mathbf{q})s_{kl}(\mathbf{p})
    \right],
\end{align}
with $\mathbf{p}\equiv\mathbf{k}-\mathbf{q}$. The trace term $s_{kl}(\mathbf{q})s_{kl}(\mathbf{p})=\left[(\hat{\mathbf{q}}\cdot\hat{\mathbf{p}})^2-1/3\right]\delta(\mathbf{p})\delta(\mathbf{q})$ is proportional to $\delta_{ij}$ and therefore cancels in $\gamma_+=\gamma_{xx}^I-\gamma_{yy}^I$ and does not contribute to $\gamma_\times=2\gamma_{xy}^I$. Expanding the remaining piece,
\begin{align}
    s_{ik}(\mathbf{q})s_{kj}(\mathbf{p})
    =
    \left[\hat q_i(\hat{\mathbf{q}}\cdot\hat{\mathbf{p}})\hat p_j
    -\frac13\hat q_i\hat q_j
    -\frac13\hat p_i\hat p_j
    +\frac19\delta^\mathrm{K}_{ij}\right]\delta(\mathbf{p})\delta(\mathbf{q}),
\end{align}
gives
\begin{align}
    \gamma_+(\mathbf{k})
    &=
    C_2
    \int\frac{d^3q}{(2\pi)^3}
    \delta(\mathbf{q})\delta(\mathbf{p})
    \Big[
        (\hat q_x\hat p_x-\hat q_y\hat p_y)
        (\hat{\mathbf{q}}\cdot\hat{\mathbf{p}})-\frac13
        \left(
            \hat q_x^2-\hat q_y^2
            +\hat p_x^2-\hat p_y^2
        \right)
    \Big],\\
    \gamma_\times(\mathbf{k})
    &=
    2C_2
    \int\frac{d^3q}{(2\pi)^3}
    \delta(\mathbf{q})\delta(\mathbf{p})
    \Big[
        \hat q_x\hat p_y
        (\hat{\mathbf{q}}\cdot\hat{\mathbf{p}})-\frac13
        \left(
            \hat q_x\hat q_y
            +\hat p_x\hat p_y
        \right)
    \Big].
\end{align}
Projecting onto the E-mode with eq.~\eqref{eq:E_modes_IA},
\begin{multline}
    \gamma^I_E(\mathbf{k})
    =
    C_2
    \int\frac{d^3q}{(2\pi)^3}
    \delta(\mathbf{q})\delta(\mathbf{p})
    \bigg\{
    \Big[
        (\hat q_x\hat p_x-\hat q_y\hat p_y)
        (\hat{\mathbf{q}}\cdot\hat{\mathbf{p}})-\frac13
        \left(
            \hat q_x^2-\hat q_y^2
            +\hat p_x^2-\hat p_y^2
        \right)
    \Big]
    (\hat k_x^2-\hat k_y^2)\\
    +4
    \Big[
        \hat q_x\hat p_y
        (\hat{\mathbf{q}}\cdot\hat{\mathbf{p}})
        -\frac13
        \left(
            \hat q_x\hat q_y
            +\hat p_x\hat p_y
        \right)
    \Big]
    \hat k_x\hat k_y
    \bigg\}.
\end{multline}
Looking at the bispectrum $\left<\delta(\mathbf{k}_1)\delta(\mathbf{k}_2)\gamma^I_E(\mathbf{k}_3)\right>$, Wick's theorem again gives two connected contractions, such that after simplification, we obtain,
\begin{multline}
    B_{\delta\delta\gamma^I_E}(\mathbf{k}_1,\mathbf{k}_2,\mathbf{k}_3)
    =
    2C_2P_L(k_1)P_L(k_2)\\
    \times\Bigg[
        \Big(
            (\hat k_{1x}\hat k_{2x}-\hat k_{1y}\hat k_{2y})
            (\hat{\mathbf{k}}_1\cdot\hat{\mathbf{k}}_2)
            -\frac13
            \left(
                \hat k_{1x}^2-\hat k_{1y}^2
                +\hat k_{2x}^2-\hat k_{2y}^2
            \right)
        \Big)
        (\hat k_{3x}^2-\hat k_{3y}^2)\\
        +
        2
        \Big(
            (\hat k_{1x}\hat k_{2y}+\hat k_{1y}\hat k_{2x})
            (\hat{\mathbf{k}}_1\cdot\hat{\mathbf{k}}_2)-\frac23
            \left(
                \hat k_{1x}\hat k_{1y}
                +\hat k_{2x}\hat k_{2y}
            \right)
        \Big)
        \hat k_{3x}\hat k_{3y}
    \Bigg].
\end{multline}

\subsection{Combined matter--matter--intrinsic alignment bispectrum}

Combining the three contributions derived above, we obtain
\begin{multline}
    B_{\delta\delta\gamma^I_E}(\mathbf{k}_1,\mathbf{k}_2,\mathbf{k}_3)
    =
    C_1
    B_{\delta\delta\delta}
    (\mathbf{k}_1,\mathbf{k}_2,\mathbf{k}_3)\\
    +
    C_{1\delta}
    P_L(k_1)
    P_L(k_2)
    \Big[
        (\hat{k}_{1x}^2-\hat{k}_{1y}^2
        +
        \hat{k}_{2x}^2-\hat{k}_{2y}^2)
        (\hat{k}_{3x}^2-\hat{k}_{3y}^2)+4(\hat{k}_{1x}\hat{k}_{1y}
        +
        \hat{k}_{2x}\hat{k}_{2y})
        \hat{k}_{3x}\hat{k}_{3y}
    \Big]\\
    +
    2C_2P_L(k_1)P_L(k_2)\\
    \times\Bigg[
        \Big(
            (\hat k_{1x}\hat k_{2x}-\hat k_{1y}\hat k_{2y})
            (\hat{\mathbf{k}}_1\cdot\hat{\mathbf{k}}_2)
            -\frac13
            \left(
                \hat k_{1x}^2-\hat k_{1y}^2
                +\hat k_{2x}^2-\hat k_{2y}^2
            \right)
        \Big)
        (\hat k_{3x}^2-\hat k_{3y}^2)\\
        +
        2
        \Big(
            (\hat k_{1x}\hat k_{2y}+\hat k_{1y}\hat k_{2x})
            (\hat{\mathbf{k}}_1\cdot\hat{\mathbf{k}}_2)-\frac23
            \left(
                \hat k_{1x}\hat k_{1y}
                +\hat k_{2x}\hat k_{2y}
            \right)
        \Big)
        \hat k_{3x}\hat k_{3y}
    \Bigg],
\end{multline}
which we further simplify to
\begin{align}
    B_{\delta\delta\gamma^I_E}&(\mathbf{k}_1,\mathbf{k}_2,\mathbf{k}_3)
    =
    C_1
    B_{\delta\delta\delta}
    (\mathbf{k}_1,\mathbf{k}_2,\mathbf{k}_3)\nonumber\\
     +&(C_{1\delta}-\frac{2}{3}C_2)P_L(k_1)P_L(k_2)\left[(\hat{k}_{1x}^2-\hat{k}_{1y}^2+\hat{k}_{2x}^2-\hat{k}_{2y}^2)(\hat{k}_{3x}^2-\hat{k}_{3y}^2)+4(\hat{k}_{1x}\hat{k}_{1y}+\hat{k}_{2x}\hat{k}_{2y})\hat{k}_{3x}\hat{k}_{3y}\right] \nonumber\\
     +&2C_2P_L(k_1)P_L(k_2)(\hat{\mathbf{k}}_1\cdot\hat{\mathbf{k}}_2)\bigg[(\hat{k}_{1x}\hat{k}_{2x}-\hat{k}_{1y}\hat{k}_{2y})(\hat{k}_{3x}^2-\hat{k}_{3y}^2)
    +2(\hat{k}_{1x}\hat{k}_{2y}+\hat{k}_{1y}\hat{k}_{2x})\hat{k}_{3x}\hat{k}_{3y}\bigg].
\end{align}
\subsection{Expression in terms of angular cosines}

The previous expression can be written in a more compact form in terms of the cosines of the angles between wavevectors, $\mu_{ij}=\hat{\mathbf{k}}_i\cdot\hat{\mathbf{k}}_j$. Since all wavevectors lie in the plane of the sky under the Limber approximation, we write $\hat{\mathbf{k}}_i=(\cos\phi_i,\sin\phi_i,0)$, so that
\begin{equation}
    \mu_{ij}=\cos\phi_i\cos\phi_j+\sin\phi_i\sin\phi_j=\cos(\phi_i-\phi_j).
\end{equation}
Using $\hat{k}_{ix}^2-\hat{k}_{iy}^2=\cos2\phi_i$ and $2\hat{k}_{ix}\hat{k}_{iy}=\sin2\phi_i$, the angular factor multiplying $C_{1\delta}-\frac{2}{3}C_2$ becomes
\begin{align}
    (\hat{k}_{1x}^2-\hat{k}_{1y}^2+\hat{k}_{2x}^2-\hat{k}_{2y}^2)&(\hat{k}_{3x}^2-\hat{k}_{3y}^2)+4(\hat{k}_{1x}\hat{k}_{1y}+\hat{k}_{2x}\hat{k}_{2y})\hat{k}_{3x}\hat{k}_{3y} \nonumber\\
    &=(\cos2\phi_1+\cos2\phi_2)\cos2\phi_3+(\sin2\phi_1+\sin2\phi_2)\sin2\phi_3 \nonumber\\
    &=\cos2(\phi_1-\phi_3)+\cos2(\phi_2-\phi_3) \nonumber\\
    &=2\left(\mu_{13}^2+\mu_{23}^2-1\right). \label{eq:app_angle1}
\end{align}
Similarly, using $\hat{k}_{1x}\hat{k}_{2x}-\hat{k}_{1y}\hat{k}_{2y}=\cos(\phi_1+\phi_2)$ and $\hat{k}_{1x}\hat{k}_{2y}+\hat{k}_{1y}\hat{k}_{2x}=\sin(\phi_1+\phi_2)$, the angular factor multiplying $C_2$ becomes
\begin{align}
    (\hat{\mathbf{k}}_1\cdot\hat{\mathbf{k}}_2)\Big[(\hat{k}_{1x}\hat{k}_{2x}-\hat{k}_{1y}\hat{k}_{2y})&(\hat{k}_{3x}^2-\hat{k}_{3y}^2)+2(\hat{k}_{1x}\hat{k}_{2y}+\hat{k}_{1y}\hat{k}_{2x})\hat{k}_{3x}\hat{k}_{3y}\Big] \nonumber\\
    &=\cos(\phi_1-\phi_2)\left[\cos(\phi_1+\phi_2)\cos2\phi_3+\sin(\phi_1+\phi_2)\sin2\phi_3\right] \nonumber\\
    &=\cos(\phi_1-\phi_2)\,\cos(\phi_1+\phi_2-2\phi_3). \label{eq:app_angle2}
\end{align}
Writing $a=\phi_1-\phi_3$ and $b=\phi_2-\phi_3$, this product is $\cos(a-b)\cos(a+b)=\frac{1}{2}\left(\cos2a+\cos2b\right)$, so that
\begin{equation}\label{eq:app_identity}
    \cos(\phi_1-\phi_2)\,\cos(\phi_1+\phi_2-2\phi_3)=\frac{1}{2}\left[\cos2(\phi_1-\phi_3)+\cos2(\phi_2-\phi_3)\right]=\mu_{13}^2+\mu_{23}^2-1 .
\end{equation}
The two angular factors of eqs.~\eqref{eq:app_angle1} and~\eqref{eq:app_angle2} are therefore proportional to each other. Combining the two contributions, the bispectrum simplifies to
\begin{equation}\label{eq:app_Bddg_final}
    B_{\delta\delta\gamma^I_E}(\mathbf{k}_1,\mathbf{k}_2,\mathbf{k}_3)=C_1B_{\delta\delta\delta}(\mathbf{k}_1,\mathbf{k}_2,\mathbf{k}_3)+2\left(C_{1\delta}+\frac{1}{3}C_2\right)\left(\mu_{13}^2+\mu_{23}^2-1\right)P_\mathrm{L}(k_1)P_\mathrm{L}(k_2),
\end{equation}
where $\mathbf{k}_3$ is the wavevector of the intrinsic alignment field.

\subsection{Extension to the remaining bispectra}

The previous result generalises to the bispectra involving two or three intrinsic alignment fields. At tree-level, each term of the bispectrum involves the second-order part of only one of the three fields, the two other fields entering through their linear part. For intrinsic alignment fields, the latter is $\gamma^I_E=C_1\delta$, as derived for the linear alignment contribution.
Each term is therefore obtained from eq.~\eqref{eq:app_Bddg_final} by assigning the second-order part to each field in turn. The matter--intrinsic alignment--intrinsic alignment bispectrum is
\begin{multline}\label{eq:app_Bdgg}
    B_{\delta\gamma^I_E\gamma^I_E}(\mathbf{k}_1,\mathbf{k}_2,\mathbf{k}_3)=C_1^2B_{\delta\delta\delta}(\mathbf{k}_1,\mathbf{k}_2,\mathbf{k}_3) \\
    +2C_1\left(C_{1\delta}+\frac{1}{3}C_2\right)\left[\left(\mu_{13}^2+\mu_{23}^2-1\right)P_\mathrm{L}(k_1)P_\mathrm{L}(k_2)+\left(\mu_{12}^2+\mu_{23}^2-1\right)P_\mathrm{L}(k_1)P_\mathrm{L}(k_3)\right],
\end{multline}
and the intrinsic alignment auto-bispectrum is
\begin{multline}\label{eq:app_Bggg}
    B_{\gamma^I_E\gamma^I_E\gamma^I_E}(\mathbf{k}_1,\mathbf{k}_2,\mathbf{k}_3)=C_1^3B_{\delta\delta\delta}(\mathbf{k}_1,\mathbf{k}_2,\mathbf{k}_3)
    +2C_1^2\left(C_{1\delta}+\frac{1}{3}C_2\right)\Big[\left(\mu_{13}^2+\mu_{23}^2-1\right)P_\mathrm{L}(k_1)P_\mathrm{L}(k_2) \\
    +\left(\mu_{12}^2+\mu_{23}^2-1\right)P_\mathrm{L}(k_1)P_\mathrm{L}(k_3)+\left(\mu_{12}^2+\mu_{13}^2-1\right)P_\mathrm{L}(k_2)P_\mathrm{L}(k_3)\Big].
\end{multline}
Eqs.~\eqref{eq:app_Bddg_final}, \eqref{eq:app_Bdgg} and~\eqref{eq:app_Bggg} all take the form of eq.~\eqref{eq:tree_level_bispectrum}, with the linear and quadratic intrinsic alignment kernels
\begin{align}
    K_1^I(\mathbf{k}) &= C_1, \label{eq:app_K1I}\\
    K_2^I(\mathbf{k}_a,\mathbf{k}_b) &= C_1F_2(\mathbf{k}_a,\mathbf{k}_b)+\left(C_{1\delta}+\frac{1}{3}C_2\right)\left(\mu_{ac}^2+\mu_{bc}^2-1\right), \label{eq:app_K2I}
\end{align}
where $\mathbf{k}_c=-(\mathbf{k}_a+\mathbf{k}_b)$ is the wavevector of the intrinsic alignment field.

\end{document}